\documentclass[prx,onecolumn,floatfix,longbibliography,notitlepage, nofootinbib]{revtex4-1}

\usepackage[letterpaper,top=2cm,bottom=2cm,left=3cm,right=3cm,marginparwidth=1.75cm]{geometry}

\usepackage{tcolorbox}
\tcbuselibrary{breakable}
\usepackage{graphicx, color, graphpap}
\usepackage{enumitem}
\usepackage{amssymb}
\usepackage{amsthm}
\usepackage{dsfont}
\usepackage{mathtools}
\usepackage{multirow}
\usepackage[colorlinks=true,citecolor=magenta,linkcolor=blue]{hyperref}
\usepackage[T1]{fontenc}
\usepackage{bbm}
\usepackage{thmtools,thm-restate}
\usepackage{verbatim}
\usepackage{mathtools}
\usepackage{titlesec}
\usepackage{amsmath}
\usepackage{subfigure}
\usepackage[normalem]{ulem}
\usepackage{circuitikz}
\usepackage{braket}

\usepackage{listings}
\usepackage{csquotes}

\long\def\ca#1\cb{} 

\renewcommand{\geq}{\geqslant}
\renewcommand{\leq}{\leqslant}

\begin{document}

\title{Thermodynamic Linear Algebra}
\author{Maxwell Aifer, Kaelan Donatella, Max Hunter Gordon, Samuel Duffield, Thomas~Ahle, Daniel~Simpson, Gavin Crooks, Patrick J. Coles}

\affiliation{Normal Computing Corporation, New York, New York, USA}

\begin{abstract}
Linear algebra is central to many algorithms in engineering, science, and machine learning; hence, accelerating it would have tremendous economic impact. Quantum computing has been proposed for this purpose, although the resource requirements are far beyond current technological capabilities. We consider an alternative physics-based computing paradigm based on classical thermodynamics, to provide a near-term approach to accelerating linear algebra. At first sight, thermodynamics and linear algebra seem to be unrelated fields. Here, we connect solving linear algebra problems to sampling from the thermodynamic equilibrium distribution of a system of coupled harmonic oscillators. We present simple thermodynamic algorithms for solving linear systems of equations, computing matrix inverses, and computing matrix determinants. Under reasonable assumptions, we rigorously establish asymptotic speedups for our algorithms, relative to digital methods, that scale linearly in matrix dimension. Our algorithms exploit thermodynamic principles like ergodicity, entropy, and equilibration, highlighting the deep connection between these two seemingly distinct fields, and opening up algebraic applications for thermodynamic computers.
\end{abstract}

\maketitle

\section{Introduction}
Basic linear algebra primitives like solving linear systems and inverting matrices are present in many modern algorithms. Such primitives are relevant to a multitude of applications, for example optimal control of dynamic systems and resource allocation. They are also a common subroutine of many artificial intelligence (AI) algorithms, and account for a substantial portion of the time and energy costs in some cases. The most common method to perform these primitives is LU decomposition, whose time-complexity scales as $O(d^3)$. Many proposals have been made to accelerate such primitives, for example using iterative methods such as the conjugate gradient method. 
In the last decade, these primitives have been accelerated by hardware improvements, notably by graphical processing units (GPUs), fueling massive parallelization. However, the scaling of these methods is still a prohibitive factor, and obtaining a good approximate solution to a dense matrix of more than a few tens of thousand dimensions remains challenging.

Exploiting physics to solve mathematical problems is a deep idea, with much focus on solving optimization problems~\cite{vadlamani2020physics,mohseni2022ising,inagaki2016coherent}. In the context of linear algebra, much attention has been paid to quantum computers~\cite{feynman1982simulating}, since the mathematics of discrete-variable quantum mechanics matches that of linear algebra. A quantum algorithm~\cite{harrow2009quantum} to solve linear systems has been proposed, which for sparse and well-conditioned matrices scales as $\log d$. However, the resource requirements~\cite{scherer2017concrete} for this algorithm are far beyond current hardware capabilities. More generally building large-scale quantum hardware has remained difficult~\cite{preskill2018quantum}, and variational quantum algorithms for linear algebra~\cite{bravo2020variational,xu2019variational,cerezo2020variationalreview} have battled with vanishing gradient issues~\cite{mcclean2018barren,cerezo2020cost,wang2020noise}. 

Therefore, the search for alternative hardware proposals that can exploit physical dynamics to accelerate linear algebra primitives has been ongoing. Notably, memristor crossbar arrays have been of interest for accelerating matrix-vector multiplications~\cite{li2018analogue,yi2023activity}. Solving linear systems has also been the subject of analog computing approaches~\cite{huang2016evaluation}.

Recently, we defined a new class of hardware, built from stochastic, analog building blocks, which is ultimately thermodynamic in nature~\cite{coles2023thermodynamic}. (See also probabilistic-bit computers~\cite{aadit2022massively,Camsari_2019,kaiser2022life} and thermodynamic neural networks~\cite{hylton2020thermodynamic,hylton2022thermodynamic,ganesh2020rebooting,8123676,lipka2023thermodynamic} for alternative approaches to thermodynamic computing~\cite{conte2019thermodynamic}). AI applications like generative modeling are a natural fit for this thermodynamic hardware, where stochastic fluctuations are exploited to generate novel samples. 

In this work, we surprisingly show that the same thermodynamic hardware from Ref.~\cite{coles2023thermodynamic} can also be used to accelerate key primitives in linear algebra. Thermodynamics is not typically associated with linear algebra, and connecting these two fields is therefore non-trivial. 
Here, we exploit the fact that the mathematics of harmonic oscillator systems is inherently affine (i.e., linear), and hence we can map linear algebraic primitives onto such systems. (See also Ref.~\cite{babbush2023exponential} for a discussion of harmonic oscillators in the context of quantum computing speedups.) We show that simply by sampling from the thermal equilibrium distribution of coupled harmonic oscillators, one can solve a variety of linear algebra problems.

Specifically we develop thermodynamic algorithms for the following linear algebraic primitives: (i) solving a linear system $Ax = b$, (ii) estimating a matrix inverse $A^{-1}$, (iii) solving Lyapunov equations \cite{parks1992lyapunov} of the form $A\Sigma+ \Sigma A^{\intercal} = \mathds{1}$ and (iv) estimating the determinant of a symmetric positive definite matrix $A$. We show that if implemented on thermodynamic hardware, these methods scale favorably with problem size compared to digital algorithms. Our numerical simulations corroborate our analytical scaling results and also provide evidence of the fast convergence of these primitives with the wall-clock time, with the speedup relative to digital methods getting more pronounced with increasing dimension and condition number.

We remark that there is a connection between our thermodynamic algorithms and digital Monte-Carlo (MC) algorithms that were developed for linear algebra~\cite{forsythe1950matrix,alexandrov1996comparison,okten2005solving,rosca2006monte,dimov2015new}. Namely, our algorithms can be viewed as a continuous-time version of these digital MC algorithms. However, we emphasize that the continuous time (i.e., physics-based rather than physics-inspired) nature of our algorithms is crucial for obtaining our predicted asymptotic speedup. Additionally, thermodynamic algorithms can be run on a single device \cite{duffield2023thermodynamic} whereas efficient digital MC linear algebra requires extensive parallelization \cite{alexandrov1996comparison}.

\section{Results}

\subsection{Algorithmic Scaling}

In Table~\ref{tab:runtime_scaling}, we summarize the asymptotic scaling results for our thermodynamic algorithms as compared to the best state-of-the-art (SOTA) digital methods for dense symmetric positive-definite matrices. The derivations of these results can be found in the Supplemental Information, and are based on bounds obtained for physical thermodynamic quantities, including correlation times, equilibration times, and free energy differences. As one can see from Table~\ref{tab:runtime_scaling}, an asymptotic speedup is predicted for our thermodynamic algorithms relative to the digital SOTA algorithms. Specifically, a speedup that is linear in $d$ is expected for each of the linear algebraic primitives (ignoring a possible dependence of $\kappa$ on $d$). We remark that the complexity of analog algorithms is subtle~\cite{valiant2023matrix} and depends, e.g., on assumptions of how the hardware size grows with problem size. The assumptions made to obtain our scaling results are detailed in the Methods section. In what follows, we systematically present our thermodynamic algorithms for various linear algebraic primitives.

\begin{table}[t]
    \centering
    \renewcommand{\arraystretch}{2}
    \begin{tabular}{c|c|c}
        \textbf{Problem} & \textbf{Digital SOTA}  
        & \textbf{Thermodynamic}  \\
        \hline
        Linear System &$O(\min\{d^\omega,d^2 \sqrt{\kappa}\})$ & $O( d \kappa^2\varepsilon^{-2})$\\
        Matrix Inverse &  $O(d^\omega)$  & $O( d^2 \kappa \varepsilon^{-2})$\\
        Lyapunov Equation &$O(d^3)$ &   $O( d^2 \kappa \varepsilon^{-2})$ \\
        Matrix Determinant & $O(d^\omega)$ &   $O(d \kappa  \ln(\kappa)^3 \varepsilon^{-2})$\\
    \end{tabular}
    \caption{\textbf{Comparison of asymptotic complexities of linear algebra algorithms.} Here, $d$ is the matrix dimension, $\kappa$ is the condition number, and $\varepsilon$ is the error. For our thermodynamic algorithms, the complexity depends on the dynamical regime. Here we display the overdamped dynamics which have marginally better complexity than the underdamped equivalents. For the digital SOTA, the complexity of solving symmetric, positive definite linear systems, matrix inverse, Lyapunov equation, and matrix determinant problems are respectively for algorithms based on: conjugate gradient method~\cite{shewchuk1994introduction}, fast matrix multiplication/inverse~\cite{robinson2005toward}, Bartels-Stewart algorithm~\cite{bartels1972solution}, and LUP decomposition~\cite{aho1974design}. $\omega\approx 2.3$ denotes the matrix multiplication constant.}
    \label{tab:runtime_scaling}
\end{table}

\subsection{Solving Linear Systems of Equations}\label{sec:linear_sys}

The celebrated linear systems problem is to find $x \in \mathbb{R}^d$ such that
\begin{equation}
\label{eq:linear-system-def}
    A x = b,
\end{equation}
given some invertible matrix $A \in \mathbb{R}^{d \times d}$ and nonzero $b \in \mathbb{R}^d$. We may assume without loss of generality that the matrix $A$ in Eq.~\eqref{eq:linear-system-def} is symmetric and positive definite (SPD); if $A$ is not SPD, then we may consider the system $A^\intercal A x = A^\intercal b$, whose solution $x=A^{-1} b$ is also the solution of $A x = b$. Note that this will affect the total runtime, but still allows for asymptotic scaling improvements with respect to digital methods, in some cases\footnote{Constructing an SPD system from a generic one in this way results in the squaring of the condition number, which influences performance.}. In what follows, we will therefore assume that $A$ is SPD.

\begin{figure}[t]
    \centering
    \includegraphics{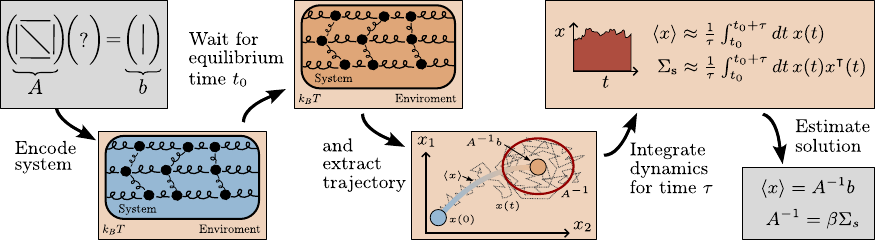}
    \caption{\textbf{Diagram of our thermodynamic algorithm for solving linear systems and inverse estimation.} The system of linear equations, or the matrix $A$, is encoded into the thermodynamic hardware, the system is then allowed to evolve until the stationary distribution has been reached, when the trajectory is then integrated to estimate the sample mean or covariance. This gives estimates of the solution of the linear system or the inverse of $A$ respectively.}
    \label{fig:LinearSystems}
\end{figure}

Now let us connect this problem to thermodynamics. We consider a macroscopic device with $d$ degrees of freedom, described by classical physics. Suppose the device has potential energy function:
\begin{equation}
    \label{eq:device-potential}
    U(x) = \frac{1}{2}x^\intercal A x - b^\intercal x,
\end{equation}
where $A \in \text{SPD}_{d}(\mathbb{R})$. Note that this is a quadratic potential that can be physically realized with a system of harmonic oscillators, where the coupling between the oscillators is determined by the matrix $A$, and the $b$ vector describes a constant force on each individual oscillator. (We remark that while Figure~\ref{fig:LinearSystems} depicts mechanical oscillators, from a practical perspective, one can build the device from electrical oscillators such as RLC circuits.)

Suppose that we allow this device to come to thermal equilibrium with its environment, whose inverse temperature is $\beta = 1/k_B T$. At thermal equilibrium, the Boltzmann distribution describes the probability for the oscillators to have a given spatial coordinate: $f(x) \propto \exp (- \beta U(x))$. Because $U(x)$ is a quadratic form, $f(x)$ corresponds to a multivariate Gaussian distribution. Thus at thermal equilibrium, the spatial coordinate $x$ is a Gaussian random variable
\begin{equation}
    \label{eq:equilibrium-dist}
    x\sim \mathcal{N}[A^{-1}b, \beta^{-1} A^{-1}].
\end{equation}

The key observation is that the unique minimum of $U(x)$ occurs where $Ax-b = 0$, which also corresponds to the unique maximum of $f(x)$. For a Gaussian distribution, the maximum of $f(x)$ is also the first moment~$\braket{x}$. Thus, we have that, at thermal equilibrium, the first moment is the solution to the linear system of equations:
\begin{equation}
    \label{eq:firstmoment}
    \braket{x} = A^{-1}b.
\end{equation}

From this analysis, we can construct a simple thermodynamic protocol for solving linear systems, which is depicted in Figure~\ref{fig:LinearSystems}. Namely, the protocol involves realizing the potential in Eq.~\eqref{eq:device-potential}, waiting for the system to come to equilibrium, and then sampling $x$ to estimate the mean $\braket{x}$ of the distribution. This mean can be approximated using a time-average, defined as
\begin{equation}
\label{eq:x-bar-def}
    \braket{x}\approx \bar{x}(\tau) = \frac{1}{\tau} \int_{t_0}^{t_0+\tau} dt' x(t'),
\end{equation}
where $t_0$ must be sufficiently large to allow for equilibration and $\tau$ must be sufficiently large for the average to converge to a desired degree of precision. The eventual convergence of this time average to the mean is the content of the ergodic hypothesis \cite{gallavotti1995ergodicity, sinai1963foundations}, which is often assumed for quite generic thermodynamic systems. It should be mentioned that the mean could also be approximated as the average of a sequence of samples; however the integration approach has the advantage that it can conveniently be implemented in a completely analog way (for example, using an integrator electrical circuit), which obviates the need for transferring data from the physical device until the end of the protocol. 

\begin{figure}
    \centering
    \includegraphics[width=0.9\linewidth]{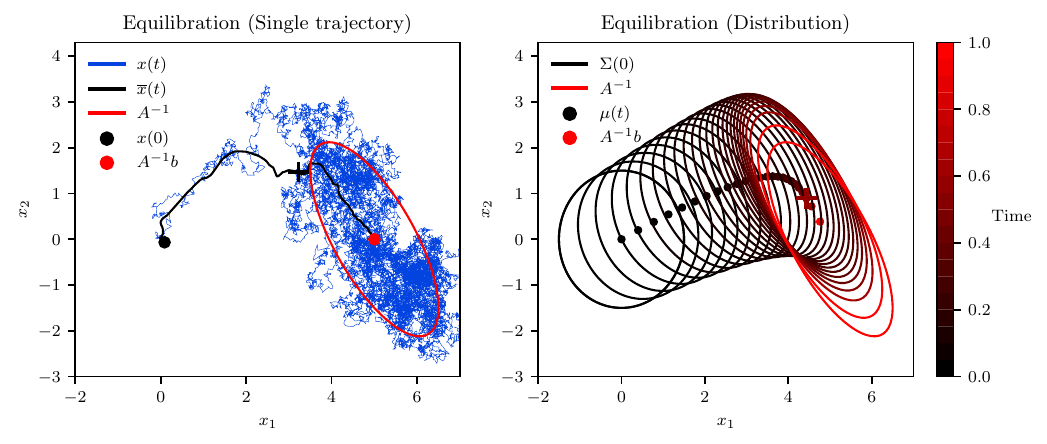}
    \caption{\textbf{Equilibration of the thermodynamic system.} The process of equilibration is depicted on the single-trajectory level (left) and on the distribution level (right). The trajectory dynamics are described by the overdamped Langevin equation and the distributional dynamics by the Fokker-Planck equation \cite{fokker1914mittlere} The system displays ergodicity, as the time average of a single trajectory (blue curve, left) approaches the ensemble average (dots, right) in the long-time limit. Time and the coordinate vector $(x_1,x_2)$ are in arbitrary units.}
    \label{fig:Ergodicity}
\end{figure}

Figure~\ref{fig:Ergodicity} shows the equilibration process for both a single trajectory (left panel) and the overall distribution (right panel). One can see the ergodic principle illustrated in this figure, since the time dynamics of a single trajectory at thermal equilibrium are representative of the overall ensemble.

The overall protocol can be summarized as follows.

\medskip
\begin{tcolorbox}[title={Linear System Protocol}, breakable]
\begin{enumerate}
\item Given a linear system $A x = b$, set the potential of the device to
\begin{equation}
    U(x) = \frac{1}{2} x^\intercal A x - b^\intercal x
\end{equation}
at time $t=0$.
\item Choose equilibration tolerance parameters $\varepsilon_{\mu0}, \varepsilon_{\Sigma 0} \in \mathbb{R}^+$, and choose the equilibration time
\begin{equation}
\label{eq:UDL-eq-time}
t_0 \geq \widehat{t}_0,
\end{equation}
where $\widehat{t}_0$ is computed from the system's physical properties or using heuristic methods based on Eqs.~\eqref{eq:t0-tau-linear-systems-odl}, \eqref{eq:t0-tau-linear-systems-udl}. Allow the system to evolve under its dynamics until $t=t_0$, which ensures that $\left\|\braket{x} - A^{-1} b\right\|/\|A^{-1} b\| \leq \varepsilon_{\mu 0}$ and $\left\|\Sigma - \beta^{-1} A^{-1} \right\|/\|\beta^{-1}A^{-1}\| \leq \varepsilon_{\Sigma0}$.
\item Choose error tolerance parameter $\varepsilon_x$ and success probability $P_\varepsilon$, and choose the integration time
\begin{equation}
    \tau \geq \widehat{\tau},
\end{equation}
where $\widehat{\tau}$ is computed from the system's physical properties, Eq.~\eqref{eq:t0-tau-linear-systems-odl} or \eqref{eq:t0-tau-linear-systems-udl}. Use an analog integrator to measure the time average
\begin{equation}
    \bar{x} = \frac{1}{\tau} \int_{t_0}^{t_0 + \tau} dt\, x(t),
\end{equation}
which satisfies $\left\|A \bar{x} - b\right\|/\|b\| \leq \varepsilon_x$ with probability at least $P_\delta$.
\end{enumerate}
\end{tcolorbox}
\medskip

In order to implement the protocol above, the necessary values of $\widehat{t}_0$ and $\widehat{\tau}$ must be identified, which requires a more quantitative description of equilibration and ergodicity. To obtain such a description, a model of the system's microscopic dynamics may be introduced. Given that the system under consideration is composed of harmonic oscillators in contact with a heat bath, it is natural to allow for damping (i.e., energy loss to the bath) and stochastic thermal noise, which always accompanies damping due to the fluctuation-dissipation theorem \cite{kubo1966fluctuation, weber1956fluctuation}. The Langevin equation accounts for these effects, and specifically we consider two common formulations, the overdamped Langevin (ODL) equation  and the underdamped Langevin (UDL) equations. In the Methods section, we provide additional details on ODL and UDL dynamics, and we provide explicit formulas for $\widehat{t}_0$ and $\widehat{\tau}$ for the overdamped and underdamped regimes.

\subsection{Estimating the Inverse of a Matrix}

The results of the previous section rely on estimating the mean of $x$, but make no use of the fluctuations in $x$ at equilibrium. By using the second moments of the equilibrium distribution, we can go beyond solving linear systems. For example it is possible to find the inverse of an SPD matrix $A$. As mentioned, the stationary distribution of $x$ is $ \mathcal{N}[A^{-1} b, \beta^{-1} A^{-1}]$, meaning the inverse of $A$ can be obtained by evaluating the covariance matrix of $x$. This can be accomplished in an entirely analog way, using a combination of analog multipliers and integrators. By setting $b=0$ for this protocol, we ensure that $\braket{x} = 0$, so the stationary covariance matrix is, by definition 
\begin{equation}
    \Sigma_\text{s} = \lim_{t \to \infty}\braket{x(t) x^\intercal(t)}.
\end{equation}
In order to estimate this, we again perform time averages after allowing the system to come to equilibrium
\begin{equation}
    \Sigma_\text{s} \approx \overline{x x^\intercal} = \frac{1}{\tau}
    \int_{t_0}^{t_0 + \tau} dt\,  x(t) x^\intercal(t).
\end{equation}
It is therefore necessary to have an analog component which evaluates the product $x_i(t) x_j(t)$ for each pair $(i,j)$, resulting in $d^2$ analog multiplier components. Each of these products is then fed into an analog integrator component, which computes one element of the time-averaged covariance matrix
\begin{equation}
    \Sigma_{\text{s},ij} \approx \frac{1}{\tau} \int_{t_0}^{t_0 + \tau}dt \, x_i(t) x_j(t).
\end{equation}
While the equilibration time is the same as for the linear system protocol, the integration time is different, because in general the covariance matrix is slower to converge than the mean. We now give a detailed description of the inverse estimation protocol, assuming ODL dynamics (the corresponding results for underdamped dynamics can be found in the Supplemental Information). In the Methods section,  we provide explicit formulas for $\widehat{t}_0$ and $\widehat{\tau}$ for the Inverse Estimation Protocol. We remark that our matrix inversion algorithm is a special case of our general algorithm for solving Lyapunov equations; the latter is presented in the Supplemental Information.
\medskip
\begin{tcolorbox}[title={Inverse Estimation Protocol},breakable]
\begin{enumerate}
\item Given a positive definite matrix $A$, set the potential of the device to
\begin{equation}
    U(x) = \frac{1}{2} x^\intercal A x
\end{equation}
at time $t=0$.
\item Choose equilibration tolerance parameter $\varepsilon_{\Sigma 0} \in \mathbb{R}^+$, and choose the equilibration time
\begin{equation}
\label{eq:UDL-eq-time}
t_0 \geq \widehat{t}_0,
\end{equation}
where $\widehat{t}_0$ is computed from the system's physical properties, Eq.~\eqref{eq:t0-tau-inverse-odl} or \eqref{eq:t0-tau-inverse-udl}. Allow the system to evolve under its dynamics until $t=t_0$, which ensures that $\left\|\Sigma - \beta^{-1} A^{-1} b\right\|/\|\beta^{-1}A^{-1}\| \leq \varepsilon_\Sigma$.
\item  Choose error tolerance parameter $\varepsilon_\Sigma$ and success probability $P_\varepsilon$, and choose the integration time
\begin{equation}
    \tau \geq \widehat{\tau},
\end{equation}
where $\widehat{\tau}$ is computed from the system's physical properties, Eq.~\eqref{eq:t0-tau-inverse-odl} or \eqref{eq:t0-tau-inverse-udl}. Use analog multipliers and integrators to measure the the time averages
\begin{equation}
    \overline{x_i x_j} = \frac{1}{\tau^2} \int_{t_0}^\tau dt\, x_i(t) x_j(t),
\end{equation}
which satisfies $\|\overline{x x^\intercal} - \beta^{-1}A^{-1}\|_F/\|\beta^{-1}A^{-1}\|_F \leq \varepsilon_A$ with probability at least $P_\varepsilon$.
\end{enumerate}
\end{tcolorbox}
\medskip

\subsection{Estimating the Determinant of a Matrix}

The determinant of the covariance matrix appears in the normalization factor of a multivariate normal distribution, whose density function is
\begin{equation}
    f_{\mu;\Sigma}(x) = (2 \pi)^{-d/2}\left| \Sigma\right|^{-1/2}
    \exp \left(-\frac{1}{2} x^\intercal \Sigma^{-1} x \right),
\end{equation}
and it is therefore natural to wonder whether hardware which is capable of preparing a Gaussian distribution may be used to somehow estimate the determinant of a matrix. This can in fact be done, as the problem is equivalent to the estimation of free energy differences, an important application of stochastic thermodynamics. Recall that the difference in free energy between equilibrium states of potentials $U_1$ and $U_2$ is \cite{christ2010basic}
\begin{equation}
    \Delta F = F_2 - F_1 = -\beta^{-1} \ln \left(\frac{\int dx \, e^{-\beta U_2(x)}}{\int dx \, e^{-\beta U_1(x)}}\right).
\end{equation}
Suppose the potentials are quadratic, with $U_1(x) = x^\intercal A_1 x$ and $U_2(x) = x^\intercal A_2 x$. Then each integral simplifies to the inverse of a Gaussian normalization factor,
\begin{equation}
    \int dx\,  e^{-\beta V_j(x)} = (2\pi)^{d/2}\sqrt{\beta^{-1}\left|A_j^{-1} \right|},
\end{equation}
so
\begin{equation}
    \Delta F = -\beta^{-1} \ln \left(\sqrt{\frac{\left| A_2^{-1}\right|}{\left| A_1^{-1}\right|}}\right) = -\beta^{-1} \ln \left(\sqrt{\frac{\left| A_1\right|}{\left| A_2\right|}}\right).
\end{equation}
This suggests that the determinant of a matrix $A_1$ can found by comparing the free energies of the equilibrium states with potentials $U_1$ and $U_2$ (where $A_2$ has known determinant), and then computing
\begin{equation}
    \left| A_1\right| = e^{-2 \beta \Delta F} \left| A_2\right|.
\end{equation}
Fortunately, the free energy difference $\Delta F$ can be found, assuming we have the ability to measure the work which is done on the system as the potential $U(x)$ is changed from $U_1$ to $U_2$. According to the Jarzynski equality \cite{jarzynski1997nonequilibrium}, the free energy difference between the (equilibrium) states in the initial and final potential is
\begin{equation}
    e^{-\beta \Delta F} = \braket{e^{-\beta W}},
\end{equation}
where $\braket{\cdot}$ denotes an average over all possible trajectories of the system between time $t=0$ and time $t=\tau$, weighed by their respective probabilities. This may be approximated by an average over $N$ repeated trials,
\begin{equation}
    e^{-\beta \Delta F}\approx \overline{e^{-\beta W}}  \equiv \frac{1}{N}\sum_{j=1}^N e^{-\beta W_j}.
\end{equation}
However, while Jarzynski's relation may be applied directly to estimate the free energy difference, this estimator has large bias and is slow to converge. Far more well-behaved estimators have been found based on work measurements. For simplicity, we here provide the expression based on Jarzynski's estimator, while in the Methods section and the Supplemental Information we refer to more suitable estimators. In summary, the determinant of $A_1$ is approximated by
\begin{equation}
\left| A_1\right| \approx \left(\overline{e^{-\beta W}}\right)^2 \left| A_2\right|.
\end{equation}
In practice we will generally be interested in the log determinant to avoid computational overflow. This is
\begin{equation}
\ln\left(\left| A_1\right|\right) \approx 2 \ln\left(\overline{e^{-\beta W}}\right)  + \ln\left( \left| A_2\right|\right).
\end{equation}
It is shown in the Supplemental Information that to estimate the log determinant to within (absolute) error $\delta_\text{LD}$ with probability at least $P_\delta$, the total amount of time required is roughly
\begin{align}
\tau \approx \frac{ d\,\ln(\kappa)^2}{\delta^2_\text{LD}(1-P_\delta)} \ln\left(\chi^2 \kappa^{3/2}\varepsilon_{\Sigma0}^{-1}\left[\frac{1}{4\zeta_\text{max}^{2}}+1\right]\right)
\tau_\text{r(UD)} = O(d \, \ln(\kappa)^3).
\end{align}
We also present numerical simulations of a protocol for determinant estimation that does not include directly measuring the work in the Supplemental Information.

\subsection{Convergence and comparison to digital algorithms}

\subsubsection{Convergence}

\begin{figure}[t]
    \centering
    \includegraphics[width=0.98\linewidth]{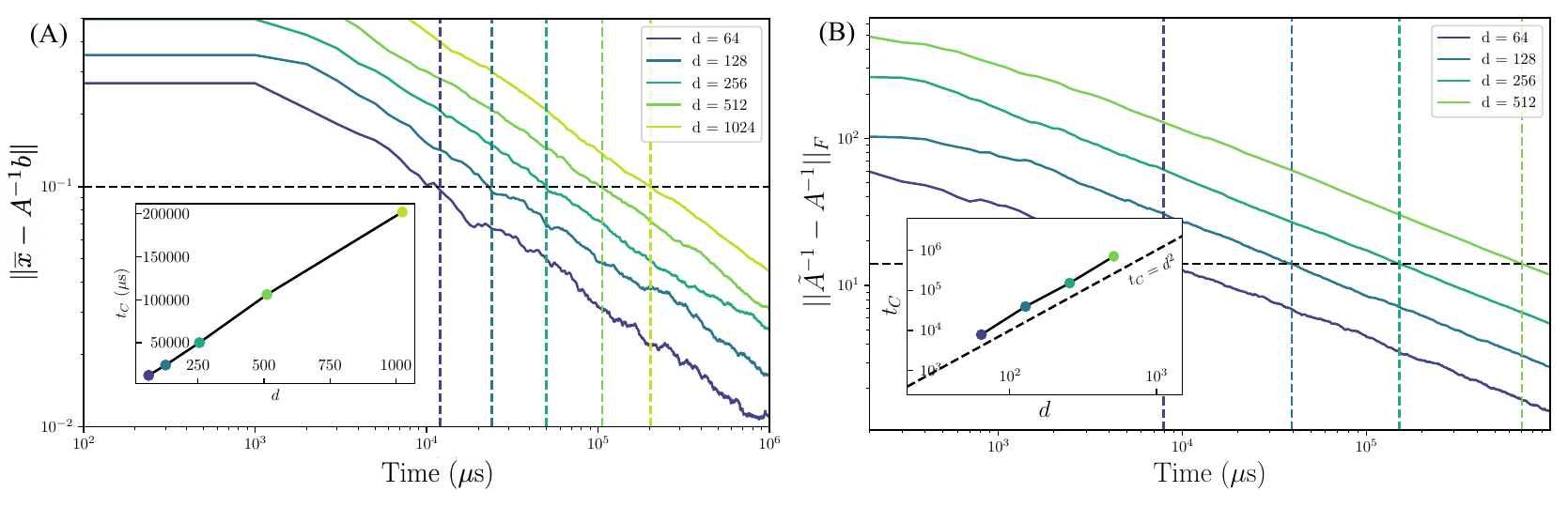}
    \caption{\textbf{Error of our thermodynamic algorithms as a function of the analog integration time for different dimensions.} Matrices $A$ are drawn from a Wishart distribution with $2d$ degrees of freedom. Vertical dashed lines are the times $t_C$ at which error goes below a threshold (horizontal dashed line). Inset: Crossing time $t_C$ as a function of dimension $d$. (A) For the linear systems algorithm, a linear relationship between dimension and the analog dynamics runtime is observed.
    (B) For the matrix inversion algorithm, a quadratic relationship between dimension and the analog dynamics runtime is observed.}
    \label{fig:ErrorVsAnalogTime}
\end{figure}

We now present several numerical experiments to corroborate our analytical results. Figure~\ref{fig:ErrorVsAnalogTime}(A) displays the convergence of the absolute error, $||\bar{x} - A^{-1}b||$ where $||.||$ denotes the 2-norm, as a function of time for our thermodynamic linear systems algorithm. This plot shows that the expected convergence time to reach a given error is linearly proportional to the dimension of the system, which is in agreement with the analytical bounds that we presented above.

Similarly, let us examine the performance of the inverse estimation protocol. We employ the absolute error on the inverse, $ \|\tilde{A}^{-1} - A^{-1}\|_F$ where $\|\cdot\|_F$ denotes the Frobenius norm. Figure~\ref{fig:ErrorVsAnalogTime}(B) shows the convergence of the error as a function of the analog dynamics time for our thermodynamic inverse estimation algorithm. We see that the expected convergence time to reach a given error is quadratic ($\propto d^2$) in the dimension, in agreement with the analytical bounds presented above.

\begin{figure}[t]
   \centering
\includegraphics[width=\linewidth]{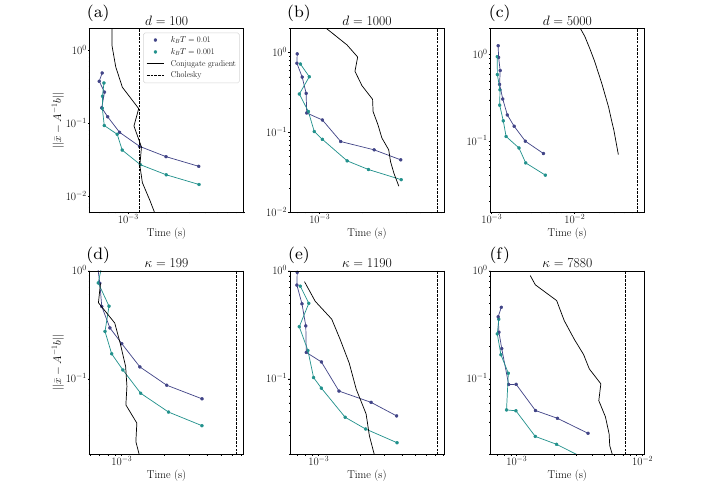}
   \caption{\textbf{Comparison of the error $||\bar{x} - A^{-1}b||$ of the thermodynamic algorithm (TA) to solve linear systems with the conjugate gradient method and Cholesky decomposition as a function of total runtime.} Panels (a)-(c): the TA is shown for different values of $k_BT$ (units of 1/$\gamma$) for each dimension in $\{100, 1000, 5000\}$. Random matrices are drawn from the Wishart distribution and then mixed with the identity such that their condition numbers are respectively 120, 1189, 5995. Panels (d)-(f): same quantities with a fixed condition number $\kappa$, respectively 199, 1190, and 7880 for fixed dimension $d = 1000$. Calculations were performed on an Nvidia RTX 6000 GPU.}
    \label{fig:linearsys_comparison_d}
\end{figure}

\subsubsection{Comparison to digital algorithms}

Another question of key importance is how the thermodynamic algorithm is expected to perform in practical scenarios, i.e., when being run on real thermodynamic hardware. Due to the hardware being analog in nature, this involves additional digital-to-analog compilation steps. To investigate this question, we consider a timing model for the thermodynamic algorithm, based on the hardware proposal described  Ref.~\cite{coles2023thermodynamic} (See the Supplemental Information for a brief summary of this hardware, whose dynamics correspond to the overdamped regime as in Eq.~\eqref{eq:ODLdynamics_main}). This model includes all the digital, digital-to-analog and analog operations needed to solve the problem, starting with a matrix $A$ stored on a digital device, and sending back the solution $x$ from the thermodynamic system to the digital device. Note that this includes a compilation step that scales as $O(d^2)$, which is absent for the digital methods\footnote{Cholesky and conjugate gradients are run on a digital computer, and the initial matrix is stored on that same computer, hence there is no transfer cost, unlike for the thermodynamic algorithm.}. Assumptions about this model are detailed in the Methods section. Note that analog imprecision is not taken into account in these experiments, and is the subject of further investigations~\cite{aifer2024error}.

Figure~\ref{fig:linearsys_comparison_d} plots the absolute error for solving linear systems as a function of time for the thermodynamic algorithm (TA), the conjugate gradient (CG) method, and the Cholesky decomposition (which is exact). In panels (a) - (c) we explore how the methods converge with varying $\kappa$ and $d$. While at low dimensions our method performs poorly with respect to the Cholesky decomposition and only slightly better than CG, it  becomes very competitive for dimensions $d = 1000$ and $d=5000$. Panels (d) - (f) show the error as a function of time for different condition numbers, at fixed dimension. One can see that as $\kappa$ grows (as conditioning is worse) our method becomes more competitive with CG. This suggests that, even in practical scenarios where we account for realistic computational overhead issues, our thermodynamic linear systems algorithm can outperform SOTA digital methods, especially for large $d$ and large $\kappa$.

Figure~\ref{fig:linearsys_comparison_d} also shows that the thermodynamic algorithm performs significantly better than the CG method at early times, although the CG method ultimately achieves a higher quality result for later times. This suggests that the thermodynamic algorithm is ideally suited to providing an approximate solution in a short amount of time. Nevertheless, we note that the effective temperature of the thermodynamic hardware is an important parameter, and one can lower this temperature to achieve higher precision solutions from the thermodynamic hardware, as can be seen from the curves in Fig.~\ref{fig:linearsys_comparison_d}.

\begin{figure}[t]
   \centering
\includegraphics[width=0.6\linewidth]{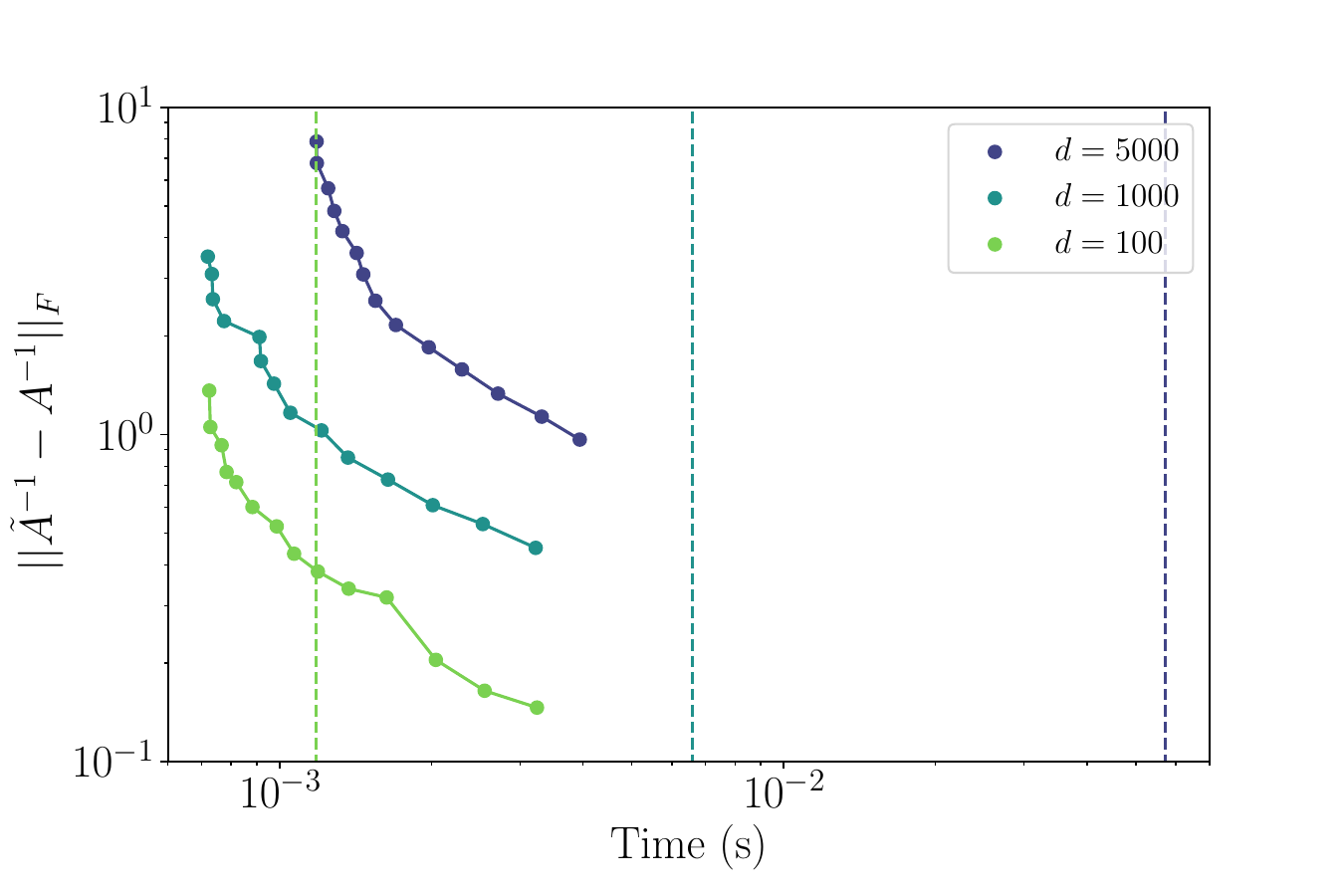}
   \caption{\textbf{Comparison of the error of the thermodynamic algorithm (TA) to invert matrices with the Cholesky decomposition as a function of total runtime.} Dimensions $d = 100, 1000, 5000$, respectively in light green, light blue, and purple, are shown for the thermodynamic algorithm (solid lines) and the Cholesky decomposition (dashed lines). Here the condition numbers are respectively $\{120, 1189, 5995\}$. Calculations were performed on an Nvidia RTX A600 GPU.}
\label{fig:inverse_estim_comparison_d}
\end{figure}

Using a timing model similar to that employed for the linear systems protocol, we performed a runtime comparison to Cholesky decomposition for the task of matrix inversion. The results are shown in Fig.~\ref{fig:inverse_estim_comparison_d}, where the error is plotted as a function of physical time for dimensions $100, 1000$, and $5000$. The dashed lines represent the corresponding times for Cholesky decomposition, for given dimensions. We see that as the dimension grows, the advantage with respect to the Cholesky decomposition also grows, thus highlighting a practical thermodynamic advantage. Our method for the inverse estimation therefore has the advantage of having well-defined convergence properties as a function of dimension and condition number (compared to other approximate methods for inverting dense matrices, which do not have well defined convergence properties), as well as leading to reasonable error values in practical settings.

Overall, these numerical experiments highlight the potential utility of thermodynamic hardware by showing the opportunity for speedup over SOTA digital methods, based on a simulated timing model of the thermodynamic device.

\section{Discussion}

Various types of physics-based computers have been devised, which are supposed to expedite calculations by using physical processes to evaluate expensive functions \cite{haensch2018next, small1993general,feynman1982simulating,preskill2018quantum, nielsen2000quantum}. These devices (which include quantum computers and a number of distinct analog architectures) have been shown to offer theoretical advantages for solving certain problems, including linear systems of equations, but they have not found common use commercially. A key obstacle to harnessing the power of physical computing is that fluctuations in the system's state tend to cause errors that compound over time, and which cannot be corrected in a straightforward way \cite{huang2017analog} (as can be done for digital computers). 

For this reason, we have considered \emph{thermodynamic} algorithms, which treat the naturally-present fluctuations as a resource, or at the very least are indifferent to them. In fact, we have introduced three distinct classes of thermodynamic algorithms: first-moment based, second-moment based, and all-moment based algorithms. Other thermodynamic algorithms will likely be discovered making use of third and higher moments, implying that such methods form a hierarchy. In some sense, using higher moments allows us to solve ``harder'' problems, for example inverting a matrix (which uses the second moments) is harder than solving a linear system of equations (which uses the first moment). Whether a precise relationship can be found between computational hardness and the hierarchy of thermodynamic algorithms is currently an open question.

Another open question concerns the optimality of these new thermodynamic algorithms. Our analysis implies that, while the time and energy costs of linear-algebraic primitives are negotiable, the product of time and energy necessary for a computation is fundamentally constrained (see Methods). It is therefore of interest to search for thermodynamic algorithms which achieve lower values of the energy-time product for these computations, and also to see whether such constraints may apply to other problems as well. We anticipate that non-equilibrium thermodynamics will be a crucial tool in exploring such resource tradeoffs for computation. For example, we have used the fact that a thermodynamic distance may be defined between equilibrium configurations of a system, and this distance determines the minimal amount of dissipated energy necessary to transition from one configuration to another in a finite time \cite{crooks2007measuring, cafaro2022thermodynamic, quevedo2007geometrothermodynamics,andresen1996finite,chen2021extrapolating}; the shorter the time of transition, the more dissipation must occur. Perhaps, then, the search for algorithms which have minimal energy-time product may be framed as a variational problem of minimizing length on the thermodynamic manifold. Although proofs of optimal algorithmic performance are notoriously hard to find in digital computing paradigms \cite{saptharishi2015survey}, unavoidable resource tradeoffs are  relatively mundane in thermodynamic analyses \cite{chiribella2022nonequilibrium, riechers2018transforming}, suggesting that computational cost may be fruitfully studied within the thermodynamic computing paradigm.

Aside from the theoretical questions mentioned, clearly the task of actually implementing our algorithms remains an important one. Recently, we created an electrical thermodynamic computing device~\cite{melanson2023thermodynamic} on a printed circuit board and used it to demonstrate the thermodynamic matrix inversion algorithm presented here, inverting 8 by 8 matrices using 8 coupled electrical oscillators. A potential next step would be to experimentally verify our predicted scaling of integration time with dimension (e.g., linear scaling for linear systems and quadratic scaling for matrix inversion), thus confirming our predicted speedup over digital methods. We anticipate that other researchers may independently seek to verify our results experimentally, leading to a rapid development of thermodynamic hardware. As a result, we predict that these methods will become appealing alternatives to digital algorithms, particularly in settings where it is desirable to trade some accuracy for better time and energy scaling.

In addition to our work's direct impact, the broader impact is laying the theoretical, mathematical foundations for the emerging paradigm of thermodynamic computing~\cite{conte2019thermodynamic}. Our work provides the first mathematical analysis, as well as the first numerical benchmarks, of potential speedups for thermodynamic hardware. Thus we have taken the somewhat vague notion of thermodynamic computing and made it concrete and precise, with a clear set of applications. Moving forward, we expect new applications to be discovered, beyond linear algebra, since one can simply modify the potential energy function $U(x)$ to solve, e.g., non-linear algebraic problems. There is also the exciting prospect of running multiple applications, such as the linear algebra ones here and the probabilistic AI ones discussed in Ref.~\cite{coles2023thermodynamic}, on the same thermodynamic hardware, providing the user with a flexible programming experience. One can envision that much of the amazing technological developments (compilers, simulators, programming languages, etc.) that have happened in quantum computing will likely happen for thermodynamic computing in the near future.

\section{Methods}

\subsection{Timing parameters for overdamped and underdamped regimes}

\subsubsection{Linear Systems}

For our linear systems algorithm, the overdamped Langevin (ODL) equation takes the form:
\begin{equation}
    \label{eq:ODLdynamics_main}
    dx = - \frac{1}{\gamma}(A x - b) dt +\mathcal{N}\left[0,2\gamma^{-1}\beta^{-1}\, dt\right],
\end{equation}
where $\gamma >0$ is called the damping constant and $\beta = 1/k_B T$ is the inverse temperature of the environment. The system has a physical timescale (which is clear from dimensional analysis) that we call the \emph{relaxation time} $\tau_\text{r} = \gamma/\|A\|$. The condition number of $A$ is $\kappa = \alpha_\text{max}/\alpha_\text{min}$, where $\alpha_1\dots\alpha_d$ are the eigenvalues of $A$. With these definitions, we arrive (see Supplemental Information for derivation) at the following formulas for $\widehat{t_0}$ and $\widehat{\tau}$ in the overdamped case, which can be used in the linear systems algorithm:
\begin{align}
\label{eq:t0-tau-linear-systems-odl}
\widehat{t}_0 =
 \max \left\{\kappa \tau_\text{r}\ln \left(\kappa \varepsilon^{-1}_{\mu 0}\right),\frac{1}{2} \kappa \tau_\text{r}\ln\left(2\kappa\varepsilon_{\Sigma 0}^{-1}\right)\right\},\quad
    \widehat{\tau} = \frac{2 \kappa^2 d\, \|A\|  }{\beta \|b\|^2 \varepsilon_x^2 (1-P_\varepsilon)}\tau_\text{r}.
\end{align}

The underdamped model is instead described by the UDL equations,
\begin{equation}
\label{eq:UDL-x}
    dx  = \frac{1}{M} p \, dt, \qquad  dp = -(A x - b) \, dt- \frac{\gamma}{M} p \, dt + \mathcal{N}[0,2 \gamma \beta^{-1}  \mathbb{I} dt].
\end{equation}
We define $\xi = \gamma/2M$, $\omega_j = \sqrt{\alpha_j/M}$, and $\zeta_j = \xi/\omega_j$. Moreover, a timescale $\tau_\text{r(UD)}$ can be identified for the underdamped system which is analogous to the quantity $\tau_\text{r}$ associated with the overdamped system. In particular, we define $\tau_{\text{r(UD)}} = \xi^{-1}$. We introduce a dimensionless quantity $\chi$ as well, which is $\chi = (1+ \xi/\omega_\text{min})^{1/2}(1- \xi/\omega_\text{min})^{-1/2}$.
With these definitions, we arrive (see Supplemental Information for derivation) at the following formulas for the timing parameters in the underdamped case:
\begin{align}
\label{eq:t0-tau-linear-systems-udl}
\widehat{t}_0 =  \max \left\{ \tau_{\text{r(UD)}} \ln \left( \kappa^{1/2}\chi \varepsilon_{\mu 0}^{-1}\right),\frac{1}{2} \tau_\text{r(UD)}\ln\left(\chi^2 \kappa^{3/2}\varepsilon_{\Sigma0}^{-1}\left[\frac{1}{4\zeta_\text{max}^{2}}+1\right]\right)\right\},\quad
    \widehat{\tau} = \frac{2\sqrt{\kappa} \chi d\|A\| }{ \beta \|b\|^2\varepsilon_x^2(1-P_\varepsilon)} \tau_{\text{r(UD)}}.
\end{align}

An important distinction between the ODL and UDL regimes is that the random variable $x$ undergoes a Markov stochastic process in the ODL case, but is non-Markovian in the UDL case \cite{doerries2021correlation,h1989colored}. The simple interpretation of this non-Markovianity is that the underdamped system exhibits inertia, which is a form of memory-dependence. This inertia has a non-trivial (and sometimes beneficial) effect on the algorithm's overall performance, which is apparent from the scaling results in Table~\ref{tab:runtime_scaling}.

\subsubsection{Matrix Inversion}

The timing parameters for the inverse estimation protocol (as derived in the Supplemental Information) are, for the overdamped case,
\begin{align}
\label{eq:t0-tau-inverse-odl}
\widehat{t}_0 =
 \frac{1}{2} \kappa \tau_\text{r}\ln\left(2\kappa\varepsilon_{\Sigma 0}^{-1}\right),\quad
    \widehat{\tau} = \frac{4 \kappa d(d+1)}{(1-P_\varepsilon)\varepsilon_\Sigma^2}\tau_\text{r},
\end{align}
and for the underdamped case
\begin{align}
\label{eq:t0-tau-inverse-udl}
\widehat{t}_0 = \frac{1}{2} \tau_\text{r(UD)}\ln\left(\chi^2 \kappa^{3/2}\varepsilon_{\Sigma0}^{-1}\left[\frac{1}{4\zeta_\text{max}^{2}}+1\right]\right),\quad
    \widehat{\tau} =\frac{4 \kappa d(d+1)}{(1-P_\varepsilon)\varepsilon_\Sigma^2}\tau_\text{r(UD)}.
\end{align}

\subsection{Energy-Time Tradeoff}

Note the appearance of the ratio $\|A\|/\|b\|^2$ in the time required to solve a linear system given by Eqs.~\eqref{eq:t0-tau-linear-systems-odl} and \eqref{eq:t0-tau-linear-systems-udl}. It is tempting to imagine that one might solve the system faster simply by multiplying $b$ by some constant $c$. Then, the time required to solve the system is apparently reduced by a factor of $c^2$, and the solution to original problem is obtained (up to a factor of $c$). A similar approach would be to multiply $A$ by a small number; of course, in practice it is not possible to solve linear systems of equations in vanishingly short periods of time, which is reflected in an energy-time tradeoff. As we explain in the Supplemental Information, there is an energy cost associated with initializing the system proportional to $b^\intercal A^{-1} b$, and this results in a re-formulation of Eqs.~\eqref{eq:t0-tau-linear-systems-odl} and \eqref{eq:t0-tau-linear-systems-udl} as lower bounds on the product of energy and time. If $\mathcal{E}$ is the energy required to solve the system $A x = b$, and $\tau$ is the necessary time, then we have, in the overdamped case:
\begin{align}
\label{eq:time-energy-odl}
    \mathcal{E}\tau &\geq \frac{2 \kappa^2 d  }{ \varepsilon_x^2 (1-P_\varepsilon)} \beta^{-1}\tau_\text{r}
\end{align}
and in the underdamped case:
\begin{align}
\label{time-energy-udl}
    \mathcal{E}\tau \geq  \frac{2\sqrt{\kappa} \chi  d}{  \varepsilon_x^2(1-P_\varepsilon)} \beta^{-1}\tau_{\text{r(UD)}}.
\end{align}
This fundamental energy-time tradeoff appears naturally within this computational model. While digital computations can often be accelerated by investing more energy (for example, via parallelization), it is generally less obvious what exact form the relationship between time and energy cost takes, suggesting that thermodynamic algorithms may offer a new and useful perspective on algorithmic complexity.

\subsection{Detailed Algorithmic Scaling}

In the main text, we presented a simplified version of the detailed table shown in Table~\ref{tab:runtime_scaling_methods}. Table~\ref{tab:runtime_scaling_methods} breaks down the scaling into the overdamped and underdamped regimes, whereas Table~\ref{tab:runtime_scaling} just takes the best scalings of our algorithms. The latter is essentially the scaling associated with the underdamped regime. However, in practice, there can be some engineering advantages to working in the overdamped regime, and hence it is useful to see the complexities of both regimes. 
\begin{table}[t]
    \centering
    \renewcommand{\arraystretch}{2}
    \begin{tabular}{c|c|c|c}
        \textbf{Problem} & \textbf{Digital SOTA}  
        & \textbf{This work (Overdamped)} & \textbf{This work (Underdamped)} \\
        \hline
        Linear System &$O(\min\{d^\omega,d^2 \sqrt{\kappa}\})$& $ O(d \kappa^2 \varepsilon^{-2})$ & $O(\gamma d \kappa^{3/2}  \varepsilon^{-2})$ \\
        Matrix Inverse &  $O(d^\omega)$ & $O(d^2 \kappa \varepsilon^{-2})$ & $ O(\gamma  d^2 \kappa^2 \varepsilon^{-2})$\\
        Lyapunov Equation &$O(d^3)$ & $O( d^2 \kappa \varepsilon^{-2})$&  $ O(\gamma d^2 \kappa^2\varepsilon^{-2})$ \\
        Matrix Determinant &  $O(d^\omega)$& $O(d  \kappa \ln(\kappa)^3 \varepsilon^{-2})$ & $ O(\gamma d \kappa  \ln(\kappa)^3 \varepsilon^{-2})$\\
    \end{tabular}
    \caption{\textbf{Asymptotic complexities of linear algebra algorithms, including overdamped and underdamped regimes.} For our thermodynamic algorithms, the complexity depends on the dynamical regime, i.e., whether the dynamics are overdamped and underdamped, as shown in this table.}
    \label{tab:runtime_scaling_methods}
\end{table}

\subsection{Assumptions}

A number of assumptions are made in the analytical derivations of the findings presented in the Results section. Certain aspects of the problem have been idealized in order to reveal the fundamental performance characteristics of the thermodynamic algorithms. The main assumptions are the following
\begin{itemize}
    \item The dynamics of the system may be described by the ODL equation or the UDL equations.
    \item The potential function $U(x)$, and in particular the matrix $A$ and vector $b$, can be switched instantaneously between different values.
    \item The potential energy function $U(x)$ can be implemented to arbitrary accuracy.
    \item Before a protocol begins, the system may be taken to be in an equilibrium distribution $\mathcal{N}[0, \beta^{-1}\|A\|^{-1}\mathbb{I}]$.
\end{itemize}

\subsection{Numerical simulations}

We outline the methods used for our numerical simulations of the thermodynamic algorithms. In general, we simulated the overdamped system dynamics because the performance is similar to the underdamped case, and the overdamped system is more numerically stable. The ODL equation, $dx = - \mathcal{A}x + \mathcal{N}[0, \mathcal{B} dt]$, is often written using an Itô integral \cite{gardiner1985handbook}:
\begin{equation}
\label{eq:ou-process-ito}
    x(t) = e^{-\mathcal{A} t}x_0 + \int_{0}^t dt' e^{-\mathcal{A}(t-t')} Ld W_t,
\end{equation}
where $L L^\intercal=\mathcal{B}$, and in this form it is apparent that the deterministic and stochastic parts of the evolution (the first and second terms above) can be evaluated separately. As this is a Gaussian process, the corresponding Fokker-Planck dynamics are fully captured by the behavior of the first and second moments, which can be evaluated directly using the well-known solution to the Ornstein-Uhlenbeck (OU) process. The JAX~\cite{frostig2018compiling} library was used to simulate the system at high dimensions, leveraging efficient implementations of matrix exponentials, diagonalization, and convolution to evaluate the various terms in the solution of the OU process. In what follows, we describe the timing model employed for benchmarking our thermodynamic algorithm, assuming an implementation using electrical circuits.

\subsubsection{Timing model}

To obtain the comparisons to other digital methods, we considered the following procedure to run the TA on electrical hardware. For more details on our model for the hardware implementation we refer the reader to the Supplemental Information. We take the $RC=1/\gamma = 1 \mu\text{s}$, which sets the characteristic timescale of the thermodynamic device. The determinant estimation procedure is excluded here for clarity, as it involves directly measuring work, which may involve a more complicated hardware proposal.
\begin{enumerate}
     \item Compute the values of the resistors $\{R_{ij}, R'_i\}$ entering the matrix $\mathcal{J}$ that encodes the $A$ matrix.
    \item Digital-to-analog (DAC) conversion of the $\mathcal{J}$ matrix and the $b$ vector with a given bit-precision.
    \item Let the dynamics run for $t_0$ (the equilibration time). Note that for simulations this time was chosen heuristically by exploring convergence in the solutions of the problem of interest.
    \item Switch on the integrators (and multipliers for the inverse estimation) and let the system evolve for time $\tau$.
    \item Analog-to-digital (ADC) conversion of the solution outputted from the integrators sent back to the digital device.
\end{enumerate}
For step 1, we measured the time for the digital operation to be performed, and for the other steps we estimated the time, based on the following assumptions:
\begin{itemize}
    \item 16 bit-precision
    \item 5000 ADC/DAC channels with a sampling rate: 250 Msamples/s.
    \item $R = 10^3 \,\Omega$, $C= 1 \,\text{nF}$, which means $RC = 1 \mu \text{s}$ is the characteristic timescale of the system.
\end{itemize}
Finally, note that in all cases that were investigated, the dominant contribution to the total runtime was the digital compilation step. This step includes $O(d^2)$ operations and involves conversion of the matrix $A$ to $\mathcal{J}$, which is detailed in the Supplemental Information. Hence some assumptions about the DAC/ADC may be relaxed and the total thermodynamic runtime would be similar. The $RC$ time constant may also be reduced to make the algorithm faster.

\bibliography{thermo_new.bib}

\begin{thebibliography}{81}%
\makeatletter
\providecommand \@ifxundefined [1]{%
 \@ifx{#1\undefined}
}%
\providecommand \@ifnum [1]{%
 \ifnum #1\expandafter \@firstoftwo
 \else \expandafter \@secondoftwo
 \fi
}%
\providecommand \@ifx [1]{%
 \ifx #1\expandafter \@firstoftwo
 \else \expandafter \@secondoftwo
 \fi
}%
\providecommand \natexlab [1]{#1}%
\providecommand \enquote  [1]{``#1''}%
\providecommand \bibnamefont  [1]{#1}%
\providecommand \bibfnamefont [1]{#1}%
\providecommand \citenamefont [1]{#1}%
\providecommand \href@noop [0]{\@secondoftwo}%
\providecommand \href [0]{\begingroup \@sanitize@url \@href}%
\providecommand \@href[1]{\@@startlink{#1}\@@href}%
\providecommand \@@href[1]{\endgroup#1\@@endlink}%
\providecommand \@sanitize@url [0]{\catcode `\\12\catcode `\$12\catcode
  `\&12\catcode `\#12\catcode `\^12\catcode `\_12\catcode `\%12\relax}%
\providecommand \@@startlink[1]{}%
\providecommand \@@endlink[0]{}%
\providecommand \url  [0]{\begingroup\@sanitize@url \@url }%
\providecommand \@url [1]{\endgroup\@href {#1}{\urlprefix }}%
\providecommand \urlprefix  [0]{URL }%
\providecommand \Eprint [0]{\href }%
\providecommand \doibase [0]{http://dx.doi.org/}%
\providecommand \selectlanguage [0]{\@gobble}%
\providecommand \bibinfo  [0]{\@secondoftwo}%
\providecommand \bibfield  [0]{\@secondoftwo}%
\providecommand \translation [1]{[#1]}%
\providecommand \BibitemOpen [0]{}%
\providecommand \bibitemStop [0]{}%
\providecommand \bibitemNoStop [0]{.\EOS\space}%
\providecommand \EOS [0]{\spacefactor3000\relax}%
\providecommand \BibitemShut  [1]{\csname bibitem#1\endcsname}%
\let\auto@bib@innerbib\@empty
\bibitem [{\citenamefont {Vadlamani}\ \emph {et~al.}(2020)\citenamefont
  {Vadlamani}, \citenamefont {Xiao},\ and\ \citenamefont
  {Yablonovitch}}]{vadlamani2020physics}%
  \BibitemOpen
  \bibfield  {author} {\bibinfo {author} {\bibfnamefont {Sri~Krishna}\
  \bibnamefont {Vadlamani}}, \bibinfo {author} {\bibfnamefont
  {Tianyao~Patrick}\ \bibnamefont {Xiao}}, \ and\ \bibinfo {author}
  {\bibfnamefont {Eli}\ \bibnamefont {Yablonovitch}},\ }\bibfield  {title}
  {\enquote {\bibinfo {title} {Physics successfully implements lagrange
  multiplier optimization},}\ }\href {\doibase 10.1073/pnas.2015192117}
  {\bibfield  {journal} {\bibinfo  {journal} {Proc. Natl. Acad. U.S.A.}\
  }\textbf {\bibinfo {volume} {117}},\ \bibinfo {pages} {26639--26650}
  (\bibinfo {year} {2020})}\BibitemShut {NoStop}%
\bibitem [{\citenamefont {Mohseni}\ \emph {et~al.}(2022)\citenamefont
  {Mohseni}, \citenamefont {McMahon},\ and\ \citenamefont
  {Byrnes}}]{mohseni2022ising}%
  \BibitemOpen
  \bibfield  {author} {\bibinfo {author} {\bibfnamefont {Naeimeh}\ \bibnamefont
  {Mohseni}}, \bibinfo {author} {\bibfnamefont {Peter~L.}\ \bibnamefont
  {McMahon}}, \ and\ \bibinfo {author} {\bibfnamefont {Tim}\ \bibnamefont
  {Byrnes}},\ }\bibfield  {title} {\enquote {\bibinfo {title} {Ising machines
  as hardware solvers of combinatorial optimization problems},}\ }\href
  {\doibase 10.1038/s42254-022-00440-8} {\bibfield  {journal} {\bibinfo
  {journal} {Nat. Rev. Phys.}\ }\textbf {\bibinfo {volume} {4}},\ \bibinfo
  {pages} {363--379} (\bibinfo {year} {2022})}\BibitemShut {NoStop}%
\bibitem [{\citenamefont {Inagaki}\ \emph {et~al.}(2016)\citenamefont
  {Inagaki}, \citenamefont {Haribara}, \citenamefont {Igarashi}, \citenamefont
  {Sonobe}, \citenamefont {Tamate}, \citenamefont {Honjo}, \citenamefont
  {Marandi}, \citenamefont {McMahon}, \citenamefont {Umeki}, \citenamefont
  {Enbutsu}, \citenamefont {Tadanaga}, \citenamefont {Takenouchi},
  \citenamefont {Aihara}, \citenamefont {Kawarabayashi}, \citenamefont {Inoue},
  \citenamefont {Utsunomiya},\ and\ \citenamefont
  {Takesue}}]{inagaki2016coherent}%
  \BibitemOpen
  \bibfield  {author} {\bibinfo {author} {\bibfnamefont {Takahiro}\
  \bibnamefont {Inagaki}}, \bibinfo {author} {\bibfnamefont {Yoshitaka}\
  \bibnamefont {Haribara}}, \bibinfo {author} {\bibfnamefont {Koji}\
  \bibnamefont {Igarashi}}, \bibinfo {author} {\bibfnamefont {Tomohiro}\
  \bibnamefont {Sonobe}}, \bibinfo {author} {\bibfnamefont {Shuhei}\
  \bibnamefont {Tamate}}, \bibinfo {author} {\bibfnamefont {Toshimori}\
  \bibnamefont {Honjo}}, \bibinfo {author} {\bibfnamefont {Alireza}\
  \bibnamefont {Marandi}}, \bibinfo {author} {\bibfnamefont {Peter~L.}\
  \bibnamefont {McMahon}}, \bibinfo {author} {\bibfnamefont {Takeshi}\
  \bibnamefont {Umeki}}, \bibinfo {author} {\bibfnamefont {Koji}\ \bibnamefont
  {Enbutsu}}, \bibinfo {author} {\bibfnamefont {Osamu}\ \bibnamefont
  {Tadanaga}}, \bibinfo {author} {\bibfnamefont {Hirokazu}\ \bibnamefont
  {Takenouchi}}, \bibinfo {author} {\bibfnamefont {Kazuyuki}\ \bibnamefont
  {Aihara}}, \bibinfo {author} {\bibfnamefont {Ken-ichi}\ \bibnamefont
  {Kawarabayashi}}, \bibinfo {author} {\bibfnamefont {Kyo}\ \bibnamefont
  {Inoue}}, \bibinfo {author} {\bibfnamefont {Shoko}\ \bibnamefont
  {Utsunomiya}}, \ and\ \bibinfo {author} {\bibfnamefont {Hiroki}\ \bibnamefont
  {Takesue}},\ }\bibfield  {title} {\enquote {\bibinfo {title} {A coherent
  ising machine for 2000-node optimization problems},}\ }\href {\doibase
  10.1126/science.aah4243} {\bibfield  {journal} {\bibinfo  {journal}
  {Science}\ }\textbf {\bibinfo {volume} {354}},\ \bibinfo {pages} {603--606}
  (\bibinfo {year} {2016})}\BibitemShut {NoStop}%
\bibitem [{\citenamefont {Feynman}(1982)}]{feynman1982simulating}%
  \BibitemOpen
  \bibfield  {author} {\bibinfo {author} {\bibfnamefont {Richard~P.}\
  \bibnamefont {Feynman}},\ }\bibfield  {title} {\enquote {\bibinfo {title}
  {Simulating physics with computers},}\ }\href {\doibase 10.1007/BF02650179}
  {\bibfield  {journal} {\bibinfo  {journal} {Int. J. Theor. Phys.}\ }\textbf
  {\bibinfo {volume} {21}},\ \bibinfo {pages} {467--488} (\bibinfo {year}
  {1982})}\BibitemShut {NoStop}%
\bibitem [{\citenamefont {Harrow}\ \emph {et~al.}(2009)\citenamefont {Harrow},
  \citenamefont {Hassidim},\ and\ \citenamefont {Lloyd}}]{harrow2009quantum}%
  \BibitemOpen
  \bibfield  {author} {\bibinfo {author} {\bibfnamefont {Aram~W.}\ \bibnamefont
  {Harrow}}, \bibinfo {author} {\bibfnamefont {Avinatan}\ \bibnamefont
  {Hassidim}}, \ and\ \bibinfo {author} {\bibfnamefont {Seth}\ \bibnamefont
  {Lloyd}},\ }\bibfield  {title} {\enquote {\bibinfo {title} {Quantum algorithm
  for linear systems of equations},}\ }\href {\doibase
  10.1103/PhysRevLett.103.150502} {\bibfield  {journal} {\bibinfo  {journal}
  {Phys. Rev. Lett.}\ }\textbf {\bibinfo {volume} {103}},\ \bibinfo {pages}
  {150502} (\bibinfo {year} {2009})}\BibitemShut {NoStop}%
\bibitem [{\citenamefont {Scherer}\ \emph {et~al.}(2017)\citenamefont
  {Scherer}, \citenamefont {Valiron}, \citenamefont {Mau}, \citenamefont
  {Alexander}, \citenamefont {Van~den Berg},\ and\ \citenamefont
  {Chapuran}}]{scherer2017concrete}%
  \BibitemOpen
  \bibfield  {author} {\bibinfo {author} {\bibfnamefont {Artur}\ \bibnamefont
  {Scherer}}, \bibinfo {author} {\bibfnamefont {Beno{\^\i}t}\ \bibnamefont
  {Valiron}}, \bibinfo {author} {\bibfnamefont {Siun-Chuon}\ \bibnamefont
  {Mau}}, \bibinfo {author} {\bibfnamefont {Scott}\ \bibnamefont {Alexander}},
  \bibinfo {author} {\bibfnamefont {Eric}\ \bibnamefont {Van~den Berg}}, \ and\
  \bibinfo {author} {\bibfnamefont {Thomas~E.}\ \bibnamefont {Chapuran}},\
  }\bibfield  {title} {\enquote {\bibinfo {title} {Concrete resource analysis
  of the quantum linear-system algorithm used to compute the electromagnetic
  scattering cross section of a {2D} target},}\ }\href {\doibase
  10.1007/s11128-016-1495-5} {\bibfield  {journal} {\bibinfo  {journal}
  {Quantum Inf. Process.}\ }\textbf {\bibinfo {volume} {16}},\ \bibinfo {pages}
  {1--65} (\bibinfo {year} {2017})}\BibitemShut {NoStop}%
\bibitem [{\citenamefont {Preskill}(2018)}]{preskill2018quantum}%
  \BibitemOpen
  \bibfield  {author} {\bibinfo {author} {\bibfnamefont {John}\ \bibnamefont
  {Preskill}},\ }\bibfield  {title} {\enquote {\bibinfo {title} {Quantum
  computing in the nisq era and beyond},}\ }\href {\doibase
  10.22331/q-2018-08-06-79} {\bibfield  {journal} {\bibinfo  {journal}
  {Quantum}\ }\textbf {\bibinfo {volume} {2}},\ \bibinfo {pages} {79} (\bibinfo
  {year} {2018})}\BibitemShut {NoStop}%
\bibitem [{\citenamefont {Bravo-Prieto}\ \emph {et~al.}(2019)\citenamefont
  {Bravo-Prieto}, \citenamefont {LaRose}, \citenamefont {Cerezo}, \citenamefont
  {Subasi}, \citenamefont {Cincio},\ and\ \citenamefont
  {Coles}}]{bravo2020variational}%
  \BibitemOpen
  \bibfield  {author} {\bibinfo {author} {\bibfnamefont {Carlos}\ \bibnamefont
  {Bravo-Prieto}}, \bibinfo {author} {\bibfnamefont {Ryan}\ \bibnamefont
  {LaRose}}, \bibinfo {author} {\bibfnamefont {M.}~\bibnamefont {Cerezo}},
  \bibinfo {author} {\bibfnamefont {Yigit}\ \bibnamefont {Subasi}}, \bibinfo
  {author} {\bibfnamefont {Lukasz}\ \bibnamefont {Cincio}}, \ and\ \bibinfo
  {author} {\bibfnamefont {Patrick}\ \bibnamefont {Coles}},\ }\bibfield
  {title} {\enquote {\bibinfo {title} {Variational quantum linear solver},}\
  }\href {https://arxiv.org/abs/1909.05820} {\bibfield  {journal} {\bibinfo
  {journal} {arXiv:1909.05820}\ } (\bibinfo {year} {2019})}\BibitemShut
  {NoStop}%
\bibitem [{\citenamefont {Xu}\ \emph {et~al.}(2021)\citenamefont {Xu},
  \citenamefont {Sun}, \citenamefont {Endo}, \citenamefont {Li}, \citenamefont
  {Benjamin},\ and\ \citenamefont {Yuan}}]{xu2019variational}%
  \BibitemOpen
  \bibfield  {author} {\bibinfo {author} {\bibfnamefont {Xiaosi}\ \bibnamefont
  {Xu}}, \bibinfo {author} {\bibfnamefont {Jinzhao}\ \bibnamefont {Sun}},
  \bibinfo {author} {\bibfnamefont {Suguru}\ \bibnamefont {Endo}}, \bibinfo
  {author} {\bibfnamefont {Ying}\ \bibnamefont {Li}}, \bibinfo {author}
  {\bibfnamefont {Simon~C.}\ \bibnamefont {Benjamin}}, \ and\ \bibinfo {author}
  {\bibfnamefont {Xiao}\ \bibnamefont {Yuan}},\ }\bibfield  {title} {\enquote
  {\bibinfo {title} {Variational algorithms for linear algebra},}\ }\href
  {\doibase 10.1016/j.scib.2021.06.023} {\bibfield  {journal} {\bibinfo
  {journal} {Sci. Bull.}\ }\textbf {\bibinfo {volume} {66}},\ \bibinfo {pages}
  {2181--2188} (\bibinfo {year} {2021})}\BibitemShut {NoStop}%
\bibitem [{\citenamefont {Cerezo}\ \emph
  {et~al.}(2021{\natexlab{a}})\citenamefont {Cerezo}, \citenamefont
  {Arrasmith}, \citenamefont {Babbush}, \citenamefont {Benjamin}, \citenamefont
  {Endo}, \citenamefont {Fujii}, \citenamefont {McClean}, \citenamefont
  {Mitarai}, \citenamefont {Yuan}, \citenamefont {Cincio},\ and\ \citenamefont
  {Coles}}]{cerezo2020variationalreview}%
  \BibitemOpen
  \bibfield  {author} {\bibinfo {author} {\bibfnamefont {M.}~\bibnamefont
  {Cerezo}}, \bibinfo {author} {\bibfnamefont {Andrew}\ \bibnamefont
  {Arrasmith}}, \bibinfo {author} {\bibfnamefont {Ryan}\ \bibnamefont
  {Babbush}}, \bibinfo {author} {\bibfnamefont {Simon~C.}\ \bibnamefont
  {Benjamin}}, \bibinfo {author} {\bibfnamefont {Suguru}\ \bibnamefont {Endo}},
  \bibinfo {author} {\bibfnamefont {Keisuke}\ \bibnamefont {Fujii}}, \bibinfo
  {author} {\bibfnamefont {Jarrod~R.}\ \bibnamefont {McClean}}, \bibinfo
  {author} {\bibfnamefont {Kosuke}\ \bibnamefont {Mitarai}}, \bibinfo {author}
  {\bibfnamefont {Xiao}\ \bibnamefont {Yuan}}, \bibinfo {author} {\bibfnamefont
  {Lukasz}\ \bibnamefont {Cincio}}, \ and\ \bibinfo {author} {\bibfnamefont
  {Patrick~J.}\ \bibnamefont {Coles}},\ }\bibfield  {title} {\enquote {\bibinfo
  {title} {Variational quantum algorithms},}\ }\href {\doibase
  10.1038/s42254-021-00348-9} {\bibfield  {journal} {\bibinfo  {journal} {Nat.
  Rev. Phys.}\ }\textbf {\bibinfo {volume} {3}},\ \bibinfo {pages} {625–644}
  (\bibinfo {year} {2021}{\natexlab{a}})}\BibitemShut {NoStop}%
\bibitem [{\citenamefont {McClean}\ \emph {et~al.}(2018)\citenamefont
  {McClean}, \citenamefont {Boixo}, \citenamefont {Smelyanskiy}, \citenamefont
  {Babbush},\ and\ \citenamefont {Neven}}]{mcclean2018barren}%
  \BibitemOpen
  \bibfield  {author} {\bibinfo {author} {\bibfnamefont {Jarrod~R.}\
  \bibnamefont {McClean}}, \bibinfo {author} {\bibfnamefont {Sergio}\
  \bibnamefont {Boixo}}, \bibinfo {author} {\bibfnamefont {Vadim~N.}\
  \bibnamefont {Smelyanskiy}}, \bibinfo {author} {\bibfnamefont {Ryan}\
  \bibnamefont {Babbush}}, \ and\ \bibinfo {author} {\bibfnamefont {Hartmut}\
  \bibnamefont {Neven}},\ }\bibfield  {title} {\enquote {\bibinfo {title}
  {Barren plateaus in quantum neural network training landscapes},}\ }\href
  {\doibase 10.1038/s41467-018-07090-4} {\bibfield  {journal} {\bibinfo
  {journal} {Nat. Commun.}\ }\textbf {\bibinfo {volume} {9}},\ \bibinfo {pages}
  {1--6} (\bibinfo {year} {2018})}\BibitemShut {NoStop}%
\bibitem [{\citenamefont {Cerezo}\ \emph
  {et~al.}(2021{\natexlab{b}})\citenamefont {Cerezo}, \citenamefont {Sone},
  \citenamefont {Volkoff}, \citenamefont {Cincio},\ and\ \citenamefont
  {Coles}}]{cerezo2020cost}%
  \BibitemOpen
  \bibfield  {author} {\bibinfo {author} {\bibfnamefont {M.}~\bibnamefont
  {Cerezo}}, \bibinfo {author} {\bibfnamefont {Akira}\ \bibnamefont {Sone}},
  \bibinfo {author} {\bibfnamefont {Tyler}\ \bibnamefont {Volkoff}}, \bibinfo
  {author} {\bibfnamefont {Lukasz}\ \bibnamefont {Cincio}}, \ and\ \bibinfo
  {author} {\bibfnamefont {Patrick~J.}\ \bibnamefont {Coles}},\ }\bibfield
  {title} {\enquote {\bibinfo {title} {Cost function dependent barren plateaus
  in shallow parametrized quantum circuits},}\ }\href {\doibase
  10.1038/s41467-021-21728-w} {\bibfield  {journal} {\bibinfo  {journal} {Nat.
  Commun.}\ }\textbf {\bibinfo {volume} {12}},\ \bibinfo {pages} {1--12}
  (\bibinfo {year} {2021}{\natexlab{b}})}\BibitemShut {NoStop}%
\bibitem [{\citenamefont {Wang}\ \emph {et~al.}(2021)\citenamefont {Wang},
  \citenamefont {Fontana}, \citenamefont {Cerezo}, \citenamefont {Sharma},
  \citenamefont {Sone}, \citenamefont {Cincio},\ and\ \citenamefont
  {Coles}}]{wang2020noise}%
  \BibitemOpen
  \bibfield  {author} {\bibinfo {author} {\bibfnamefont {Samson}\ \bibnamefont
  {Wang}}, \bibinfo {author} {\bibfnamefont {Enrico}\ \bibnamefont {Fontana}},
  \bibinfo {author} {\bibfnamefont {M.}~\bibnamefont {Cerezo}}, \bibinfo
  {author} {\bibfnamefont {Kunal}\ \bibnamefont {Sharma}}, \bibinfo {author}
  {\bibfnamefont {Akira}\ \bibnamefont {Sone}}, \bibinfo {author}
  {\bibfnamefont {Lukasz}\ \bibnamefont {Cincio}}, \ and\ \bibinfo {author}
  {\bibfnamefont {Patrick~J.}\ \bibnamefont {Coles}},\ }\bibfield  {title}
  {\enquote {\bibinfo {title} {Noise-induced barren plateaus in variational
  quantum algorithms},}\ }\href {\doibase 10.1038/s41467-021-27045-6}
  {\bibfield  {journal} {\bibinfo  {journal} {Nat. Commun.}\ }\textbf {\bibinfo
  {volume} {12}},\ \bibinfo {pages} {1--11} (\bibinfo {year}
  {2021})}\BibitemShut {NoStop}%
\bibitem [{\citenamefont {Li}\ \emph {et~al.}(2018)\citenamefont {Li},
  \citenamefont {Hu}, \citenamefont {Li}, \citenamefont {Jiang}, \citenamefont
  {Ge}, \citenamefont {Montgomery}, \citenamefont {Zhang}, \citenamefont
  {Song}, \citenamefont {D{\'a}vila}, \citenamefont {Graves}, \citenamefont
  {Li}, \citenamefont {Strachan}, \citenamefont {Lin}, \citenamefont {Wang},
  \citenamefont {Barnell}, \citenamefont {Wu}, \citenamefont {Williams},\ and\
  \citenamefont {Yang}}]{li2018analogue}%
  \BibitemOpen
  \bibfield  {author} {\bibinfo {author} {\bibfnamefont {Can}\ \bibnamefont
  {Li}}, \bibinfo {author} {\bibfnamefont {Miao}\ \bibnamefont {Hu}}, \bibinfo
  {author} {\bibfnamefont {Yunning}\ \bibnamefont {Li}}, \bibinfo {author}
  {\bibfnamefont {Hao}\ \bibnamefont {Jiang}}, \bibinfo {author} {\bibfnamefont
  {Ning}\ \bibnamefont {Ge}}, \bibinfo {author} {\bibfnamefont {Eric}\
  \bibnamefont {Montgomery}}, \bibinfo {author} {\bibfnamefont {Jiaming}\
  \bibnamefont {Zhang}}, \bibinfo {author} {\bibfnamefont {Wenhao}\
  \bibnamefont {Song}}, \bibinfo {author} {\bibfnamefont {Noraica}\
  \bibnamefont {D{\'a}vila}}, \bibinfo {author} {\bibfnamefont {Catherine~E.}\
  \bibnamefont {Graves}}, \bibinfo {author} {\bibfnamefont {Zhiyong}\
  \bibnamefont {Li}}, \bibinfo {author} {\bibfnamefont {John~Paul}\
  \bibnamefont {Strachan}}, \bibinfo {author} {\bibfnamefont {Peng}\
  \bibnamefont {Lin}}, \bibinfo {author} {\bibfnamefont {Zhongrui}\
  \bibnamefont {Wang}}, \bibinfo {author} {\bibfnamefont {Mark}\ \bibnamefont
  {Barnell}}, \bibinfo {author} {\bibfnamefont {Qing}\ \bibnamefont {Wu}},
  \bibinfo {author} {\bibfnamefont {R.~Stanley}\ \bibnamefont {Williams}}, \
  and\ \bibinfo {author} {\bibfnamefont {J.~Joshua}\ \bibnamefont {Yang}},\
  }\bibfield  {title} {\enquote {\bibinfo {title} {Analogue signal and image
  processing with large memristor crossbars},}\ }\href {\doibase
  doi.org/10.1038/s41928-017-0002-z} {\bibfield  {journal} {\bibinfo  {journal}
  {Nat. Electron.}\ }\textbf {\bibinfo {volume} {1}},\ \bibinfo {pages}
  {52--59} (\bibinfo {year} {2018})}\BibitemShut {NoStop}%
\bibitem [{\citenamefont {Yi}\ \emph {et~al.}(2023)\citenamefont {Yi},
  \citenamefont {Kendall}, \citenamefont {Williams},\ and\ \citenamefont
  {Kumar}}]{yi2023activity}%
  \BibitemOpen
  \bibfield  {author} {\bibinfo {author} {\bibfnamefont {Su-in}\ \bibnamefont
  {Yi}}, \bibinfo {author} {\bibfnamefont {Jack~D.}\ \bibnamefont {Kendall}},
  \bibinfo {author} {\bibfnamefont {R.~Stanley}\ \bibnamefont {Williams}}, \
  and\ \bibinfo {author} {\bibfnamefont {Suhas}\ \bibnamefont {Kumar}},\
  }\bibfield  {title} {\enquote {\bibinfo {title} {Activity-difference training
  of deep neural networks using memristor crossbars},}\ }\href {\doibase
  10.1038/s41928-022-00869-w} {\bibfield  {journal} {\bibinfo  {journal} {Nat.
  Electron.}\ }\textbf {\bibinfo {volume} {6}},\ \bibinfo {pages} {45--51}
  (\bibinfo {year} {2023})}\BibitemShut {NoStop}%
\bibitem [{\citenamefont {Huang}\ \emph {et~al.}(2016)\citenamefont {Huang},
  \citenamefont {Guo}, \citenamefont {Seok}, \citenamefont {Tsividis},\ and\
  \citenamefont {Sethumadhavan}}]{huang2016evaluation}%
  \BibitemOpen
  \bibfield  {author} {\bibinfo {author} {\bibfnamefont {Yipeng}\ \bibnamefont
  {Huang}}, \bibinfo {author} {\bibfnamefont {Ning}\ \bibnamefont {Guo}},
  \bibinfo {author} {\bibfnamefont {Mingoo}\ \bibnamefont {Seok}}, \bibinfo
  {author} {\bibfnamefont {Yannis}\ \bibnamefont {Tsividis}}, \ and\ \bibinfo
  {author} {\bibfnamefont {Simha}\ \bibnamefont {Sethumadhavan}},\ }\bibfield
  {title} {\enquote {\bibinfo {title} {Evaluation of an analog accelerator for
  linear algebra},}\ }\href@noop {} {\bibfield  {journal} {\bibinfo  {journal}
  {Comput. Archit. News}\ }\textbf {\bibinfo {volume} {44}},\ \bibinfo {pages}
  {570--582} (\bibinfo {year} {2016})}\BibitemShut {NoStop}%
\bibitem [{\citenamefont {Coles}\ \emph {et~al.}(2023)\citenamefont {Coles},
  \citenamefont {Szczepanski}, \citenamefont {Melanson}, \citenamefont
  {Donatella}, \citenamefont {Martinez},\ and\ \citenamefont
  {Sbahi}}]{coles2023thermodynamic}%
  \BibitemOpen
  \bibfield  {author} {\bibinfo {author} {\bibfnamefont {Patrick~J.}\
  \bibnamefont {Coles}}, \bibinfo {author} {\bibfnamefont {Collin}\
  \bibnamefont {Szczepanski}}, \bibinfo {author} {\bibfnamefont {Denis}\
  \bibnamefont {Melanson}}, \bibinfo {author} {\bibfnamefont {Kaelan}\
  \bibnamefont {Donatella}}, \bibinfo {author} {\bibfnamefont {Antonio~J.}\
  \bibnamefont {Martinez}}, \ and\ \bibinfo {author} {\bibfnamefont {Faris}\
  \bibnamefont {Sbahi}},\ }\href@noop {} {\enquote {\bibinfo {title}
  {Thermodynamic {AI} and the fluctuation frontier},}\ } (\bibinfo {year}
  {2023}),\ \Eprint {http://arxiv.org/abs/2302.06584} {arXiv:2302.06584
  [cs.ET]} \BibitemShut {NoStop}%
\bibitem [{\citenamefont {Aadit}\ \emph {et~al.}(2022)\citenamefont {Aadit},
  \citenamefont {Grimaldi}, \citenamefont {Carpentieri}, \citenamefont
  {Theogarajan}, \citenamefont {Martinis}, \citenamefont {Finocchio},\ and\
  \citenamefont {Camsari}}]{aadit2022massively}%
  \BibitemOpen
  \bibfield  {author} {\bibinfo {author} {\bibfnamefont {Navid~Anjum}\
  \bibnamefont {Aadit}}, \bibinfo {author} {\bibfnamefont {Andrea}\
  \bibnamefont {Grimaldi}}, \bibinfo {author} {\bibfnamefont {Mario}\
  \bibnamefont {Carpentieri}}, \bibinfo {author} {\bibfnamefont {Luke}\
  \bibnamefont {Theogarajan}}, \bibinfo {author} {\bibfnamefont {John~M.}\
  \bibnamefont {Martinis}}, \bibinfo {author} {\bibfnamefont {Giovanni}\
  \bibnamefont {Finocchio}}, \ and\ \bibinfo {author} {\bibfnamefont
  {Kerem~Y.}\ \bibnamefont {Camsari}},\ }\bibfield  {title} {\enquote {\bibinfo
  {title} {Massively parallel probabilistic computing with sparse {I}sing
  machines},}\ }\href {\doibase 10.1038/s41928-022-00774-2} {\bibfield
  {journal} {\bibinfo  {journal} {Nat. Electron.}\ }\textbf {\bibinfo {volume}
  {5}},\ \bibinfo {pages} {460--468} (\bibinfo {year} {2022})}\BibitemShut
  {NoStop}%
\bibitem [{\citenamefont {Camsari}\ \emph {et~al.}(2019)\citenamefont
  {Camsari}, \citenamefont {Sutton},\ and\ \citenamefont
  {Datta}}]{Camsari_2019}%
  \BibitemOpen
  \bibfield  {author} {\bibinfo {author} {\bibfnamefont {Kerem~Y.}\
  \bibnamefont {Camsari}}, \bibinfo {author} {\bibfnamefont {Brian~M.}\
  \bibnamefont {Sutton}}, \ and\ \bibinfo {author} {\bibfnamefont {Supriyo}\
  \bibnamefont {Datta}},\ }\bibfield  {title} {\enquote {\bibinfo {title}
  {p-bits for probabilistic spin logic},}\ }\href {\doibase 10.1063/1.5055860}
  {\bibfield  {journal} {\bibinfo  {journal} {Appl. Phys. Rev.}\ }\textbf
  {\bibinfo {volume} {6}},\ \bibinfo {pages} {011305} (\bibinfo {year}
  {2019})}\BibitemShut {NoStop}%
\bibitem [{\citenamefont {Kaiser}\ \emph {et~al.}(2022)\citenamefont {Kaiser},
  \citenamefont {Datta},\ and\ \citenamefont {Behin-Aein}}]{kaiser2022life}%
  \BibitemOpen
  \bibfield  {author} {\bibinfo {author} {\bibfnamefont {J.}~\bibnamefont
  {Kaiser}}, \bibinfo {author} {\bibfnamefont {S.}~\bibnamefont {Datta}}, \
  and\ \bibinfo {author} {\bibfnamefont {B.}~\bibnamefont {Behin-Aein}},\
  }\bibfield  {title} {\enquote {\bibinfo {title} {Life is probabilistic—why
  should all our computers be deterministic? computing with p-bits: Ising
  solvers and beyond},}\ }in\ \href@noop {} {\emph {\bibinfo {booktitle} {2022
  International Electron Devices Meeting (IEDM)}}}\ (\bibinfo {organization}
  {IEEE},\ \bibinfo {year} {2022})\ pp.\ \bibinfo {pages} {21--4}\BibitemShut
  {NoStop}%
\bibitem [{\citenamefont {Hylton}(2020)}]{hylton2020thermodynamic}%
  \BibitemOpen
  \bibfield  {author} {\bibinfo {author} {\bibfnamefont {Todd}\ \bibnamefont
  {Hylton}},\ }\bibfield  {title} {\enquote {\bibinfo {title} {Thermodynamic
  neural network},}\ }\href {\doibase 10.3390/e22030256} {\bibfield  {journal}
  {\bibinfo  {journal} {Entropy}\ }\textbf {\bibinfo {volume} {22}},\ \bibinfo
  {pages} {256} (\bibinfo {year} {2020})}\BibitemShut {NoStop}%
\bibitem [{\citenamefont {Hylton}(2022)}]{hylton2022thermodynamic}%
  \BibitemOpen
  \bibfield  {author} {\bibinfo {author} {\bibfnamefont {Todd}\ \bibnamefont
  {Hylton}},\ }\bibfield  {title} {\enquote {\bibinfo {title} {Thermodynamic
  state machine network},}\ }\href {\doibase 10.3390/e24060744} {\bibfield
  {journal} {\bibinfo  {journal} {Entropy}\ }\textbf {\bibinfo {volume} {24}},\
  \bibinfo {pages} {744} (\bibinfo {year} {2022})}\BibitemShut {NoStop}%
\bibitem [{\citenamefont {Ganesh}(2020)}]{ganesh2020rebooting}%
  \BibitemOpen
  \bibfield  {author} {\bibinfo {author} {\bibfnamefont {Natesh}\ \bibnamefont
  {Ganesh}},\ }\bibfield  {title} {\enquote {\bibinfo {title} {Rebooting
  neuromorphic design-a complexity engineering approach},}\ }in\ \href@noop {}
  {\emph {\bibinfo {booktitle} {2020 International Conference on Rebooting
  Computing (ICRC)}}}\ (\bibinfo {organization} {IEEE},\ \bibinfo {year}
  {2020})\ pp.\ \bibinfo {pages} {80--89}\BibitemShut {NoStop}%
\bibitem [{\citenamefont {Ganesh}(2017)}]{8123676}%
  \BibitemOpen
  \bibfield  {author} {\bibinfo {author} {\bibfnamefont {Natesh}\ \bibnamefont
  {Ganesh}},\ }\bibfield  {title} {\enquote {\bibinfo {title} {A thermodynamic
  treatment of intelligent systems},}\ }in\ \href {\doibase
  10.1109/ICRC.2017.8123676} {\emph {\bibinfo {booktitle} {2017 IEEE
  International Conference on Rebooting Computing (ICRC)}}}\ (\bibinfo {year}
  {2017})\ pp.\ \bibinfo {pages} {1--4}\BibitemShut {NoStop}%
\bibitem [{\citenamefont {Lipka-Bartosik}\ \emph {et~al.}(2023)\citenamefont
  {Lipka-Bartosik}, \citenamefont {Perarnau-Llobet},\ and\ \citenamefont
  {Brunner}}]{lipka2023thermodynamic}%
  \BibitemOpen
  \bibfield  {author} {\bibinfo {author} {\bibfnamefont {Patryk}\ \bibnamefont
  {Lipka-Bartosik}}, \bibinfo {author} {\bibfnamefont {Mart{\'\i}}\
  \bibnamefont {Perarnau-Llobet}}, \ and\ \bibinfo {author} {\bibfnamefont
  {Nicolas}\ \bibnamefont {Brunner}},\ }\bibfield  {title} {\enquote {\bibinfo
  {title} {Thermodynamic computing via autonomous quantum thermal machines},}\
  }\href {https://arxiv.org/abs/2308.15905} {\bibfield  {journal} {\bibinfo
  {journal} {arXiv preprint arXiv:2308.15905}\ } (\bibinfo {year}
  {2023})}\BibitemShut {NoStop}%
\bibitem [{\citenamefont {Conte}\ \emph {et~al.}(2019)\citenamefont {Conte},
  \citenamefont {DeBenedictis}, \citenamefont {Ganesh}, \citenamefont {Hylton},
  \citenamefont {Strachan}, \citenamefont {Williams}, \citenamefont {Alemi},
  \citenamefont {Altenberg}, \citenamefont {Crooks}, \citenamefont
  {Crutchfield} \emph {et~al.}}]{conte2019thermodynamic}%
  \BibitemOpen
  \bibfield  {author} {\bibinfo {author} {\bibfnamefont {Tom}\ \bibnamefont
  {Conte}}, \bibinfo {author} {\bibfnamefont {Erik}\ \bibnamefont
  {DeBenedictis}}, \bibinfo {author} {\bibfnamefont {Natesh}\ \bibnamefont
  {Ganesh}}, \bibinfo {author} {\bibfnamefont {Todd}\ \bibnamefont {Hylton}},
  \bibinfo {author} {\bibfnamefont {John~Paul}\ \bibnamefont {Strachan}},
  \bibinfo {author} {\bibfnamefont {R~Stanley}\ \bibnamefont {Williams}},
  \bibinfo {author} {\bibfnamefont {Alexander}\ \bibnamefont {Alemi}}, \bibinfo
  {author} {\bibfnamefont {Lee}\ \bibnamefont {Altenberg}}, \bibinfo {author}
  {\bibfnamefont {Gavin~E.}\ \bibnamefont {Crooks}}, \bibinfo {author}
  {\bibfnamefont {James}\ \bibnamefont {Crutchfield}},  \emph {et~al.},\
  }\bibfield  {title} {\enquote {\bibinfo {title} {Thermodynamic computing},}\
  }\href {https://arxiv.org/abs/1911.01968} {\bibfield  {journal} {\bibinfo
  {journal} {arXiv preprint arXiv:1911.01968}\ } (\bibinfo {year}
  {2019})}\BibitemShut {NoStop}%
\bibitem [{\citenamefont {Babbush}\ \emph {et~al.}(2023)\citenamefont
  {Babbush}, \citenamefont {Berry}, \citenamefont {Kothari}, \citenamefont
  {Somma},\ and\ \citenamefont {Wiebe}}]{babbush2023exponential}%
  \BibitemOpen
  \bibfield  {author} {\bibinfo {author} {\bibfnamefont {Ryan}\ \bibnamefont
  {Babbush}}, \bibinfo {author} {\bibfnamefont {Dominic~W}\ \bibnamefont
  {Berry}}, \bibinfo {author} {\bibfnamefont {Robin}\ \bibnamefont {Kothari}},
  \bibinfo {author} {\bibfnamefont {Rolando~D}\ \bibnamefont {Somma}}, \ and\
  \bibinfo {author} {\bibfnamefont {Nathan}\ \bibnamefont {Wiebe}},\ }\bibfield
   {title} {\enquote {\bibinfo {title} {Exponential quantum speedup in
  simulating coupled classical oscillators},}\ }\href
  {https://arxiv.org/abs/2303.13012} {\bibfield  {journal} {\bibinfo  {journal}
  {arXiv preprint arXiv:2303.13012}\ } (\bibinfo {year} {2023})}\BibitemShut
  {NoStop}%
\bibitem [{\citenamefont {Parks}(1992)}]{parks1992lyapunov}%
  \BibitemOpen
  \bibfield  {author} {\bibinfo {author} {\bibfnamefont {Patrick~C}\
  \bibnamefont {Parks}},\ }\bibfield  {title} {\enquote {\bibinfo {title} {Am
  lyapunov's stability theory—100 years on},}\ }\href@noop {} {\bibfield
  {journal} {\bibinfo  {journal} {IMA journal of Mathematical Control and
  Information}\ }\textbf {\bibinfo {volume} {9}},\ \bibinfo {pages} {275--303}
  (\bibinfo {year} {1992})}\BibitemShut {NoStop}%
\bibitem [{\citenamefont {Forsythe}\ and\ \citenamefont
  {Leibler}(1950)}]{forsythe1950matrix}%
  \BibitemOpen
  \bibfield  {author} {\bibinfo {author} {\bibfnamefont {George~E}\
  \bibnamefont {Forsythe}}\ and\ \bibinfo {author} {\bibfnamefont {Richard~A}\
  \bibnamefont {Leibler}},\ }\bibfield  {title} {\enquote {\bibinfo {title}
  {Matrix inversion by a monte carlo method},}\ }\href@noop {} {\bibfield
  {journal} {\bibinfo  {journal} {Mathematics of Computation}\ }\textbf
  {\bibinfo {volume} {4}},\ \bibinfo {pages} {127--129} (\bibinfo {year}
  {1950})}\BibitemShut {NoStop}%
\bibitem [{\citenamefont {Alexandrov}\ and\ \citenamefont
  {Lakka}(1996)}]{alexandrov1996comparison}%
  \BibitemOpen
  \bibfield  {author} {\bibinfo {author} {\bibfnamefont {Vassil~N}\
  \bibnamefont {Alexandrov}}\ and\ \bibinfo {author} {\bibfnamefont
  {S}~\bibnamefont {Lakka}},\ }\bibfield  {title} {\enquote {\bibinfo {title}
  {Comparison of three monte carlo methods for matrix inversion},}\ }in\
  \href@noop {} {\emph {\bibinfo {booktitle} {Euro-Par'96 Parallel Processing:
  Second International Euro-Par Conference Lyon, France, August 26--29, 1996
  Proceedings, Volume II 2}}}\ (\bibinfo {organization} {Springer},\ \bibinfo
  {year} {1996})\ pp.\ \bibinfo {pages} {72--80}\BibitemShut {NoStop}%
\bibitem [{\citenamefont {{\"O}kten}(2005)}]{okten2005solving}%
  \BibitemOpen
  \bibfield  {author} {\bibinfo {author} {\bibfnamefont {Giray}\ \bibnamefont
  {{\"O}kten}},\ }\bibfield  {title} {\enquote {\bibinfo {title} {Solving
  linear equations by monte carlo simulation},}\ }\href@noop {} {\bibfield
  {journal} {\bibinfo  {journal} {SIAM Journal on Scientific Computing}\
  }\textbf {\bibinfo {volume} {27}},\ \bibinfo {pages} {511--531} (\bibinfo
  {year} {2005})}\BibitemShut {NoStop}%
\bibitem [{\citenamefont {Rosca}(2006)}]{rosca2006monte}%
  \BibitemOpen
  \bibfield  {author} {\bibinfo {author} {\bibfnamefont {Natalia}\ \bibnamefont
  {Rosca}},\ }\bibfield  {title} {\enquote {\bibinfo {title} {Monte carlo
  methods for systems of linear equations},}\ }\href@noop {} {\bibfield
  {journal} {\bibinfo  {journal} {Studia Univ. BABES-BOLYAI Mathematica}\
  }\textbf {\bibinfo {volume} {51}} (\bibinfo {year} {2006})}\BibitemShut
  {NoStop}%
\bibitem [{\citenamefont {Dimov}\ \emph {et~al.}(2015)\citenamefont {Dimov},
  \citenamefont {Maire},\ and\ \citenamefont {Sellier}}]{dimov2015new}%
  \BibitemOpen
  \bibfield  {author} {\bibinfo {author} {\bibfnamefont {Ivan}\ \bibnamefont
  {Dimov}}, \bibinfo {author} {\bibfnamefont {Sylvain}\ \bibnamefont {Maire}},
  \ and\ \bibinfo {author} {\bibfnamefont {Jean~Michel}\ \bibnamefont
  {Sellier}},\ }\bibfield  {title} {\enquote {\bibinfo {title} {A new walk on
  equations monte carlo method for solving systems of linear algebraic
  equations},}\ }\href@noop {} {\bibfield  {journal} {\bibinfo  {journal}
  {Applied Mathematical Modelling}\ }\textbf {\bibinfo {volume} {39}},\
  \bibinfo {pages} {4494--4510} (\bibinfo {year} {2015})}\BibitemShut {NoStop}%
\bibitem [{\citenamefont {Duffield}\ \emph {et~al.}(2023)\citenamefont
  {Duffield}, \citenamefont {Aifer}, \citenamefont {Crooks}, \citenamefont
  {Ahle},\ and\ \citenamefont {Coles}}]{duffield2023thermodynamic}%
  \BibitemOpen
  \bibfield  {author} {\bibinfo {author} {\bibfnamefont {Samuel}\ \bibnamefont
  {Duffield}}, \bibinfo {author} {\bibfnamefont {Maxwell}\ \bibnamefont
  {Aifer}}, \bibinfo {author} {\bibfnamefont {Gavin}\ \bibnamefont {Crooks}},
  \bibinfo {author} {\bibfnamefont {Thomas}\ \bibnamefont {Ahle}}, \ and\
  \bibinfo {author} {\bibfnamefont {Patrick~J}\ \bibnamefont {Coles}},\
  }\bibfield  {title} {\enquote {\bibinfo {title} {Thermodynamic matrix
  exponentials and thermodynamic parallelism},}\ }\href
  {https://arxiv.org/abs/2311.12759} {\bibfield  {journal} {\bibinfo  {journal}
  {arXiv preprint arXiv:2311.12759}\ } (\bibinfo {year} {2023})}\BibitemShut
  {NoStop}%
\bibitem [{\citenamefont {Valiant}(2023)}]{valiant2023matrix}%
  \BibitemOpen
  \bibfield  {author} {\bibinfo {author} {\bibfnamefont {Gregory}\ \bibnamefont
  {Valiant}},\ }\bibfield  {title} {\enquote {\bibinfo {title} {Matrix
  multiplication in quadratic time and energy? towards a fine-grained
  energy-centric church-turing thesis},}\ }\href
  {https://arxiv.org/abs/2311.16342} {\bibfield  {journal} {\bibinfo  {journal}
  {arXiv preprint arXiv:2311.16342}\ } (\bibinfo {year} {2023})}\BibitemShut
  {NoStop}%
\bibitem [{\citenamefont {Shewchuk}\ \emph {et~al.}(1994)\citenamefont
  {Shewchuk} \emph {et~al.}}]{shewchuk1994introduction}%
  \BibitemOpen
  \bibfield  {author} {\bibinfo {author} {\bibfnamefont {Jonathan~Richard}\
  \bibnamefont {Shewchuk}} \emph {et~al.},\ }\href@noop {} {\enquote {\bibinfo
  {title} {An introduction to the conjugate gradient method without the
  agonizing pain},}\ } (\bibinfo {year} {1994})\BibitemShut {NoStop}%
\bibitem [{\citenamefont {Robinson}(2005)}]{robinson2005toward}%
  \BibitemOpen
  \bibfield  {author} {\bibinfo {author} {\bibfnamefont {Sara}\ \bibnamefont
  {Robinson}},\ }\bibfield  {title} {\enquote {\bibinfo {title} {Toward an
  optimal algorithm for matrix multiplication},}\ }\href@noop {} {\bibfield
  {journal} {\bibinfo  {journal} {SIAM news}\ }\textbf {\bibinfo {volume}
  {38}},\ \bibinfo {pages} {1--3} (\bibinfo {year} {2005})}\BibitemShut
  {NoStop}%
\bibitem [{\citenamefont {Bartels}\ and\ \citenamefont
  {Stewart}(1972)}]{bartels1972solution}%
  \BibitemOpen
  \bibfield  {author} {\bibinfo {author} {\bibfnamefont {Richard~H.}\
  \bibnamefont {Bartels}}\ and\ \bibinfo {author} {\bibfnamefont {George~W}\
  \bibnamefont {Stewart}},\ }\bibfield  {title} {\enquote {\bibinfo {title}
  {Solution of the matrix equation ax+ xb= c [f4]},}\ }\href@noop {} {\bibfield
   {journal} {\bibinfo  {journal} {Communications of the ACM}\ }\textbf
  {\bibinfo {volume} {15}},\ \bibinfo {pages} {820--826} (\bibinfo {year}
  {1972})}\BibitemShut {NoStop}%
\bibitem [{\citenamefont {Aho}\ \emph {et~al.}(1974)\citenamefont {Aho},
  \citenamefont {Hopcroft},\ and\ \citenamefont {Ullman}}]{aho1974design}%
  \BibitemOpen
  \bibfield  {author} {\bibinfo {author} {\bibfnamefont {Alfred~V.}\
  \bibnamefont {Aho}}, \bibinfo {author} {\bibfnamefont {John~E.}\ \bibnamefont
  {Hopcroft}}, \ and\ \bibinfo {author} {\bibfnamefont {Jeffrey~D.}\
  \bibnamefont {Ullman}},\ }\href@noop {} {\emph {\bibinfo {title} {{The Design
  and Analysis of Computer Algorithms}}}}\ (\bibinfo  {publisher}
  {Addison-Wesley},\ \bibinfo {year} {1974})\BibitemShut {NoStop}%
\bibitem [{\citenamefont {Gallavotti}(1995)}]{gallavotti1995ergodicity}%
  \BibitemOpen
  \bibfield  {author} {\bibinfo {author} {\bibfnamefont {Giovanni}\
  \bibnamefont {Gallavotti}},\ }\bibfield  {title} {\enquote {\bibinfo {title}
  {Ergodicity, ensembles, irreversibility in boltzmann and beyond},}\
  }\href@noop {} {\bibfield  {journal} {\bibinfo  {journal} {Journal of
  Statistical Physics}\ }\textbf {\bibinfo {volume} {78}},\ \bibinfo {pages}
  {1571--1589} (\bibinfo {year} {1995})}\BibitemShut {NoStop}%
\bibitem [{\citenamefont {Sinai}(1963)}]{sinai1963foundations}%
  \BibitemOpen
  \bibfield  {author} {\bibinfo {author} {\bibfnamefont {Yakov~Grigor'evich}\
  \bibnamefont {Sinai}},\ }\bibfield  {title} {\enquote {\bibinfo {title} {On
  the foundations of the ergodic hypothesis for a dynamical system of
  statistical mechanics},}\ }in\ \href@noop {} {\emph {\bibinfo {booktitle}
  {Doklady Akademii Nauk}}},\ Vol.\ \bibinfo {volume} {153}\ (\bibinfo
  {organization} {Russian Academy of Sciences},\ \bibinfo {year} {1963})\ pp.\
  \bibinfo {pages} {1261--1264}\BibitemShut {NoStop}%
\bibitem [{\citenamefont {Fokker}(1914)}]{fokker1914mittlere}%
  \BibitemOpen
  \bibfield  {author} {\bibinfo {author} {\bibfnamefont {Adriaan~Dani{\"e}l}\
  \bibnamefont {Fokker}},\ }\bibfield  {title} {\enquote {\bibinfo {title} {Die
  mittlere energie rotierender elektrischer dipole im strahlungsfeld},}\
  }\href@noop {} {\bibfield  {journal} {\bibinfo  {journal} {Annalen der
  Physik}\ }\textbf {\bibinfo {volume} {348}},\ \bibinfo {pages} {810--820}
  (\bibinfo {year} {1914})}\BibitemShut {NoStop}%
\bibitem [{\citenamefont {Kubo}(1966)}]{kubo1966fluctuation}%
  \BibitemOpen
  \bibfield  {author} {\bibinfo {author} {\bibfnamefont {Rep}\ \bibnamefont
  {Kubo}},\ }\bibfield  {title} {\enquote {\bibinfo {title} {The
  fluctuation-dissipation theorem},}\ }\href@noop {} {\bibfield  {journal}
  {\bibinfo  {journal} {Reports on progress in physics}\ }\textbf {\bibinfo
  {volume} {29}},\ \bibinfo {pages} {255} (\bibinfo {year} {1966})}\BibitemShut
  {NoStop}%
\bibitem [{\citenamefont {Weber}(1956)}]{weber1956fluctuation}%
  \BibitemOpen
  \bibfield  {author} {\bibinfo {author} {\bibfnamefont {J}~\bibnamefont
  {Weber}},\ }\bibfield  {title} {\enquote {\bibinfo {title} {Fluctuation
  dissipation theorem},}\ }\href@noop {} {\bibfield  {journal} {\bibinfo
  {journal} {Physical Review}\ }\textbf {\bibinfo {volume} {101}},\ \bibinfo
  {pages} {1620} (\bibinfo {year} {1956})}\BibitemShut {NoStop}%
\bibitem [{\citenamefont {Christ}\ \emph {et~al.}(2010)\citenamefont {Christ},
  \citenamefont {Mark},\ and\ \citenamefont {Van~Gunsteren}}]{christ2010basic}%
  \BibitemOpen
  \bibfield  {author} {\bibinfo {author} {\bibfnamefont {Clara~D}\ \bibnamefont
  {Christ}}, \bibinfo {author} {\bibfnamefont {Alan~E}\ \bibnamefont {Mark}}, \
  and\ \bibinfo {author} {\bibfnamefont {Wilfred~F}\ \bibnamefont
  {Van~Gunsteren}},\ }\bibfield  {title} {\enquote {\bibinfo {title} {Basic
  ingredients of free energy calculations: a review},}\ }\href@noop {}
  {\bibfield  {journal} {\bibinfo  {journal} {Journal of computational
  chemistry}\ }\textbf {\bibinfo {volume} {31}},\ \bibinfo {pages} {1569--1582}
  (\bibinfo {year} {2010})}\BibitemShut {NoStop}%
\bibitem [{\citenamefont {Jarzynski}(1997)}]{jarzynski1997nonequilibrium}%
  \BibitemOpen
  \bibfield  {author} {\bibinfo {author} {\bibfnamefont {Christopher}\
  \bibnamefont {Jarzynski}},\ }\bibfield  {title} {\enquote {\bibinfo {title}
  {Nonequilibrium equality for free energy differences},}\ }\href@noop {}
  {\bibfield  {journal} {\bibinfo  {journal} {Physical Review Letters}\
  }\textbf {\bibinfo {volume} {78}},\ \bibinfo {pages} {2690} (\bibinfo {year}
  {1997})}\BibitemShut {NoStop}%
\bibitem [{\citenamefont {Aifer}\ \emph {et~al.}(2024)\citenamefont {Aifer},
  \citenamefont {Melanson}, \citenamefont {Donatella}, \citenamefont {Crooks},
  \citenamefont {Ahle},\ and\ \citenamefont {Coles}}]{aifer2024error}%
  \BibitemOpen
  \bibfield  {author} {\bibinfo {author} {\bibfnamefont {Maxwell}\ \bibnamefont
  {Aifer}}, \bibinfo {author} {\bibfnamefont {Denis}\ \bibnamefont {Melanson}},
  \bibinfo {author} {\bibfnamefont {Kaelan}\ \bibnamefont {Donatella}},
  \bibinfo {author} {\bibfnamefont {Gavin}\ \bibnamefont {Crooks}}, \bibinfo
  {author} {\bibfnamefont {Thomas}\ \bibnamefont {Ahle}}, \ and\ \bibinfo
  {author} {\bibfnamefont {Patrick~J}\ \bibnamefont {Coles}},\ }\bibfield
  {title} {\enquote {\bibinfo {title} {Error mitigation for thermodynamic
  computing},}\ }\href {https://arxiv.org/abs/2401.16231} {\bibfield  {journal}
  {\bibinfo  {journal} {arXiv preprint arXiv:2401.16231}\ } (\bibinfo {year}
  {2024})}\BibitemShut {NoStop}%
\bibitem [{\citenamefont {Haensch}\ \emph {et~al.}(2018)\citenamefont
  {Haensch}, \citenamefont {Gokmen},\ and\ \citenamefont
  {Puri}}]{haensch2018next}%
  \BibitemOpen
  \bibfield  {author} {\bibinfo {author} {\bibfnamefont {Wilfried}\
  \bibnamefont {Haensch}}, \bibinfo {author} {\bibfnamefont {Tayfun}\
  \bibnamefont {Gokmen}}, \ and\ \bibinfo {author} {\bibfnamefont {Ruchir}\
  \bibnamefont {Puri}},\ }\bibfield  {title} {\enquote {\bibinfo {title} {The
  next generation of deep learning hardware: Analog computing},}\ }\href@noop
  {} {\bibfield  {journal} {\bibinfo  {journal} {Proceedings of the IEEE}\
  }\textbf {\bibinfo {volume} {107}},\ \bibinfo {pages} {108--122} (\bibinfo
  {year} {2018})}\BibitemShut {NoStop}%
\bibitem [{\citenamefont {Small}(1993)}]{small1993general}%
  \BibitemOpen
  \bibfield  {author} {\bibinfo {author} {\bibfnamefont {James~S}\ \bibnamefont
  {Small}},\ }\bibfield  {title} {\enquote {\bibinfo {title} {General-purpose
  electronic analog computing: 1945-1965},}\ }\href@noop {} {\bibfield
  {journal} {\bibinfo  {journal} {IEEE Annals of the History of Computing}\
  }\textbf {\bibinfo {volume} {15}},\ \bibinfo {pages} {8--18} (\bibinfo {year}
  {1993})}\BibitemShut {NoStop}%
\bibitem [{\citenamefont {Nielsen}\ and\ \citenamefont
  {Chuang}(2000)}]{nielsen2000quantum}%
  \BibitemOpen
  \bibfield  {author} {\bibinfo {author} {\bibfnamefont {Michael~A.}\
  \bibnamefont {Nielsen}}\ and\ \bibinfo {author} {\bibfnamefont {Isaac~L.}\
  \bibnamefont {Chuang}},\ }\href@noop {} {\emph {\bibinfo {title} {Quantum
  Computation and Quantum Information}}}\ (\bibinfo  {publisher} {Cambridge
  University Press},\ \bibinfo {address} {Cambridge},\ \bibinfo {year}
  {2000})\BibitemShut {NoStop}%
\bibitem [{\citenamefont {Huang}\ \emph {et~al.}(2017)\citenamefont {Huang},
  \citenamefont {Guo}, \citenamefont {Seok}, \citenamefont {Tsividis},\ and\
  \citenamefont {Sethumadhavan}}]{huang2017analog}%
  \BibitemOpen
  \bibfield  {author} {\bibinfo {author} {\bibfnamefont {Yipeng}\ \bibnamefont
  {Huang}}, \bibinfo {author} {\bibfnamefont {Ning}\ \bibnamefont {Guo}},
  \bibinfo {author} {\bibfnamefont {Mingoo}\ \bibnamefont {Seok}}, \bibinfo
  {author} {\bibfnamefont {Yannis}\ \bibnamefont {Tsividis}}, \ and\ \bibinfo
  {author} {\bibfnamefont {Simha}\ \bibnamefont {Sethumadhavan}},\ }\bibfield
  {title} {\enquote {\bibinfo {title} {Analog computing in a modern context: A
  linear algebra accelerator case study},}\ }\href@noop {} {\bibfield
  {journal} {\bibinfo  {journal} {IEEE Micro}\ }\textbf {\bibinfo {volume}
  {37}},\ \bibinfo {pages} {30--38} (\bibinfo {year} {2017})}\BibitemShut
  {NoStop}%
\bibitem [{\citenamefont {Crooks}(2007)}]{crooks2007measuring}%
  \BibitemOpen
  \bibfield  {author} {\bibinfo {author} {\bibfnamefont {Gavin~E.}\
  \bibnamefont {Crooks}},\ }\bibfield  {title} {\enquote {\bibinfo {title}
  {Measuring thermodynamic length},}\ }\href {\doibase
  10.1103/PhysRevLett.99.100602} {\bibfield  {journal} {\bibinfo  {journal}
  {Phys. Rev. Lett.}\ }\textbf {\bibinfo {volume} {99}},\ \bibinfo {pages}
  {100602} (\bibinfo {year} {2007})}\BibitemShut {NoStop}%
\bibitem [{\citenamefont {Cafaro}\ \emph {et~al.}(2022)\citenamefont {Cafaro},
  \citenamefont {Luongo}, \citenamefont {Mancini},\ and\ \citenamefont
  {Quevedo}}]{cafaro2022thermodynamic}%
  \BibitemOpen
  \bibfield  {author} {\bibinfo {author} {\bibfnamefont {Carlo}\ \bibnamefont
  {Cafaro}}, \bibinfo {author} {\bibfnamefont {Orlando}\ \bibnamefont
  {Luongo}}, \bibinfo {author} {\bibfnamefont {Stefano}\ \bibnamefont
  {Mancini}}, \ and\ \bibinfo {author} {\bibfnamefont {Hernando}\ \bibnamefont
  {Quevedo}},\ }\bibfield  {title} {\enquote {\bibinfo {title} {Thermodynamic
  length, geometric efficiency and legendre invariance},}\ }\href@noop {}
  {\bibfield  {journal} {\bibinfo  {journal} {Physica A: Statistical Mechanics
  and its Applications}\ }\textbf {\bibinfo {volume} {590}},\ \bibinfo {pages}
  {126740} (\bibinfo {year} {2022})}\BibitemShut {NoStop}%
\bibitem [{\citenamefont {Quevedo}(2007)}]{quevedo2007geometrothermodynamics}%
  \BibitemOpen
  \bibfield  {author} {\bibinfo {author} {\bibfnamefont {Hernando}\
  \bibnamefont {Quevedo}},\ }\bibfield  {title} {\enquote {\bibinfo {title}
  {Geometrothermodynamics},}\ }\href@noop {} {\bibfield  {journal} {\bibinfo
  {journal} {Journal of Mathematical Physics}\ }\textbf {\bibinfo {volume}
  {48}} (\bibinfo {year} {2007})}\BibitemShut {NoStop}%
\bibitem [{\citenamefont {Andresen}(1996)}]{andresen1996finite}%
  \BibitemOpen
  \bibfield  {author} {\bibinfo {author} {\bibfnamefont {Bjarne}\ \bibnamefont
  {Andresen}},\ }\bibfield  {title} {\enquote {\bibinfo {title} {Finite-time
  thermodynamics and thermodynamic length},}\ }\href@noop {} {\bibfield
  {journal} {\bibinfo  {journal} {Revue g{\'e}n{\'e}rale de thermique}\
  }\textbf {\bibinfo {volume} {35}},\ \bibinfo {pages} {647--650} (\bibinfo
  {year} {1996})}\BibitemShut {NoStop}%
\bibitem [{\citenamefont {Chen}\ \emph {et~al.}(2021)\citenamefont {Chen},
  \citenamefont {Sun},\ and\ \citenamefont {Dong}}]{chen2021extrapolating}%
  \BibitemOpen
  \bibfield  {author} {\bibinfo {author} {\bibfnamefont {Jin-Fu}\ \bibnamefont
  {Chen}}, \bibinfo {author} {\bibfnamefont {CP}~\bibnamefont {Sun}}, \ and\
  \bibinfo {author} {\bibfnamefont {Hui}\ \bibnamefont {Dong}},\ }\bibfield
  {title} {\enquote {\bibinfo {title} {Extrapolating the thermodynamic length
  with finite-time measurements},}\ }\href@noop {} {\bibfield  {journal}
  {\bibinfo  {journal} {Physical Review E}\ }\textbf {\bibinfo {volume}
  {104}},\ \bibinfo {pages} {034117} (\bibinfo {year} {2021})}\BibitemShut
  {NoStop}%
\bibitem [{\citenamefont {Saptharishi}(2015)}]{saptharishi2015survey}%
  \BibitemOpen
  \bibfield  {author} {\bibinfo {author} {\bibfnamefont {Ramprasad}\
  \bibnamefont {Saptharishi}},\ }\bibfield  {title} {\enquote {\bibinfo {title}
  {A survey of lower bounds in arithmetic circuit complexity},}\ }\href@noop {}
  {\bibfield  {journal} {\bibinfo  {journal} {Github survey}\ }\textbf
  {\bibinfo {volume} {95}} (\bibinfo {year} {2015})}\BibitemShut {NoStop}%
\bibitem [{\citenamefont {Chiribella}\ \emph {et~al.}(2022)\citenamefont
  {Chiribella}, \citenamefont {Meng}, \citenamefont {Renner},\ and\
  \citenamefont {Yung}}]{chiribella2022nonequilibrium}%
  \BibitemOpen
  \bibfield  {author} {\bibinfo {author} {\bibfnamefont {Giulio}\ \bibnamefont
  {Chiribella}}, \bibinfo {author} {\bibfnamefont {Fei}\ \bibnamefont {Meng}},
  \bibinfo {author} {\bibfnamefont {Renato}\ \bibnamefont {Renner}}, \ and\
  \bibinfo {author} {\bibfnamefont {Man-Hong}\ \bibnamefont {Yung}},\
  }\bibfield  {title} {\enquote {\bibinfo {title} {The nonequilibrium cost of
  accurate information processing},}\ }\href@noop {} {\bibfield  {journal}
  {\bibinfo  {journal} {Nature Communications}\ }\textbf {\bibinfo {volume}
  {13}},\ \bibinfo {pages} {7155} (\bibinfo {year} {2022})}\BibitemShut
  {NoStop}%
\bibitem [{\citenamefont {Riechers}(2018)}]{riechers2018transforming}%
  \BibitemOpen
  \bibfield  {author} {\bibinfo {author} {\bibfnamefont {Paul~M}\ \bibnamefont
  {Riechers}},\ }\bibfield  {title} {\enquote {\bibinfo {title} {Transforming
  metastable memories: The nonequilibrium thermodynamics of computation},}\
  }\href@noop {} {\bibfield  {journal} {\bibinfo  {journal} {arXiv preprint
  arXiv:1808.03429}\ } (\bibinfo {year} {2018})}\BibitemShut {NoStop}%
\bibitem [{\citenamefont {Melanson}\ \emph {et~al.}(2023)\citenamefont
  {Melanson}, \citenamefont {Khater}, \citenamefont {Aifer}, \citenamefont
  {Donatella}, \citenamefont {Gordon}, \citenamefont {Ahle}, \citenamefont
  {Crooks}, \citenamefont {Martinez}, \citenamefont {Sbahi},\ and\
  \citenamefont {Coles}}]{melanson2023thermodynamic}%
  \BibitemOpen
  \bibfield  {author} {\bibinfo {author} {\bibfnamefont {Denis}\ \bibnamefont
  {Melanson}}, \bibinfo {author} {\bibfnamefont {Mohammad~Abu}\ \bibnamefont
  {Khater}}, \bibinfo {author} {\bibfnamefont {Maxwell}\ \bibnamefont {Aifer}},
  \bibinfo {author} {\bibfnamefont {Kaelan}\ \bibnamefont {Donatella}},
  \bibinfo {author} {\bibfnamefont {Max~Hunter}\ \bibnamefont {Gordon}},
  \bibinfo {author} {\bibfnamefont {Thomas}\ \bibnamefont {Ahle}}, \bibinfo
  {author} {\bibfnamefont {Gavin}\ \bibnamefont {Crooks}}, \bibinfo {author}
  {\bibfnamefont {Antonio~J}\ \bibnamefont {Martinez}}, \bibinfo {author}
  {\bibfnamefont {Faris}\ \bibnamefont {Sbahi}}, \ and\ \bibinfo {author}
  {\bibfnamefont {Patrick~J}\ \bibnamefont {Coles}},\ }\bibfield  {title}
  {\enquote {\bibinfo {title} {Thermodynamic computing system for {AI}
  applications},}\ }\href {https://arxiv.org/abs/2312.04836} {\bibfield
  {journal} {\bibinfo  {journal} {arXiv preprint arXiv:2312.04836}\ } (\bibinfo
  {year} {2023})}\BibitemShut {NoStop}%
\bibitem [{\citenamefont {Doerries}\ \emph {et~al.}(2021)\citenamefont
  {Doerries}, \citenamefont {Loos},\ and\ \citenamefont
  {Klapp}}]{doerries2021correlation}%
  \BibitemOpen
  \bibfield  {author} {\bibinfo {author} {\bibfnamefont {Timo~J}\ \bibnamefont
  {Doerries}}, \bibinfo {author} {\bibfnamefont {Sarah~AM}\ \bibnamefont
  {Loos}}, \ and\ \bibinfo {author} {\bibfnamefont {Sabine~HL}\ \bibnamefont
  {Klapp}},\ }\bibfield  {title} {\enquote {\bibinfo {title} {Correlation
  functions of non-markovian systems out of equilibrium: Analytical expressions
  beyond single-exponential memory},}\ }\href@noop {} {\bibfield  {journal}
  {\bibinfo  {journal} {Journal of Statistical Mechanics: Theory and
  Experiment}\ }\textbf {\bibinfo {volume} {2021}},\ \bibinfo {pages} {033202}
  (\bibinfo {year} {2021})}\BibitemShut {NoStop}%
\bibitem [{\citenamefont {H'walisz}\ \emph {et~al.}(1989)\citenamefont
  {H'walisz}, \citenamefont {Jung}, \citenamefont {H{\"a}nggi}, \citenamefont
  {Talkner},\ and\ \citenamefont {Schimansky-Geier}}]{h1989colored}%
  \BibitemOpen
  \bibfield  {author} {\bibinfo {author} {\bibfnamefont {Lutz}\ \bibnamefont
  {H'walisz}}, \bibinfo {author} {\bibfnamefont {Peter}\ \bibnamefont {Jung}},
  \bibinfo {author} {\bibfnamefont {Peter}\ \bibnamefont {H{\"a}nggi}},
  \bibinfo {author} {\bibfnamefont {Peter}\ \bibnamefont {Talkner}}, \ and\
  \bibinfo {author} {\bibfnamefont {Lutz}\ \bibnamefont {Schimansky-Geier}},\
  }\bibfield  {title} {\enquote {\bibinfo {title} {Colored noise driven systems
  with inertia},}\ }\href@noop {} {\bibfield  {journal} {\bibinfo  {journal}
  {Zeitschrift f{\"u}r Physik B Condensed Matter}\ }\textbf {\bibinfo {volume}
  {77}},\ \bibinfo {pages} {471--483} (\bibinfo {year} {1989})}\BibitemShut
  {NoStop}%
\bibitem [{\citenamefont {Gardiner}(1985)}]{gardiner1985handbook}%
  \BibitemOpen
  \bibfield  {author} {\bibinfo {author} {\bibfnamefont {Crispin~W.}\
  \bibnamefont {Gardiner}},\ }\href@noop {} {\emph {\bibinfo {title} {Handbook
  of stochastic methods}}},\ Vol.~\bibinfo {volume} {3}\ (\bibinfo  {publisher}
  {springer Berlin},\ \bibinfo {year} {1985})\BibitemShut {NoStop}%
\bibitem [{\citenamefont {Frostig}\ \emph {et~al.}(2018)\citenamefont
  {Frostig}, \citenamefont {Johnson},\ and\ \citenamefont
  {Leary}}]{frostig2018compiling}%
  \BibitemOpen
  \bibfield  {author} {\bibinfo {author} {\bibfnamefont {Roy}\ \bibnamefont
  {Frostig}}, \bibinfo {author} {\bibfnamefont {Matthew~James}\ \bibnamefont
  {Johnson}}, \ and\ \bibinfo {author} {\bibfnamefont {Chris}\ \bibnamefont
  {Leary}},\ }\bibfield  {title} {\enquote {\bibinfo {title} {Compiling machine
  learning programs via high-level tracing},}\ }\href@noop {} {\bibfield
  {journal} {\bibinfo  {journal} {Systems for Machine Learning}\ }\textbf
  {\bibinfo {volume} {4}} (\bibinfo {year} {2018})}\BibitemShut {NoStop}%
\bibitem [{\citenamefont {Callen}(1998)}]{callen1998thermodynamics}%
  \BibitemOpen
  \bibfield  {author} {\bibinfo {author} {\bibfnamefont {Herbert~B}\
  \bibnamefont {Callen}},\ }\href@noop {} {\enquote {\bibinfo {title}
  {Thermodynamics and an introduction to thermostatistics},}\ } (\bibinfo
  {year} {1998})\BibitemShut {NoStop}%
\bibitem [{\citenamefont {Zwanzig}(2001)}]{zwanzig2001nonequilibrium}%
  \BibitemOpen
  \bibfield  {author} {\bibinfo {author} {\bibfnamefont {Robert}\ \bibnamefont
  {Zwanzig}},\ }\href@noop {} {\emph {\bibinfo {title} {Nonequilibrium
  statistical mechanics}}}\ (\bibinfo  {publisher} {Oxford university press},\
  \bibinfo {year} {2001})\BibitemShut {NoStop}%
\bibitem [{\citenamefont {Doob}(1942)}]{doob1942brownian}%
  \BibitemOpen
  \bibfield  {author} {\bibinfo {author} {\bibfnamefont {Joseph~L}\
  \bibnamefont {Doob}},\ }\bibfield  {title} {\enquote {\bibinfo {title} {The
  brownian movement and stochastic equations},}\ }\href@noop {} {\bibfield
  {journal} {\bibinfo  {journal} {Annals of Mathematics}\ ,\ \bibinfo {pages}
  {351--369}} (\bibinfo {year} {1942})}\BibitemShut {NoStop}%
\bibitem [{\citenamefont {Lemons}\ and\ \citenamefont
  {Gythiel}(1997)}]{lemons1997paul}%
  \BibitemOpen
  \bibfield  {author} {\bibinfo {author} {\bibfnamefont {Don~S}\ \bibnamefont
  {Lemons}}\ and\ \bibinfo {author} {\bibfnamefont {Anthony}\ \bibnamefont
  {Gythiel}},\ }\bibfield  {title} {\enquote {\bibinfo {title} {Paul
  langevin’s 1908 paper “on the theory of brownian motion”[“sur la
  th{\'e}orie du mouvement brownien,” cr acad. sci.(paris) 146, 530--533
  (1908)]},}\ }\href@noop {} {\bibfield  {journal} {\bibinfo  {journal}
  {American Journal of Physics}\ }\textbf {\bibinfo {volume} {65}},\ \bibinfo
  {pages} {1079--1081} (\bibinfo {year} {1997})}\BibitemShut {NoStop}%
\bibitem [{\citenamefont {Hatano}\ and\ \citenamefont
  {Sasa}(2001)}]{hatano2001steady}%
  \BibitemOpen
  \bibfield  {author} {\bibinfo {author} {\bibfnamefont {Takahiro}\
  \bibnamefont {Hatano}}\ and\ \bibinfo {author} {\bibfnamefont {Shin-ichi}\
  \bibnamefont {Sasa}},\ }\bibfield  {title} {\enquote {\bibinfo {title}
  {Steady-state thermodynamics of langevin systems},}\ }\href@noop {}
  {\bibfield  {journal} {\bibinfo  {journal} {Physical review letters}\
  }\textbf {\bibinfo {volume} {86}},\ \bibinfo {pages} {3463} (\bibinfo {year}
  {2001})}\BibitemShut {NoStop}%
\bibitem [{\citenamefont {MacKay}\ \emph {et~al.}(1998)\citenamefont {MacKay}
  \emph {et~al.}}]{mackay1998introduction}%
  \BibitemOpen
  \bibfield  {author} {\bibinfo {author} {\bibfnamefont {David~JC}\
  \bibnamefont {MacKay}} \emph {et~al.},\ }\bibfield  {title} {\enquote
  {\bibinfo {title} {Introduction to gaussian processes},}\ }\href@noop {}
  {\bibfield  {journal} {\bibinfo  {journal} {NATO ASI series F computer and
  systems sciences}\ }\textbf {\bibinfo {volume} {168}},\ \bibinfo {pages}
  {133--166} (\bibinfo {year} {1998})}\BibitemShut {NoStop}%
\bibitem [{\citenamefont {Seifert}(2008)}]{seifert2008stochastic}%
  \BibitemOpen
  \bibfield  {author} {\bibinfo {author} {\bibfnamefont {Udo}\ \bibnamefont
  {Seifert}},\ }\bibfield  {title} {\enquote {\bibinfo {title} {Stochastic
  thermodynamics: principles and perspectives},}\ }\href@noop {} {\bibfield
  {journal} {\bibinfo  {journal} {The European Physical Journal B}\ }\textbf
  {\bibinfo {volume} {64}},\ \bibinfo {pages} {423--431} (\bibinfo {year}
  {2008})}\BibitemShut {NoStop}%
\bibitem [{\citenamefont {Jarzynski}(2011)}]{jarzynski2011equalities}%
  \BibitemOpen
  \bibfield  {author} {\bibinfo {author} {\bibfnamefont {Christopher}\
  \bibnamefont {Jarzynski}},\ }\bibfield  {title} {\enquote {\bibinfo {title}
  {Equalities and inequalities: Irreversibility and the second law of
  thermodynamics at the nanoscale},}\ }\href@noop {} {\bibfield  {journal}
  {\bibinfo  {journal} {Annu. Rev. Condens. Matter Phys.}\ }\textbf {\bibinfo
  {volume} {2}},\ \bibinfo {pages} {329--351} (\bibinfo {year}
  {2011})}\BibitemShut {NoStop}%
\bibitem [{\citenamefont {Shenfeld}\ \emph {et~al.}(2009)\citenamefont
  {Shenfeld}, \citenamefont {Xu}, \citenamefont {Eastwood}, \citenamefont
  {Dror},\ and\ \citenamefont {Shaw}}]{shenfeld2009minimizing}%
  \BibitemOpen
  \bibfield  {author} {\bibinfo {author} {\bibfnamefont {Daniel~K.}\
  \bibnamefont {Shenfeld}}, \bibinfo {author} {\bibfnamefont {Huafeng}\
  \bibnamefont {Xu}}, \bibinfo {author} {\bibfnamefont {Michael~P.}\
  \bibnamefont {Eastwood}}, \bibinfo {author} {\bibfnamefont {Ron~O.}\
  \bibnamefont {Dror}}, \ and\ \bibinfo {author} {\bibfnamefont {David~E.}\
  \bibnamefont {Shaw}},\ }\bibfield  {title} {\enquote {\bibinfo {title}
  {Minimizing thermodynamic length to select intermediate states for
  free-energy calculations and replica-exchange simulations},}\ }\href
  {\doibase 10.1103/PhysRevE.80.046705} {\bibfield  {journal} {\bibinfo
  {journal} {Phys. Rev. E}\ }\textbf {\bibinfo {volume} {80}},\ \bibinfo
  {pages} {046705} (\bibinfo {year} {2009})}\BibitemShut {NoStop}%
\bibitem [{\citenamefont {Amari}\ and\ \citenamefont
  {Nagaoka}(2000)}]{amari2000methods}%
  \BibitemOpen
  \bibfield  {author} {\bibinfo {author} {\bibfnamefont {Shun-ichi}\
  \bibnamefont {Amari}}\ and\ \bibinfo {author} {\bibfnamefont {Hiroshi}\
  \bibnamefont {Nagaoka}},\ }\href@noop {} {\emph {\bibinfo {title} {Methods of
  information geometry}}},\ Vol.\ \bibinfo {volume} {191}\ (\bibinfo
  {publisher} {American Mathematical Soc.},\ \bibinfo {year}
  {2000})\BibitemShut {NoStop}%
\bibitem [{\citenamefont {Jarzynski}(2006)}]{jarzynski2006rare}%
  \BibitemOpen
  \bibfield  {author} {\bibinfo {author} {\bibfnamefont {Christopher}\
  \bibnamefont {Jarzynski}},\ }\bibfield  {title} {\enquote {\bibinfo {title}
  {Rare events and the convergence of exponentially averaged work values},}\
  }\href@noop {} {\bibfield  {journal} {\bibinfo  {journal} {Physical Review
  E}\ }\textbf {\bibinfo {volume} {73}},\ \bibinfo {pages} {046105} (\bibinfo
  {year} {2006})}\BibitemShut {NoStop}%
\bibitem [{\citenamefont {Shirts}\ \emph {et~al.}(2003)\citenamefont {Shirts},
  \citenamefont {Bair}, \citenamefont {Hooker},\ and\ \citenamefont
  {Pande}}]{shirts2003equilibrium}%
  \BibitemOpen
  \bibfield  {author} {\bibinfo {author} {\bibfnamefont {Michael~R.}\
  \bibnamefont {Shirts}}, \bibinfo {author} {\bibfnamefont {Eric}\ \bibnamefont
  {Bair}}, \bibinfo {author} {\bibfnamefont {Giles}\ \bibnamefont {Hooker}}, \
  and\ \bibinfo {author} {\bibfnamefont {Vijay~S.}\ \bibnamefont {Pande}},\
  }\bibfield  {title} {\enquote {\bibinfo {title} {Equilibrium free energies
  from nonequilibrium measurements using maximum-likelihood methods},}\ }\href
  {\doibase 10.1103/PhysRevLett.91.140601} {\bibfield  {journal} {\bibinfo
  {journal} {Phys. Rev. Lett.}\ }\textbf {\bibinfo {volume} {91}},\ \bibinfo
  {pages} {140601} (\bibinfo {year} {2003})}\BibitemShut {NoStop}%
\bibitem [{\citenamefont {Bennett}(1976)}]{bennett1976efficient}%
  \BibitemOpen
  \bibfield  {author} {\bibinfo {author} {\bibfnamefont {Charles~H}\
  \bibnamefont {Bennett}},\ }\bibfield  {title} {\enquote {\bibinfo {title}
  {Efficient estimation of free energy differences from monte carlo data},}\
  }\href@noop {} {\bibfield  {journal} {\bibinfo  {journal} {Journal of
  Computational Physics}\ }\textbf {\bibinfo {volume} {22}},\ \bibinfo {pages}
  {245--268} (\bibinfo {year} {1976})}\BibitemShut {NoStop}%
\bibitem [{\citenamefont {Shirts}\ and\ \citenamefont
  {Chodera}(2008)}]{shirts2008statistically}%
  \BibitemOpen
  \bibfield  {author} {\bibinfo {author} {\bibfnamefont {Michael~R}\
  \bibnamefont {Shirts}}\ and\ \bibinfo {author} {\bibfnamefont {John~D}\
  \bibnamefont {Chodera}},\ }\bibfield  {title} {\enquote {\bibinfo {title}
  {Statistically optimal analysis of samples from multiple equilibrium
  states},}\ }\href@noop {} {\bibfield  {journal} {\bibinfo  {journal} {The
  Journal of chemical physics}\ }\textbf {\bibinfo {volume} {129}} (\bibinfo
  {year} {2008})}\BibitemShut {NoStop}%
\bibitem [{\citenamefont {Koopmans}(1995)}]{koopmans1995spectral}%
  \BibitemOpen
  \bibfield  {author} {\bibinfo {author} {\bibfnamefont {Lambert~H}\
  \bibnamefont {Koopmans}},\ }\href@noop {} {\emph {\bibinfo {title} {The
  spectral analysis of time series}}}\ (\bibinfo  {publisher} {Elsevier},\
  \bibinfo {year} {1995})\BibitemShut {NoStop}%
\bibitem [{\citenamefont {Cram{\'e}r}(1999)}]{cramer1999mathematical}%
  \BibitemOpen
  \bibfield  {author} {\bibinfo {author} {\bibfnamefont {Harald}\ \bibnamefont
  {Cram{\'e}r}},\ }\href@noop {} {\emph {\bibinfo {title} {Mathematical methods
  of statistics}}},\ Vol.~\bibinfo {volume} {43}\ (\bibinfo  {publisher}
  {Princeton university press},\ \bibinfo {year} {1999})\BibitemShut {NoStop}%
\bibitem [{\citenamefont {Nielsen}(2023)}]{nielsen2023simple}%
  \BibitemOpen
  \bibfield  {author} {\bibinfo {author} {\bibfnamefont {Frank}\ \bibnamefont
  {Nielsen}},\ }\bibfield  {title} {\enquote {\bibinfo {title} {A simple
  approximation method for the {Fisher--Rao} distance between multivariate
  normal distributions},}\ }\href {\doibase 10.3390/e25040654} {\bibfield
  {journal} {\bibinfo  {journal} {Entropy}\ }\textbf {\bibinfo {volume} {25}},\
  \bibinfo {pages} {654} (\bibinfo {year} {2023})}\BibitemShut {NoStop}%
\end{thebibliography}%

\pagebreak

\begin{appendix}

\begin{center}
\Large Supplemental Information for ``Thermodynamic Linear Algebra''
\end{center}

In this Supplemental Information we provide a derivation of the stationary distributions and equilibration times for systems governed by overdamped and underdamped Langevin equations. We also formulate rigorous statements of ergodicity (the time average of an observable converging to its ensemble average in the long-time limit). Ergodicity is proved for the first and second moments of the stationary distribution, and the integration times necessary for ergodicity of these observables are identified. In addition, we provide further details on our algorithms for determinant estimation and solving the Lyapunov equation.

\section{Thermodynamic Framework}

The algorithms detailed in the main text were developed using simple thermodynamic arguments, and the analysis of their performance makes use of ideas from the field of stochastic thermodynamics. We provide a brief summary of the relevant concepts here.

\subsection{Equilibrium}
Suppose that a system's state can be completely described (on a microscopic level) by a vector of generalized coordinates $x = (x_1 \dots x_d)^\intercal$ and canonically conjugate momenta $p = (p_1 \dots p_d)^\intercal$, and that the energy of the system is given by a Hamiltonian function $H(x,p)$
\begin{equation}
    E = H(x,p).
\end{equation}
A statistical ensemble is an imaginary collection of copies of this system, each of which has its own coordinates $(x,p)$ and energy $E$. The relative density of copies in different parts of the coordinate space is described by a non-negative density function $f(x,p)$, satisfying a normalization constraint $\int dr f(x,p) = 1$. Ensembles describe macroscopic states (macrostates) of complex systems, where the microscopic state (microstate) is not precisely known. An observable quantity which depends on the microstate $O(r)$ has an \emph{ensemble average} $\braket{O}$, defined as
\begin{equation}
    \braket{O} = \int dx \, dp \, f(x,p) O(x,p).
\end{equation}
One such ensemble, called the \emph{canonical ensemble}, has the density function \cite{callen1998thermodynamics}
\begin{equation}
f_\beta(r) = \frac{1}{Z} \exp\left(-\beta H(x,p)\right),
\end{equation}
where $\beta = 1/k_B T$ is the inverse temperature and $Z$ is partition function
\begin{equation}
    Z = \int dx\, dp \, \exp\left( -\beta H(x,p)\right).
\end{equation}
The canonical ensemble is the macrostate of a system in \emph{thermal equilibrium} with an environment at temperature $T$, which is large and does not change in response to changes in the system, a heat reservoir. 

In nature, we do not encounter anywhere the infinite collection of copies implied by the statistical ensemble; instead we have access to one particular instance of a system at different moments in time. The usage of the ensemble concept is justified by the property of \emph{ergodicity} observed empirically in many systems. A system is ergodic insofar as its observable quantities may be averaged over time to yield the appropriate ensemble averages, that is \cite{zwanzig2001nonequilibrium}
\begin{equation}
  \overline{O} =   \frac{1}{\tau} \int_{t_0}^\tau dt \, O(t) = \braket{O}
\end{equation}
for any $t_0$ and sufficiently long $\tau$. This characterization of ergodicity should not be taken as a definition, but more of a template; in practice, we will only claim (and prove) that some particular observable $O$ is ergodic when averaged over a sufficiently long length of time, which depends on the observable.

Of interest in this work are Hamiltonians of the form
\begin{equation}
    H(x,p) = U(x) + \frac{1}{2M} p^\intercal p,
\end{equation}
where $M\in \mathbb{R}^+$ and the potential is of the form
\begin{equation}
\label{eq:full-potential}
U(x) = \frac{1}{2} x^\intercal A x - b^\intercal x,
\end{equation}
for a symmetric, positive-definite $A\in \mathbb{R}^{d\times d}$ and $b\in \mathbb{R}^d$. The above can be written
\begin{equation}
\label{eq:full-hamiltonian}
    H(x,p) = \frac{1}{2}(x - A^{-1}b)^\intercal A(x - A^{-1} b) - \frac{1}{2}b^\intercal A^{-1} b + \frac{1}{2M}p^\intercal p.
\end{equation}
As a rule, the Hamiltonian may be increased or decreased by a constant without any observable effect, and equivalently any term appearing in the Hamiltonian that does not depend on $x$ or $p$ may be absorbed into the partition function. In this case, we may take
\begin{equation}
\label{eq:full-hamiltonian-shifted}
   H(x,p) = \frac{1}{2}(x - A^{-1} b)^\intercal A(x - A^{-1} b)+\frac{1}{2M}p^\intercal p,
\end{equation}
so the density function of the canonical ensemble is
\begin{equation}
f(x,p) = \frac{1}{Z}\exp \left( -\frac{\beta}{2}(x - A^{-1} b)^\intercal A(x - A^{-1} b) + \frac{1}{2M} p^\intercal p \right),
\end{equation}
which we often write as
\begin{equation}
\label{eq:full-hamiltonian-gibbs-dist}
    x \sim \mathcal{N}[A^{-1}b, \beta^{-1}A^{-1}], \: \: p \sim  \mathcal{N}\left[0, \beta^{-1} M \mathbb{I}\right].
\end{equation}

\subsection{Stochasticity}
The property of ergodicity outlined previously is very strict, and is generally not valid on small timescales, where the system may display rich behavior which is not captured by the long time averages. This behavior is the province of stochastic thermodynamics. Due to the largeness of the heat reservoir, its interactions with the system result in effectively random displacements of the coordinate vector, which we model by Brownian motion. In the absence of any other dynamics, (anisotropic) Brownian motion is given by the equation \cite{doob1942brownian}
\begin{equation}
\label{eq:pure-brownian-sde}
    dx = \mathcal{N}[0, B dt],
\end{equation}
where $B\in \mathbb{R}^{d\times d}$ is symmetric and positive definite, and we abuse notation by writing $\mathcal{N}[\mu, \Sigma]$ to denote a vector drawn from this distribution. Equation \eqref{eq:pure-brownian-sde} is self-consistent in that subdividing the time step $dt$ into two equal parts gives
\begin{equation}
    x(t+dt) = x(t) + \mathcal{N}[0,B\,dt/2] +\mathcal{N}[0,
    B\,dt/2], 
\end{equation}
and because adding two random normal vectors can be accomplished by summing their means and covariance matrices, we recover Eq. \eqref{eq:pure-brownian-sde}. If $x(0)$ is a random normal vector with mean $\mu_0$ and covariance matrix $\Sigma_0$ then, under Eq. \eqref{eq:pure-brownian-sde}, the distribution at a later time $t$ is 
\begin{equation}
    x(t) \sim \mathcal{N}\left[\mu_0, \Sigma_0 + t B\right].
\end{equation}
Note that this implies that the limit $\lim_{t\to \infty} \Sigma(t)$ does not exist, meaning there is no stationary distribution for the pure Brownian process. The overdamped Langevin (ODL) equation is a useful generalization of Brownian motion which includes a drift term determined by the potential \cite{lemons1997paul, hatano2001steady}:
\begin{equation}
\label{eq:general-odl}
    dx = -\frac{1}{\gamma} \nabla U(x) dt +\frac{1}{\gamma}\mathcal{N}[0, B dt],
\end{equation}
where $\gamma \in \mathbb{R}^+$ is called the damping constant. When $V$ is given by Eq. \eqref{eq:full-potential}, the ODL equation can be put in a particularly simple form by making a change of variables $y = x - A^{-1} b$, in terms of which the potential is $V(y) = \frac{1}{2} y^\intercal A y$. We also define $\mathcal{A} = \gamma^{-1}A$, as well as $\mathcal{B} = \gamma^{-2}B$, and now Eq. \eqref{eq:general-odl} reads
\begin{equation}
\label{eq:odl-ou-process}
    dy = - \mathcal{A}y + \mathcal{N}[0, \mathcal{B} dt],
\end{equation}
which is the defining equation of the Ornstein-Uhlenbeck (OU) process (with the requirement that $A$ is positive definite). The OU process is a Gaussian process \cite{mackay1998introduction}, meaning that if $y$ is initially a Gaussian random variable then it will be Gaussian at all times. Suppose this is the case, and $x_0 \sim \mathcal{N}[\mu_0,\Sigma_0]$. The mean $\mu(t) = \braket{y(t)}$ is given by
\begin{equation}
\label{eq:ou-solution-mu}
    \mu(t) = e^{-\mathcal{A} t}\mu_0.
\end{equation}
The second moments of $y(t)$ can also be found explicitly. Because the process is indexed both by the components of $y$ and by the time, the covariances are collected in a matrix-valued function of two time parameters,
\begin{equation}
    G(t,s) = \braket{y(t)y^\intercal(s)},
\end{equation}
which is called the correlation function. For the OU process,
\begin{equation}
\label{eq:ou-solution-G}
    G(t,s) = e^{-\mathcal{A}t}\Sigma_0 e^{-\mathcal{A}^\intercal t} + \int_{0}^{\text{min}(t,s)} dt' \, e^{-\mathcal{A} (t-t')} \mathcal{B} e^{- \mathcal{A} (s - t')},
\end{equation}
and upon setting $s=t$, we find that the covariance matrix is given by
\begin{equation}
    \Sigma(t) = e^{-\mathcal{A} t}\Sigma_0 e^{-\mathcal{A}^\intercal t} + \int_{0}^{t} dt' \, e^{-\mathcal{A} (t-t')} \mathcal{B} e^{- \mathcal{A} (t - t')}.
\end{equation}
The OU process has a unique stationary distribution, which is a Gaussian $\mathcal{N}[0,\Sigma_\text{s}]$, where $\Sigma_\text{s}$ is the unique solution to the equation
\begin{equation}
\label{eq:ou-stationary-sigma}
 \mathcal{A} \Sigma_\text{s} + \Sigma_\text{s} \mathcal{A}^\intercal= \mathcal{B}.
\end{equation}
Notice that if $\mathcal{B} = \gamma^{-2}B = 2\gamma^{-1}\beta^{-1}\mathbb{I}$, then Eq. \eqref{eq:ou-stationary-sigma} implies that the stationary solution to the ODL equation is
\begin{equation}
   x \sim \mathcal{N}[A^{-1} b, \beta^{-1}A^{-1}],
\end{equation}
agreeing with Eq. \eqref{eq:full-hamiltonian-gibbs-dist}. When the system is in its stationary distribution, the correlation function can be simplified
\begin{equation}
    G_\text{s}(t, s) = e^{-\mathcal{A} (s-t)}\Sigma_\text{s},
\end{equation}
when $s \geq t$, and $G_\text{s}(t,s) = G_\text{s}(s,t)^\intercal$. These results are quite useful due to the wide variety of cases where the OU process is applicable. For example, Eq. \eqref{eq:general-odl} may be modified to account for inertial effects, resulting in the underdamped Langevin (UDL) equations, which are
\begin{equation}
\label{eq:udl-x-general}
    dx  = \frac{1}{M} p \, dt,
\end{equation}
\begin{equation}
\label{eq:udl-p-general}
    dp = -\nabla V \, dt- \frac{\gamma}{M} p \, dt + \mathcal{N}[0,B dt],
\end{equation}
where $M, \gamma \in \mathbb{R}^+$. Taking the potential in \eqref{eq:full-potential}, Eq. \eqref{eq:udl-p-general} reads
\begin{equation}
\label{eq:udl-p-device}
    dp = -(A x - b)\, dt- \frac{\gamma}{M} p \, dt + \mathcal{N}[0,B dt].
\end{equation}
In order to put the UDL equations into the form of an OU process, we first define dimensionless coordinates
\begin{equation}
    \tilde{y} =\sqrt{\frac{\gamma^2 \beta}{M}}y, \: \: \: \tilde{p} = \sqrt{\frac{\beta}{M}}p,
\end{equation}
and define the vector $r$ as the concatenation of the dimensionless position and momentum vectors
\begin{equation}
\label{r-def-app}
    r = \left(\tilde{y}_1, \dots \tilde{y}_d, \dots \tilde{p}_1, \dots \tilde{p}_d\right)^\intercal.
\end{equation}
Then we have
\begin{equation}
    \label{UDL-vec}
    dr = -\mathcal{A} r\, dt + \mathcal{N}[0, \mathcal{B} dt],
\end{equation}
where the matrices $\mathcal{A}$ and $\mathcal{B}$ may be written as block matrices, each with four $d \times d$ blocks
\begin{equation}
    \mathcal{A} = \begin{pmatrix}
    0 & -\gamma M^{-1}\mathbb{I} \\
    \gamma^{-1}A & \gamma M^{-1} \mathbb{I}
    \end{pmatrix}, \: \: \: \:
    \mathcal{B} = \begin{pmatrix}
    0 & 0  \\
    0 & \beta M^{-1} B
    \end{pmatrix}.
\end{equation}
If we take $B = 2 \gamma \beta^{-1} \mathbb{I}$, then Eq. \eqref{eq:ou-stationary-sigma} gives
\begin{equation}
    x \sim \mathcal{N}[A^{-1}b, \beta^{-1}A^{-1}], \: \: p \sim  \mathcal{N}\left[0, \beta^{-1}M \mathbb{I}\right],
\end{equation}
which again agrees with Eq. \eqref{eq:full-hamiltonian-gibbs-dist}.

\subsection{Thermodynamics}
Thermodynamics is concerned with the relationships between macroscopically observable quantities which are well-characterized by their equilibrium ensemble averages. For example, the average energy is defined as
\begin{equation}
    \braket{E} = \int dx H(x,p) f_\beta(x,p).\\
\end{equation}
For the Hamiltonian given in Eq. \eqref{eq:full-hamiltonian}, a Gaussian integral yields
\begin{equation}
\label{eq:average-energy}
    \braket{E} =  d \, \beta^{-1}-\frac{1}{2}b^\intercal A^{-1} b.
\end{equation}
According to the results of the last section, the system asymptotically approaches the canonical equilibrium distribution (in both the overdamped and underdamped models) if $B = 2\gamma \beta^{-1}\mathbb{I}$, which can be used to eliminate temperature from Eq. \eqref{eq:average-energy}, giving
\begin{equation}
\label{eq:average-energy-2}
    \braket{E} = \frac{d\left|B\right|}{2 \gamma} -\frac{1}{2}b^\intercal A^{-1} b.
\end{equation}
This relationship between temperature, damping, and noise variance is a statement of the fluctuation-dissipation relation which occurs frequently in stochastic thermodynamics. It is often explained with reference to the fact that both damping and noise arise from interactions with the bath, and stronger coupling between the system and bath tends to increase both noise and damping. Equation \eqref{eq:average-energy-2} provides an intuitive view of the fluctuation-dissipation relation, namely that energy increases with more noise and decreases with more damping. Having evaluated the average energy, we next consider changes in energy, which are separated into work and heat. The work is defined as the part of the change in energy which is due to the variation in time of one or more control parameters. If $\lambda_1 \dots \lambda_k$ are the complete set of control parameters, then the work done on the system in a time interval $dt$ is \cite{seifert2008stochastic} given by:
\begin{equation}
    dW = \sum_{i=1}^k \frac{\partial H}{\partial \lambda_i} \frac{d \lambda_i}{d t} dt = -\sum_{i=1}^k X_i \dot{\lambda}_i,
\end{equation}
where we have defined the forces conjugate to the control parameters as 
\begin{equation}
    X_i = - \frac{\partial H}{\partial \lambda_i}.
\end{equation}
The heat is defined as the remaining part of the energy change, associated with the change in the system's coordinates
\begin{equation}
    dQ = \sum_{i=1}^d \frac{\partial H}{\partial x_i} \dot{x}_i dt + \frac{\partial H}{\partial p_i} \dot{p}_i dt ,
 \end{equation}
and in terms of these we can state the first law of thermodynamics,
\begin{equation}
    \frac{d \braket{E}}{dt} = \frac{\braket{dW}}{dt} + \frac{\braket{dQ}}{dt}.
\end{equation}
The second law can be expressed in terms of the Helmholtz free energy, which is defined as \cite{callen1998thermodynamics}
\begin{equation}
    F = -\beta^{-1} \ln(Z),
\end{equation}
and can be interpreted as the amount of energy in the system that can be converted to work at constant temperature $\beta^{-1}$. For the Hamiltonian given in Eq. \eqref{eq:full-hamiltonian}, we get
\begin{equation}
    F = -\frac{1}{2} b^\intercal A^{-1} b+ \frac{1}{2\beta} \ln \left(\left| A\right| \right) -\frac{d}{\beta} \ln\left(\frac{\beta}{2\pi \sqrt{M}}\right),
\end{equation}
and so two states with the same temperature have free energy difference
\begin{equation}
    \Delta F = F_2 - F_1 = \frac{1}{2}(b_1^\intercal A_1^{-1}b_1-b_2^\intercal A_2^{-1}b_2) + \frac{1}{2\beta} \ln \left( \frac{\left|A_2\right|}{\left|A_1\right|}\right).
\end{equation}
The second law reads
\begin{equation}
\label{eq:second-law-helmholtz}
    \braket{W} \geq \Delta F.
\end{equation}
Whereas Eq. \eqref{eq:second-law-helmholtz} is an inequality, Jarzynski identifies the following equality \cite{jarzynski1997nonequilibrium}
\begin{equation}
\label{eq:jarzynski-relation}
    e^{-\beta \Delta F} = \Braket{e^{-\beta W}},
\end{equation}
and it follows from Jensen's inequality \cite{jarzynski2011equalities} that the latter implies Eq. \eqref{eq:second-law-helmholtz}. In the limit of an infinitely slow (quasistatic) process we would have $\braket{W} = W =  \Delta F$, but in general there will be some excess (dissipated) work,
\begin{equation}
    \braket{W_\text{ex}} = \braket{W} - \Delta F.
\end{equation}
Intuitively, there is more dissipation when the system is driven between very different states in a short period of time, as friction-like effects are more significant when the system changes quickly. This notion is made rigorous by introducing the thermodynamic metric tensor $g$ \cite{crooks2007measuring, shenfeld2009minimizing}
\begin{equation}
    \label{eq:friction-tensor-def}
    g_{ij}(\lambda) = \Braket{\left(X_i - \Braket{X_i}\right)\left(X_j - \Braket{X_j}\right)},
\end{equation}
where the average is over the canonical ensemble with inverse temperature $\beta$ and control parameters $\lambda$. The thermodynamic length $\mathcal{L}$ of a trajectory in the control parameter space $\lambda(t)$ is then given by
\begin{equation}
\mathcal{L} = \int_0^\tau dt \sqrt{\dot{\lambda}^\intercal g \dot{\lambda}}.
\end{equation}
Suppose that a system is driven from one state to another, and then back again by the reverse process, and that it is allowed to come to equilibrium $N$ times during both the forward and reverse process. It has been shown \cite{crooks2007measuring} that the excess work during the combined forward and reverse process (called the hysteresis) is lower bounded as
\begin{equation}
    W_\text{ex} \geq \frac{\mathcal{L}^2}{N}.
\end{equation}
Interestingly, the thermodynamic length is also equivalent to the Fisher-Rao distance, which is \cite{amari2000methods}
\begin{equation}
    \mathcal{L} = \beta \int_{0}^\tau dt \sqrt{ \int dx \frac{\dot{f}(x)^2}{f(x)}}.
\end{equation}
The above relations imply that the more distinguishable the initial and final distributions, and the less time allowed to evolve between them, the more work will be dissipated.

Equation \eqref{eq:jarzynski-relation} suggests that the free energy difference can be estimated by repeatedly transforming one potential $U_1$ into another $U_2$ and measuring the work done in the process ($n_J$ times, say), then  taking an average. The estimate of the free energy obtained in this way is called the Jarzynski estimator $\Delta \hat{F}_J$
\begin{equation}
    \Delta F \approx \Delta \hat{F}_J = \frac{1}{n_J}\sum_{j=1}^{n_J} e^{-\beta W_j}.
\end{equation}
The Jarzynski estimator is theoretically sound, but in practice it can be very slow to converge as it is dominated by terms with large negative work, and there are very few such terms in the average because the occurrence of large negative work is unlikely \cite{jarzynski2006rare}. This problem is mitigated somewhat by using an estimator which makes use of both the forward and reverse process. The Bennett Acceptance Ratio (BAR) estimator $\Delta \hat{F}_\text{BAR}$ is defined implicitly as the solution to \cite{shirts2003equilibrium, bennett1976efficient}
\begin{equation}
   \sum_{i=1}^{N} \frac{1}{1 +  e^{ \beta W_{\text{(f)}i} - \beta \Delta \hat{F}_\text{BAR}}}  =   \sum_{i=1}^{N} \frac{1}{1 +  e^{ -\beta W_{\text{(r)}i} +\beta \Delta \hat{F}_\text{BAR}}},
\end{equation}
where it is assumed that there are $N$ realizations of both the forward and reverse process, resulting in work measurements of $W_{\text{(f)}i}$ and $W_{\text{(r)}i}$. This estimator can be generalized to the Multistate Bennett Acceptance Ratio \cite{shirts2008statistically} (MBAR), where intermediate equilibrium states are used between the initial and final states \cite{shirts2008statistically}. Interestingly, the variance of any unbiased estimator of the free energy difference is lower bounded in terms of the thermodynamic length between the terminal states and the total number of observations $n$, which corresponds to the number of times that equilibrium must be reached during the protocol \cite{shenfeld2009minimizing}
\begin{equation}
    \text{Var}(\Delta \hat{F}) \geq \frac{\beta^{-2} \mathcal{L}^2}{n}.
\end{equation}

\section{Analysis of the Overdamped Regime}

\subsection{Stationary Distribution}
The overdamped Langevin equation reads
\begin{equation}
    \label{eq:ODL-app}
    dx = - \frac{1}{\gamma}(A x - b) dt + \frac{1}{\gamma}\mathcal{N}\left[0,B\, dt\right].
\end{equation}
We set $B = 2 \gamma \beta^{-1} \mathbb{I}$ (see Methods), where $\beta =(k_B T)^{-1}$ is the inverse temperature of the environment. We then change variables to $y = x - A^{-1}b$, and have the transformed equation
\begin{equation}
    \label{ODL-transformed-app}
    dy = - \frac{1}{\gamma}A y dt + \mathcal{N}\left[0, \frac{2}{\beta \gamma} \mathbb{I}\, dt\right],
\end{equation}
which can be written as a vector Ornstein-Uhlenbeck (OU) process \cite{gardiner1985handbook}
\begin{equation}
    \label{ODL-transformed-app}
    dy = - \mathcal{A} y dt + \mathcal{N}[0, \mathcal{B}\, dt],
\end{equation}
where $\mathcal{A} = \gamma^{-1}A$ and $\mathcal{B} = 2 \gamma^{-1}\beta^{-1}\mathbb{I}$. The stationary distribution for $y$ then has mean zero (which follows from Eq. \eqref{eq:ou-solution-mu}) and its variance $\Sigma_\text{s}$ satisfies Eq. \eqref{eq:ou-stationary-sigma},
\begin{equation}
    \mathcal{A} \Sigma_\text{s} + \Sigma_\text{s} \mathcal{A} = \mathcal{B}.
\end{equation}
In this case, we have
\begin{equation}
    \frac{1}{\gamma}A \Sigma_\text{s} + \frac{1}{\gamma}\Sigma_\text{s} A = \frac{2}{\beta \gamma}\mathbb{I},
\end{equation}
which is satisfied for the choice
\begin{equation}
    \label{ODL-stationary-sigma}
    \Sigma_\text{s} = \beta^{-1} A^{-1}.
\end{equation}
The uniqueness of this solution is guaranteed, because Eq. \eqref{ODL-stationary-sigma} always has a unique solution when $A$ is positive definite. Transforming back to the original coordinates, we see that at equilibrium $x$ is distributed as
\begin{equation}
\label{eq:stationary-x-linear-system}
    x \sim \mathcal{N}[A^{-1}b, \beta^{-1} A^{-1}].
\end{equation}
Alternatively we may take $B = 2\gamma\beta^{-1}R$ (or $\mathcal{B} = 2\gamma^{-1} \beta^{-1}R$) for some symmetric positive-definite matrix $R$. Then we would have stationary covariance matrix satisfying
\begin{equation}
    A \Sigma_\text{s} +\Sigma_\text{s} A =2\beta^{-1}R,
\end{equation}
which is a Lyapunov equation. In the latter case, the stationary mean of $x$ is still $A^{-1} b$.
\subsection{Equilibration and Correlation Time}
We assume that the initial distribution is Gaussian. After some time has passed, the system will be in the equilibrium distribution described by Eq. \eqref{eq:stationary-x-linear-system}. Here we quantify the amount of time one should wait (called the equilibration time) to ensure that the distribution is arbitrarily close to the equilibrium distribution. Note that because $A$ is symmetric positive definite, the spectral norm of $e^{-A t/\gamma}$ is
\begin{equation}
\label{exp-A-norm-bound-app}
    \left\|e^{-A t/\gamma}\right\| = e^{-\alpha_\text{min} t/\gamma}.
\end{equation}
As Eq. \eqref{ODL-transformed-app} describes an OU process, the mean evolves in time according to
\begin{equation}
    \braket{y(t)} = e^{-A t/\gamma}\braket{y(0)}.
\end{equation}
Let $\alpha_1 \dots \alpha_d$ be the eigenvalues of $A$. The norm of $\braket{y}$ is exponentially bounded,
\begin{equation}
\label{odl-mean-exp-bound-app}
\left\|\braket{y(t)}\right\| =\left \|e^{- A t/\gamma} \braket{ y(0)}\right\| \leq e^{-\alpha_\text{min} t/\gamma} \left\| \braket{y(0)}\right\|.
\end{equation} 
Given $\delta_{\mu 0}>0$, in order to have $\left\|\braket{x(t_0)} -A^{-1} b\right \| \leq \delta_{\mu 0}$, it is sufficient to require
\begin{equation}
    t_0 \geq \frac{\gamma}{\alpha_\text{min}} \ln \left( \frac{\left\|\braket{x(0)} - A^{-1} b\right\|}{\delta_{\mu0}}\right).
\end{equation}
We assume the initial distribution has mean zero, $\braket{x(0)}=0$. Defining the condition number $\kappa = \alpha_\text{max}/\alpha_\text{min}$, and the relative error tolerance $\varepsilon_{\mu 0} = \delta_{\mu 0 }/\|A^{-1} b\|$ it is sufficient to require
\begin{equation}
\label{eq:odl-mean-equilibration-time-app}
    t_0 \geq \kappa \gamma \|A\|^{-1} \ln \left(\kappa \varepsilon^{-1}_{\mu 0}\right).
\end{equation}
The correlation matrix is defined by
\begin{equation}
    G(t,s) = \braket{[x(t) - \braket{x(t)}][x(s) - \braket{x(s)}]^\intercal}
\end{equation}
For $t=s$, the correlation matrix reduces to the covariance matrix
\begin{equation}
    G(t,t) = \braket{[x(t) - \braket{x(t)}][x(t) - \braket{x(t)}]^\intercal} = \Sigma(t)
\end{equation}
For an OU process, the dynamics of the correlation matrix (and the covariance matrix in particular) can be expressed in terms of the following propagator function
\begin{equation}
    \mathcal{P}_t(X) = e^{-A t/\gamma} X e^{-A^\intercal t/\gamma}.
\end{equation}
The covariance matrix for $y$ is then given by
\begin{equation}
\label{sigma-time-dep-app}
    \Sigma(t) = \mathcal{P}_t(\Sigma_0) + \int_0^t dt' \mathcal{P}_{t'} (\mathcal{B}),
\end{equation}
and for any $t > t_0$ we have
\begin{equation}
\label{sigma-error-time-dep-app}
    \Sigma(t) - \Sigma(t_0)=\mathcal{P}_{t}(\Sigma_0) -\mathcal{P}_{t_0}(\Sigma_0) + \int_{t_0}^{t} dt' \mathcal{P}_{t'} (\mathcal{B}),
\end{equation}
Taking the limit as $t \to \infty$, we evaluate
\begin{align}
    \left\|\Sigma_\text{s} - \Sigma(t_0) \right\| &= \left\| \Sigma_\text{s} -\mathcal{P}_{t_0}(\Sigma_0) + \int_{t_0}^{\infty} dt' \mathcal{P}_{t'} (\mathcal{B})\right\| \\
    & =  \left\|\mathcal{P}_{t_0}( \Sigma_\text{s} - \Sigma_0) + \int_{t_0}^{\infty} dt' \mathcal{P}_{t'} (\mathcal{B})\right\| \\
    & \leq
 \left\| \mathcal{P}_{t_0} (\Sigma_\text{s} - \Sigma_0) \right\|+ \int_{t_0}^\infty dt' \left\| \mathcal{P}_{t'}(\mathcal{B})\right\|\\
 & \leq 
  \left\| e^{-A t/\gamma}\right\|^2 \cdot \left\| \Sigma_0 - \Sigma_\text{s}\right\|+ \left\|\mathcal{B}\right\|\int_{t_0}^\infty dt' \left\| e^{-A t'/\gamma}\right\|^2 \\
  & \leq     e^{- 2\alpha_\text{min} t_0/\gamma }\left\| \Sigma_0 - \Sigma_\text{s}\right\|+ \left\|\mathcal{B}\right\|\int_{t_0}^\infty dt' e^{-2 \alpha_\text{min} t'/\gamma} \\
  & =  e^{- 2\alpha_\text{min} t_0 }\left\| \Sigma_0 - \Sigma_\text{s}\right\|+\frac{\gamma}{2\alpha_\text{min}}e^{-2 \alpha_\text{min}t_0/\gamma} \left\|\mathcal{B}\right\| \\
  & =  e^{- 2\alpha_\text{min} t_0/\gamma }\left(\left\| \Sigma_0 - \Sigma_\text{s}\right\|+\frac{\gamma}{2  \alpha_\text{min}}\|\mathcal{B}\|\right).
\end{align} 
So in order to have $\|\Sigma(t_0) - \Sigma_\text{s}\| \leq \delta_
\Sigma$, we take
\begin{equation}
    t_0   \geq  \frac{\gamma}{2 \alpha_\text{min}}\ln\left(\frac{\|\Sigma_0 - \Sigma_\text{s}\| + (\gamma / 2\alpha_\text{min})\|\mathcal{B}\|}{\delta_{\Sigma 0}}\right).
\end{equation}
We use the fact that $\Sigma_\text{s} = \beta^{-1} A^{-1}$, and $\mathcal{B} = 2 \gamma^{-1} \beta^{-1} \mathbb{I}$. Also, we assume the system is initially at equilibrium in a potential proportional to identity $A_0 = \|A\|\mathbb{I}$, so $\Sigma_0 = \beta^{-1} \|A\|^{-1}\mathbb{I}$. We have
\begin{align}
    t_0   &\geq  \frac{\kappa\gamma}{2 \|A\|}\ln\left(\frac{\beta^{-1}\|(\|A^{-1}\|\mathbb{I}) - A^{-1}\| + \kappa \beta^{-1}/\|A\|}{\delta_{\Sigma 0}}\right)
\end{align}
We can give a looser requirement
\begin{align}
    t_0   &\geq  \frac{\kappa\gamma}{2\|A\| }\ln\left(\frac{2\kappa\beta^{-1}\|A^{-1}\|}{\delta_\Sigma }\right).
\end{align}
Defining the variance relative error tolerance $\varepsilon_{\Sigma 0} = \delta_{\Sigma 0}/\beta^{-1} \|A^{-1}\|$, we have
\begin{align}
\label{eq:odl-variance-equilibration-time-app}
    t_0   &\geq  \frac{\kappa\gamma}{2 \|A\|}\ln\left(\frac{2\kappa}{\varepsilon_{\Sigma 0}}\right).
\end{align}
Combining Eqs.~\eqref{eq:odl-mean-equilibration-time-app} and \eqref{eq:odl-variance-equilibration-time-app} gives 
\begin{equation}
t_0 \geq \max \left\{\frac{\kappa \gamma}{\|A\|} \ln \left(\kappa \varepsilon^{-1}_{\mu 0}\right),\frac{\kappa\gamma}{2\|A\| }\ln\left(2\kappa\varepsilon_{\Sigma 0}^{-1}\right)\right\},
\end{equation}
Finally, we define the relaxation time $\tau_\text{r} = \gamma/\|A\|$, and we have
\begin{equation}
t_0 \geq \max \left\{\kappa \tau_\text{r}\ln \left(\kappa \varepsilon^{-1}_{\mu 0}\right),\frac{1}{2} \kappa \tau_\text{r}\ln\left(2\kappa\varepsilon_{\Sigma 0}^{-1}\right)\right\}.
\end{equation}

More generally we consider the case where $\mathcal{B} = 2 \gamma^{-1} \beta^{-1} R$, and $\|R\|=1$. Now the variance equilibration time is
\begin{align}
t_0  & \geq  \frac{\gamma}{2 \alpha_\text{min}}\ln\left(\frac{\|\Sigma_0 - \Sigma_\text{s}\| + (\gamma \kappa / 2\|A\|)\|\mathcal{B}\|}{\delta_{\Sigma 0}}\right)\\
&= \frac{1}{2}\kappa\tau_\text{r}\ln\left(\frac{\|\Sigma_0 - \Sigma_\text{s}\| + \kappa \beta^{-1} \|A\|^{-1}
}{\delta_{\Sigma 0}}\right)
.
\end{align}
However, without knowing the solution to the Lyapunov equation the above cannot be sharply bounded, so in this case we are left with the form
\begin{equation}
t_0 \geq \max \left\{\kappa \tau_\text{r}\ln \left(\kappa \varepsilon^{-1}_{\mu 0}\right),\frac{1}{2}\kappa\tau_\text{r}\ln\left(\frac{\|\Sigma_0 - \Sigma_\text{s}\| + \kappa \beta^{-1} \|A\|^{-1}
}{\delta_{\Sigma 0}}\right)
\right\}.
\end{equation}
Once the equilibration time has passed, we may consider the correlation matrix of the stationary system
\begin{equation}
    G_\text{s}(\tau) = \lim_{t \to \infty} G(t, t+\tau),
\end{equation}
which is given by 
\begin{equation}
    G_\text{s}(\tau) = e^{-A \tau/\gamma} \Sigma_\text{s}.
\end{equation}
Therefore
\begin{equation}
    \|G_\text{s}(\tau)\| \leq \|e^{-A t/\gamma}\| \cdot \|\Sigma_\text{s}\| \leq e^{- \alpha_\text{min} \tau/\gamma}  \left\|\Sigma_\text{s}\right\|.
\end{equation}
Suppose we sample once per interval $\tau_c$. To obtain samples with bounded correlation $\left\| G_\text{s} \right \| \leq \delta_G $, we need
\begin{equation}
    \tau_c \geq \frac{\gamma}{\alpha_\text{min}} \ln \left(\frac{\|\Sigma_\text{s}\|}{\delta_G}\right).
\end{equation}
Defining $\varepsilon_G = \delta_G/\|\Sigma_\text{s}\|$, we get
\begin{equation}
    \tau_c \geq\kappa \tau_\text{r} \ln \left(\kappa \varepsilon_G^{-1}\right).
\end{equation}
Interestingly, the correlation time is dimension-independent, ignoring any incidental dependence of condition number on dimension.

\subsection{Ergodicity of Mean}

Define the time-average of $y$
\begin{equation}
\bar{y} = \frac{1}{\tau} \int_{t_0}^{t_0 + \tau} dt y(t).
\end{equation}
We assume that $t_0$ is large enough that, to a good approximation, the system has reached equilibrium by time $t_0$. Therefore
\begin{equation}
\braket{\bar{y}} = \frac{1}{\tau} \int_{t_0}^{t_0 + \tau} dt \braket{y(t)} = 0.
\end{equation}
We now bound the covariance matrix of $\bar{y}$
\begin{align}
\left\|\braket{\bar{y}\bar{y}^\intercal}\right\| &=\left \| \frac{1}{\tau^2}\int_{t_0}^{t_0+\tau}\int_{t_0}^{t_0+\tau} dt''dt' \braket{y(t')y(t'')^\intercal}\right \| \\
&=  \left\|\frac{1}{\tau^2}\int_{t_0}^{t_0+\tau}\int_{t_0}^{t_0+\tau}  dt''dt' G_\text{s}(t''-t') \right\|\\
&=  \left\|\frac{2}{\tau^2}\int_{t_0}^{t_0 + \tau}dt''\int_{t_0}^{t''} dt' G_\text{s}(t''-t') \right\|\\
&\leq 
\frac{2}{\tau^2}\int_{t_0}^{t_0 + \tau}dt''\int_{t_0}^{t''} dt' \left\|G_\text{s}(t''-t') \right\|\\
&\leq 
\frac{2}{\tau^2}\int_{t_0}^{t_0 + \tau}dt''\int_{t_0}^{t''} dt' e^{-\alpha_\text{min}(t''-t')/\gamma} \left\| \Sigma_\text{s}\right\|\\
&=
\frac{2\gamma}{\alpha_\text{min}\tau^2} \left\| \Sigma_\text{s}\right\|\int_{t_0}^{t_0 + \tau}dt'' 1-e^{-\alpha_\text{min}(t'' - t_0)/\gamma} \\
& \leq 
 \frac{2\gamma}{\alpha_\text{min}\tau^2} \left\| \Sigma_\text{s}\right\|\int_{t_0}^{t_0 + \tau}dt''1 \\
 &= 
 \frac{2\gamma}{\alpha_\text{min}\tau} \left\| \Sigma_\text{s}\right\|
 \\ & = \frac{2 \gamma \kappa^2}{\beta\|A\|^2\tau}\\
 & =  \frac{2  \kappa^2 \tau_\text{r}}{\beta\|A\|\tau}.
\end{align}
According to Chebyshev's inequality
\begin{equation}
    P(y^\intercal \braket{\bar{y} \bar{y}^\intercal}^{-1} y > k^2 ) \leq \frac{d}{k^2}.
\end{equation}
Now note that the eigenvalues of $\braket{\bar{y} \bar{y}^\intercal}^{-1}$ are the inverses of the eigenvalues of $\braket{\bar{y} \bar{y}^\intercal}$, which are the same as the singular values as the covariance matrix is positive definite. Therefore because the eigenvalues of $\braket{\bar{y} \bar{y}^\intercal}$ are at most $\left\|\braket{\bar{y}\bar{y}^\intercal}\right\|$, the eigenvalues of $\braket{\bar{y} \bar{y}^\intercal}^{-1}$ are at least $ \left\|\braket{\bar{y}\bar{y}^\intercal}\right\| ^{-1}$. This implies
\begin{equation}
y^\intercal \braket{\bar{y} \bar{y}^\intercal}^{-1} y \geq \|y\|^2\frac{\beta\|A\|^2\tau}{2 \gamma \kappa^2 },
\end{equation}
and so we have the following proposition
\begin{equation}
    \|y\|^2 > \frac{2 \gamma\kappa^2}{\beta\|A\|^2\tau}k^2 \Rightarrow
    y^\intercal \braket{\bar{y} \bar{y}^\intercal}^{-1}y > k^2.
\end{equation}
Moreover, $\|Ay\|^2 \leq \|A\|^2 \|y\|^2$ so the following also holds
\begin{equation}
    \|A y\|^2 > \frac{2 \gamma\kappa^2}{\beta\tau}k^2 \Rightarrow
    y^\intercal \braket{\bar{y} \bar{y}^\intercal}^{-1}y > k^2.
\end{equation}
This means that 
\begin{equation}
    P\left(  \frac{\|A y\|^2}{\|b\|^2} > \frac{2 \gamma\kappa^2}{\|b\|^2\beta\tau}k^2\right) \leq P( y^\intercal \braket{\bar{y} \bar{y}^\intercal}^{-1}y \geq k^2) \leq \frac{d}{k^2}.
\end{equation}
Now let
\begin{equation}
\varepsilon_y = \sqrt{\frac{2 \gamma \kappa^2}{\beta\|b\|^2\tau}} k,
\end{equation}
so $k = \sqrt{\beta \|b\|^2 \tau/2\gamma \kappa^2}\delta_y$. Then we have
\begin{equation}
    P\left(  \frac{\|A y\|}{\|b\|} \geq \varepsilon_y\right) \leq \frac{d}{k^2} = \frac{2 \gamma \kappa^2d}{\beta \|b\|^2 \tau \varepsilon_y^2}.
\end{equation}
We change back to the original coordinates. If we would like to have
\begin{equation}
    \left \|A \bar{x} - b \right\| \leq \varepsilon_x \|b\| 
\end{equation}
with probability at least $P_\varepsilon$, then we can require the integration time is at least
\begin{align}
    \tau &\geq \frac{2 \gamma\kappa^2 d  }{\beta \|b\|^2 \varepsilon_x^2 (1-P_\varepsilon)} \\
    &=\frac{2 \kappa^2 d\, \|A\| \tau_\text{r} }{\beta \|b\|^2 \varepsilon_x^2 (1-P_\varepsilon)} .
\end{align}
Here, we note that the energy cost $\mathcal{E}$ can be estimated as $\mathcal{E} \geq b^\intercal A^{-1} b \geq \|A\|^{-1}\|b\|^2$, which is the depth of the bottom of the potential well relative to the point $x=0$. Therefore $\mathcal{E}^{-1} \leq \|A\|\|b\|^{-2}$, allowing us to write
\begin{align}
    \mathcal{E}\tau &\geq \frac{2 \kappa^2 d  }{ \varepsilon_x^2 (1-P_\varepsilon)} \beta^{-1}\tau_\text{r}.
\end{align}

\subsection{Ergodicity of Covariance Matrix}
The covariance matrix can also be estimated by a finite time integral
\begin{equation}
\overline{yy^\intercal} =\frac{1}{\tau} \int_{t_0}^{t_0 + \tau} dt' y(t') y^\intercal(t').
\end{equation}
\begin{equation}
    \text{var}(\overline{y_iy_j}) = \braket{\overline{y_i y_j}^2} - \Sigma^2_{\text{s}, ij}
\end{equation}

\begin{equation}
     \braket{\overline{y_i y_j}^2}  = \frac{1}{\tau^2}\int_{t_0}^{t_0 + \tau} dt'' \int_{t_0}^{t_0 + \tau} dt' \braket{y_i(t') y_j(t') y_i(t'') y_j(t'')}.
\end{equation}
Using Isserlis's theorem \cite{koopmans1995spectral},
\begin{align}
     \braket{y_i(t') y_j(t') y_i(t'') y_j(t'')} &= \braket{y_i(t')y_j(t')}\braket{y_i(t'')y_j(t'')} +\braket{y_i(t')y_i(t'')}\braket{y_j(t')y_j(t'')}\\ & +\braket{y_i(t')y_j(t'')}\braket{y_j(t')y_i(t'')}\\
     & = (\Sigma_{\text{s},ij})^2 + G_{\text{s},ii}(t''-t')G_{\text{s},jj}(t''-t') + G_{\text{s},ij}(t''-t')^2,
\end{align}
so
\begin{align}
    \text{var}(\overline{y_i y_j})
    &=\braket{\overline{y_i y_j}^2} -\Sigma_{\text{s},ij}^2  \\
    &=  \frac{1}{\tau^2}\int_{t_0}^{t_0 + \tau} dt'' \int_{t_0}^{t_0 + \tau} dt' G_{\text{s},ii}(t''-t')G_{\text{s}.jj}(t''-t') + G_{\text{s},ij}(t''-t')^2.
\end{align}
Summing over $i$ and $j$,
\begin{align}
    \sum_{ij}\text{var}(\overline{y_i y_j})
    &=  \frac{1}{\tau^2}\int_{t_0}^{t_0 + \tau} dt'' \int_{t_0}^{t_0 + \tau} dt'\sum_{ij} G_{\text{s},ii}(t''-t')G_{\text{s}.jj}(t''-t') + G_{\text{s},ij}(t''-t')^2 \\
    &\leq  \frac{2}{\tau^2}\int_{t_0}^{t_0 + \tau} dt'' \int_{t_0}^{t_0 + \tau} dt'\,(d+1)\|G_\text{s}(t'' - t')\|_\text{F}^2
\end{align}
where $\|\cdot\|_\text{F}$ is the Frobenius norm. As the Frobenius norm is bounded by $\sqrt{d}$ times the spectral norm, we have 
\begin{align}
    \sum_{ij}\text{var}(\overline{y_i y_j})
    &\leq \frac{2d(d+1)}{\tau^2}\int_{t_0}^{t_0 + \tau} dt'' \int_{t_0}^{t_0 + \tau} dt'\|G_\text{s}(t'' - t')\|^2\\
    & =  \frac{4d(d+1)}{\tau^2}\int_{t_0}^{t_0 + \tau} dt'' \int_{t_0}^{t''} dt'\|G_\text{s}(t'' - t')\|^2 \\ 
    & \leq   \frac{4d(d+1)}{\tau^2}\int_{t_0}^{t_0 + \tau} dt'' \int_{t_0}^{t''} dt'e^{-2\alpha_\text{min} (t'' - t')/\gamma} \left\| \Sigma_\text{s}\right\|^2 \\
    &=
\frac{4d(d+1)}{\tau^2} \left\| \Sigma_\text{s}\right\|^2\int_{t_0}^{t_0 + \tau}dt'' \frac{\gamma-\gamma e^{-2\alpha_\text{min}(t'' - t_0)/\gamma}}{2\alpha_\text{min}} \\
& \leq 
\frac{4d(d+1)}{\tau^2} \left\| \Sigma_\text{s}\right\|^2\int_{t_0}^{t_0 + \tau}dt'' \frac{\gamma}{2\alpha_\text{min}} \\
& = \frac{4 d (d+1) \,\gamma\left\| \Sigma_\text{s}\right\|^2}{\tau \alpha_\text{min}}. \\
\end{align}
Recall the generalization of Chebyshev's inequality, for an arbitrary norm $\|\cdot\|_\nu$,
\begin{equation}
    P\left( \|X-\mu\|_\nu \geq k \sigma_\nu\right) \leq \frac{1}{k^2},
\end{equation}
where
\begin{equation}
    \sigma_\nu^2 = \text{E}[\|X - \mu\|_\nu^2].
\end{equation}
We use the Frobenius norm for $X = \overline{y y^\intercal}$, resulting in
\begin{align}
    \sigma_\text{F}^2 = \text{E}\left[ \left\|
    \overline{yy^\intercal} - \braket{\overline{y y^\intercal}}
    \right\|^2_\text{F}\right] &= 
    \text{E}\left[ \left\|
    \overline{yy^\intercal} - \Sigma_\text{s}
    \right\|_\text{F}^2\right]\\
    &= 
    \text{E}\left[ \sum_{ij} (\overline{y_i y_j} - \braket{y_i y_j})^2\right]\\
    & = \sum_{ij} \text{var}(\overline{y_i y_j})\\
    &\leq \frac{4 d (d+1) \,\gamma\left\| \Sigma_\text{s}\right\|^2}{\tau \alpha_\text{min}}
.
\end{align}
If we set
\begin{equation}
    \delta_\Sigma^2 = k^2\frac{4 d (d+1) \,\gamma\left\| \Sigma_\text{s}\right\|^2}{\tau \alpha_\text{min}},
\end{equation}
then
\begin{align}
    P\left( \|\overline{y y^\intercal} - \Sigma_\text{s}\|_\text{F} \geq \delta_\Sigma \right)
    &\leq \frac{1}{k^2}\\
    & = \frac{4 d (d+1) \gamma \|\Sigma_\text{s}\|^2}{\tau \alpha_\text{min} \delta_\Sigma^2}.
\end{align}
We define the relative error $\varepsilon_\Sigma = \delta_\Sigma/\|\Sigma_\text{s}\|$, and have 
\begin{align}
    P\left( \frac{\|\overline{y y^\intercal} - \Sigma_\text{s}\|_\text{F}}{\|\Sigma_\text{s}\|} \geq \varepsilon_\Sigma \right)
    & \leq \frac{4 d (d+1) \gamma \|\Sigma_\text{s}\|^2}{\tau \alpha_\text{min} \varepsilon_\Sigma^2 \|\Sigma_\text{s}\|^2}\\
    & =  \frac{4 d (d+1) \kappa \tau_\text{r}}{  \varepsilon_\Sigma^2 \tau }\\
\end{align}
So if we would like to have  (Frobenius) relative error of at most $\varepsilon_\Sigma$ with probability $P_\varepsilon$ then it is sufficient to allow integration time of at least
\begin{equation}
    \tau \geq \frac{4 \kappa d(d+1)}{(1-P_\varepsilon)\varepsilon_\Sigma^2}\tau_\text{r}.
\end{equation}

\subsection{Hardware implementation}\label{app:hw}
Here we describe an electronic device comprised of $d$ coupled $RC$ cells that maps to the overdamped Langevin process described previously. The equation of motion for voltages $\mathbf{v} = (v_1, v_2, \ldots, v_d)$ across the capacitor of each cell is given by
\begin{equation}d\mathbf{v} = \mathbf{C}^{-1}(-\mathbf{J}\mathbf{v}dt+\mathbf{R}^{-1}d\mathbf{w})\end{equation}
where $\mathbf{C} = \mathrm{diag}(C_1, C_2, \ldots, C_d)$, $\mathbf{R} = \mathrm{diag}(R_1, R_2, \ldots, R_d)$, $\mathbf{w}$ is uncorrelated Brownian motion and the elements of $\mathbf{J}$ are given by
\begin{equation}
    \begin{cases}
    J_{ij} = -\frac{1}{R_{ij}} \text{ if } i\neq j \\
    J_{ij} = \frac{1}{R_{i}} + \frac{1}{R'_{i}} + \frac{1}{R_{ij}} \text{ if } i= j.
\end{cases}
\end{equation}
Hence there are two in-cell resistors, allowing for freedom on the diagonal elements of the $\mathbf{J}$ matrix independently of the $\mathbf{R}$ matrix, and one resistor coupling each cell. One can set $\mathcal{J} = \mathbf{J} \mathbf{R}$, and $\gamma = 1/RC$ as $\mathbf{R},\mathbf{C}$ are diagonal matrices. By denoting $x = \mathbf{v}$ and choosing $\mathcal{J} = A$, this leads to:

\begin{equation}
   dx =  \frac{1}{\gamma}(-Axdt + d\mathbf{w})
\end{equation}
which reduces to overdamped dynamics of the linear systems solver provided the mean of the noise is $\gamma b$, and its variance is $2\gamma/\beta$. For more details on this implementation, see ref.~\cite{coles2023thermodynamic}.

Note that to implement this in hardware the values of the in-cell and out-of-cell resistors must be computed based on the $A$ matrix, therefore this incurs a cost of $O(d^2)$ operations at the initialization. For electrical hardware, the effective temperature is related to Johnson-Nyquist noise as $\overline{v_n^2} = \sqrt{4k_BTR\Delta f}$, with $\Delta f$ the bandwidth of the system. For a bandwidth of $1 \text{MHz}$, we obtain $\Bar{v_n^2} = 0.1 \mu V$ for $T = 300K$.

\section{Analysis of the Underdamped Regime}
\label{app:UDL}

\subsection{Stationary Distribution }

Consider the following underdamped Langevin equations
\begin{equation}
\label{UDL-x-app}
    dx  = \frac{1}{M} p \, dt,
\end{equation}
\begin{equation}
\label{UDL-p-app}
    dp = -(A x - b) \, dt- \frac{\gamma}{M} p \, dt + \mathcal{N}[0,2 \gamma \beta^{-1}  \mathbb{I} dt],
\end{equation}
where $M, \gamma, \beta \in \mathbb{R}^+$. As before, define $y = x - A^{-1} b$, resulting in the transformed equations
\begin{equation}
\label{UDL-x-app-transformed}
    dy  = \frac{1}{M} p \, dt,
\end{equation}
\begin{equation}
\label{UDL-p-app-transformed}
    dp = -A y \, dt- \frac{\gamma}{M}p \, dt + \mathcal{N}[0,2 \gamma \beta^{-1} \mathbb{I} dt].
\end{equation}
Equations \eqref{UDL-x-app-transformed} and \eqref{UDL-p-app-transformed} can be made dimensionless by defining $\tilde{x}$ and $\tilde{p}$ as follows
\begin{equation}
    \tilde{y} =\sqrt{\frac{\gamma^2 \beta}{M}}y, \: \: \: \tilde{p} = \sqrt{\frac{\beta}{M}}p.
\end{equation}
and note that these choices imply that $\tilde{p} = M \gamma^{-1} \dot{\tilde{x}}$. The underdamped Langevin equations can now be formulated as a single dimensionless equation by concatenating the vectors $\tilde{y}$ and $\tilde{p}$ into a single vector $r$,
\begin{equation}
\label{r-def-app}
    r = \left(\tilde{y}_1, \dots \tilde{y}_d, \dots \tilde{p}_1, \dots \tilde{p}_d\right)^\intercal,
\end{equation}
As in the Methods section, we write the UDL equations as a single matrix equation
\begin{equation}
    dr = -\mathcal{A} r\, dt + \mathcal{N}[0, \mathcal{B} dt],
\end{equation}
with
\begin{equation}
    \mathcal{A} = \begin{pmatrix}
    0 & -\gamma M^{-1}\mathbb{I} \\
    \gamma^{-1}A & \gamma M^{-1} \mathbb{I}
    \end{pmatrix}, \: \: \: \:
    \mathcal{B} = \begin{pmatrix}
    0 & 0  \\
    0 & \beta M^{-1} B
    \end{pmatrix}.
\end{equation}
Following the analysis from the Methods section further, we find that the stationary distribution, in the original coordinates, is
\begin{equation}
    x \sim \mathcal{N}[A^{-1}b, \beta^{-1}A^{-1}], \: \: p \sim  \mathcal{N}\left[0, \beta^{-1}M \mathbb{I}\right].
\end{equation}
If instead we had $B = 2\gamma\beta^{-1}R$ for an SPD matrix $R$, the stationary covariance matrix for $x$ would be the solution to the Lyapunov equation
\begin{equation}
    A \Sigma_\text{s} +\Sigma_\text{s} A =2\beta^{-1}R,
\end{equation}
and the mean would still be $A^{-1}b$.

\subsection{Transient Behavior}
We review some results about the transient behavior of a system of coupled damped harmonic oscillators. Consider Eqs. \eqref{UDL-x-app-transformed} and \eqref{UDL-p-app-transformed} with the noise term removed. These can be combined into a single second order equation
\begin{equation}
 M   \ddot{y} + \gamma \dot{y} + A y =0.
\end{equation}
The system may be decoupled into $d$ independent equations by expanding $y$ in the (orthonormal) eigenbasis of $A$,
\begin{equation}
y = \sum_j c_j a_j,
\end{equation}
where $a_j$ are the eigenvectors of $A$ with eigenvalues $\alpha_j$. Now we have the set of uncoupled equations
\begin{equation}
\label{decoupled-DHO-app}
  \ddot{c}_j + \frac{\gamma}{M} \dot{c}_j+ \frac{\alpha_j}{M} c_j =0.
\end{equation}
We define $\omega_j = \sqrt{\alpha_j/M}$, $\xi = \gamma/2M$, and $\zeta_j = \xi/\omega_j $. The general solution to Eq. \eqref{decoupled-DHO-app} is
\begin{equation}
\label{DHO-solution-app}
    c_j(t) = c_{j+} e^{\lambda_{j+} t} + c_{j-}e^{\lambda_{j-}t},
\end{equation}
where 
\begin{equation}
\label{DHO-eigenvalues-app}
    \lambda_{j\pm}=  - \xi \pm \sqrt{\xi^2 - \omega_j^2}.
\end{equation}
If $\xi < \omega_j$, the square root in Eq. \eqref{DHO-eigenvalues-app} is imaginary, in which case the system is said to be underdamped, and when $\xi > \omega_j$ the system is overdamped. For $\xi = \omega_j$ Eq. \eqref{DHO-solution-app} is invalid, which is the critically damped case. In what follows, we assume that $\xi < \omega_j$. In this case, the solution is
\begin{equation}
    c_j(t) = c_{j0}e^{-\xi t}\sin(\omega_j t + \phi_j),
\end{equation}
We define the energy functions
\begin{equation}
    E_j(t) = \frac{1}{2} \alpha_j c_j^2 + \frac{1}{2} M \dot{c}_j^2.
\end{equation}
It can be shown that the energy is
\begin{equation}
    E_j(t) =  \frac{1}{2} M c_{j0}^2\omega_j^2e^{-2 \xi t} \left(1 + \frac{\xi  }{\omega_j}\cos(2 \omega_j t + \phi_j) \right),
\end{equation}
which can be bounded as
\begin{equation}
   \frac{1}{2} M c_{j0}^2\omega_j^2e^{-2 \xi t} \left(1 - \frac{\xi  }{\omega_j} \right) \leq  E_j(t) \leq  \frac{1}{2} M c_{j0}^2\omega_j^2e^{-2 \xi t} \left(1 + \frac{\xi  }{\omega_j} \right).
\end{equation}
This implies that the energy decays exponentially,
\begin{equation}
    E_j(t) \leq \frac{1+\xi/\omega_j}{1-\xi/\omega_j}e^{-2 \xi t} E_j(0).
\end{equation}
Therefore the total energy, which is the sum of the mode energies $E_j$, decays at least as fast as the mode whose energy decays slowest,
\begin{equation}
    E(t) = \frac{1}{2} y^\intercal A y +\frac{1}{2M} p^2 \leq  \frac{1+\xi/\omega_\text{min}}{1-\xi/\omega_\text{min}}e^{-2 \xi t} E(0)
\end{equation}
When regarded as a function of the state vector $r\in \mathbb{R}^{2d}$, the square root of the energy is a norm, which we denote by $\|r\|_E = \sqrt{\beta E(r)}$.
\begin{align}
    \|r\|_E^2 &= \frac{\beta}{2}y^\intercal A y + \frac{\beta}{2M}p^\intercal p\\
    &= \frac{M}{2\gamma^2}\tilde{y}^\intercal A \tilde{y} + \frac{1}{2 } \tilde{p}^\intercal \tilde{p}
\end{align}
Note that $y^\intercal A y \geq \alpha_\text{min} \|y\|^2$, so the energy norm gives the following upper bound on the Euclidean norm of the vector $y$
\begin{equation}
    \|y\| \leq  \sqrt{\frac{y^\intercal A y}{\alpha_\text{min}}} \leq \sqrt{\frac{2}{\beta \alpha_\text{min}}} \left \|r \right\|_E.
\end{equation}
We may also consider the matrix norm induced by the energy norm, which we also denote $\| \cdot\|_E$
\begin{equation}
    \|O\|_E = \max_{r\in \mathbb{R}^{2d}} \sqrt{\frac{\beta E(Or)}{\beta E(r)}}.
\end{equation}
It is instructive to look at the upper left $d \times d$ submatrix of $O$, which we call $O_{y}$. In particular, observe that
\begin{align}
     \|O\|_E &= \max_{r\in \mathbb{R}^{2d}} \sqrt{\frac{\beta E(Or)}{\beta E(r)}} \\
     & \geq \max_{y \in \mathbb{R}^{d}}\sqrt{\frac{(O_{y}y)^\intercal A (O_{y}y)}{y^\intercal A y}} \\
     & \geq 
      \max_{y \in \mathbb{R}^{d}}\sqrt{\frac{\alpha_\text{min}(O_{y}y)^\intercal (O_{y}y)}{\alpha_\text{max}y^\intercal  y}} \\
      & = \kappa^{-1/2} \|O_{y}\|,
\end{align}
so
\begin{equation}
    \|O_y\| \leq \sqrt{\kappa}\|O\|_E,
\end{equation}
a fact which will be used later. The earlier derivation shows that in the underdamped limit,
\begin{equation}
    E\left(e^{-\mathcal{A} t} r \right) \leq \frac{1+\xi/\omega_\text{min}}{1-\xi/\omega_\text{min}} e^{-2 \xi t} E(r),
\end{equation}
and so the operator exponential $e^{-\mathcal{A} t}$ is bounded in the energy norm
\begin{align}
    \left\|e^{-\mathcal{A} t}\right\|_E\leq  \sqrt{\frac{1+\xi/\omega_\text{min}}{1-\xi/\omega_\text{min}}}e^{- \xi t} \equiv \chi e^{-\xi t},
\end{align}
where we have defined
\begin{equation}
    \chi = \sqrt{\frac{1+\xi/\omega_\text{min}}{1-\xi/\omega_\text{min}}}.
\end{equation}

\subsection{Equilibration and Correlation Time}
We proceed in much the same way as in the overdamped case. We now have a bound on the energy norm of $e^{- \mathcal{A} t}$,
\begin{align}
    \left\|e^{-\mathcal{A} t}\right\|_E\leq  \chi e^{- \xi t}.
\end{align}
The mean of the distribution is governed by
\begin{equation}
    \braket{r(t)} = e^{-\mathcal{A} t}\braket{r(0)}.
\end{equation}
We get a bound on the Euclidean norm of $\braket{y}$
\begin{align}
\left\|\braket{y(t)}\right\| & \leq \sqrt{\frac{2}{\beta \alpha_\text{min}}} \left\|\braket{r(t)}\right\|_E \\
& \leq \sqrt{\frac{2}{\beta\alpha_\text{min}}}
\chi e^{- \xi t} \left\|\braket{r(0)} \right\|_E \\
& = 
\sqrt{\frac{2E(\braket{r(0)})}{\alpha_\text{min}}}\chi e^{-\xi t}\\
& = \sqrt{\frac{2\kappa E(\braket{r(0)})}{\|A\|}}\chi e^{-\xi t},
\end{align}
where $E_0$ is the energy of the state $\braket{r_0}$. Define the relative error tolerance as $\varepsilon_{\mu 0} > \|A x - b\|/\|b\| = \|A y\|/\|b\|$. Also assume that $\braket{x(0)} = \braket{p(0)} = 0$, so $E(\braket{r(0)}) = \frac{1}{2} b^\intercal A^{-1} b$. We then see that
\begin{align}
   \frac{\|Ay\|}{\|b\|} & \leq\frac{ \|A\|}{\|b\|} \sqrt{\frac{2\kappa E(\braket{r(0)})}{\|A\|}}\chi e^{-\xi t}\\
    &=  \sqrt{\frac{2\|A\|\kappa E(\braket{r(0)})}{\|b\|^2}}\chi e^{-\xi t}\\
    & \leq  \sqrt{\kappa}\chi e^{-\xi t}\\
\end{align}
In order to ensure that the relative error is less than $\varepsilon_{\mu 0}$ , we require
\begin{equation}
    t_0 \geq \frac{1}{\xi} \ln \left( \frac{\sqrt{\kappa} \chi}{\varepsilon_{\mu 0}}\right).
\end{equation}
Defining the underdamped relaxation time $\tau_\text{r(UD)}=1/\xi$, we have
which may be written
\begin{equation}
    t_0 \geq \tau_{\text{r(UD)}} \ln \left( \sqrt{\kappa}\chi \varepsilon_{\mu 0}^{-1}\right).
\end{equation}
As in the overdamped case, we proceed to determine the convergence rate of the (dimensionless) covariance matrix. The covariance matrix for $r$ obeys
\begin{align}
    \left\|\Sigma_\text{s} - \Sigma(t_0) \right\|_E
 & \leq 
  \left\| e^{-\mathcal{A} t}\right\|^2 \cdot \left\| \Sigma_0 - \Sigma_\text{s}\right\|_E+ \left\|\mathcal{B}\right\|_E\int_{t_0}^\infty dt' \left\| e^{-\mathcal{A} t'}\right\|_E^2 \\
  & \leq    \chi^2 e^{- 2\xi t_0 }\left\| \Sigma_0 - \Sigma_\text{s}\right\|_E+ \left\|\mathcal{B}\right\|_E\int_{t_0}^\infty dt'\chi^2 e^{-2\xi t'} \\
  & =  \chi^2 e^{- 2\xi t_0 }\left\| \Sigma_0 - \Sigma_\text{s}\right\|_E+\frac{\chi^2}{2\xi}e^{-2 \xi t_0} \left\|\mathcal{B}\right\|_E \\
  & = \chi^2 e^{- 2\xi t_0 }\left(\left\| \Sigma_0 - \Sigma_\text{s}\right\|_E+\frac{\gamma }{M \xi}\right)\\
  & =  \chi^2 e^{- 2\xi t_0 }\left(\left\| \Sigma_0 - \Sigma_\text{s}\right\|_E+2\right)
\end{align} 
Now we assume the initial covariance matrix in the dimensionless coordinates is $\Sigma_{0,r} = \gamma^2 M^{-1} \|A^{-1}\| \mathbb{I}\oplus \mathbb{I}$, which corresponds to a potential $V_0(x) = \|A\| x^\intercal x$. Therefore, $\Sigma_{\text{s},r} - \Sigma_{0,r} = \gamma^2 M^{-1}( A^{-1} - \|A^{-1}\|\mathbb{I}) \oplus 0$, which implies
\begin{align}
    \|\Sigma_{\text{s},r} - \Sigma_{0,r}\|_E &\leq \|\gamma^2 M^{-1} A^{-1}\oplus 0\|_E \\
    & =\gamma^2 M^{-1} \max_{v} \sqrt{\frac{v^\intercal A^{-1}AA^{-1} v}{v^\intercal A v}} \\
    & \leq \gamma^2 M^{-1} \|A\|^{-1} \kappa.
\end{align}
This implies that the spectral norm error of the covariance matrix for $\tilde{y}$ is bounded by
\begin{align}
    \left\|\Sigma_{\text{s},\tilde{y}} - \Sigma_{\tilde{y}}(t_0)\right\| &\leq
    \sqrt{\kappa}\chi^2 e^{- 2\xi t_0 }\left(\left\| \Sigma_0 - \Sigma_\text{s}\right\|_E+2\right)\\
    & \leq \sqrt{\kappa}\chi^2 e^{- 2\xi t_0 }\left(\gamma^2 M^{-1} \|A\|^{-1} \kappa+2\right)
\end{align}
which gives the corresponding result for the covariance matrix of $y$
\begin{equation}
    \left\|\Sigma_{\text{s},y} - \Sigma_{y}(t_0)\right\| \leq \beta^{-1}
   \sqrt{\kappa}\chi^2 e^{- 2\xi t_0 }\left(\kappa\|A\|^{-1}+M \gamma^{-2}\right).
\end{equation}
We want the relative error to be less than tolerance $\varepsilon_{\Sigma0}$
\begin{equation}
    \frac{\left\|\Sigma_{\text{s},y} - \Sigma_{y}(t_0)\right\|}{\beta^{-1} \|A^{-1}\|}\leq \varepsilon_{\Sigma0},
\end{equation}
which is satisfied if we require
\begin{equation}
    t_0 \geq \frac{1}{2} \tau_\text{r(UD)}\ln\left(\chi^2 \varepsilon_{\Sigma0}^{-1}\left[\kappa^{3/2}+\kappa^{1/2}M \gamma^{-2} \|A\|\right]\right),
\end{equation}
or
\begin{equation}
    t_0 \geq \frac{1}{2} \tau_\text{r(UD)}\ln\left(\chi^2 \kappa^{3/2}\varepsilon_{\Sigma0}^{-1}\left[\frac{1}{4\zeta_\text{max}^{2}}+1\right]\right).
\end{equation}
Combining the equilibration times for mean and variance gives
\begin{equation}
t_0 \geq \max \left\{ \tau_{\text{r(UD)}} \ln \left( \kappa^{1/2}\chi \varepsilon_{\mu 0}^{-1}\right),\frac{1}{2} \tau_\text{r(UD)}\ln\left(\chi^2 \kappa^{3/2}\varepsilon_{\Sigma0}^{-1}\left[\frac{1}{4\zeta_\text{max}^{2}}+1\right]\right)\right\}.
\end{equation}

\subsection{Ergodicity of Mean}
Once equilibrium has been reached, we may consider the correlation matrix of the stationary system
\begin{equation}
    G_\text{s}(\tau) = \lim_{t\to\infty}G(t,t+\tau),
\end{equation}
which is given by
\begin{equation}
    G_\text{s}(\tau) = e^{-\mathcal{A} \tau}\Sigma_\text{s}.
\end{equation}
We define the $A$-norm of a vector as $\|v\|_A ^2 = v^\intercal A v$, which induces an $A$-norm on operators. It clearly follows from the definitions that for all vectors $r = (u, v)^\intercal$, $\|u\|_A \leq \|r\|_E$, and similarly if $X_{y}$ is the upper left $d \times 2$ submatrix of a $2  d \times 2d$ matrix $X$, we have $\|X_{y}\|_A \leq \|X\|_E$.
Therefore
\begin{align}
    \|G_{\text{s},\tilde{y}}(\tau)\|_A &\leq \|G_\text{s}(\tau)\|_E\\
    & \leq \|e^{-\mathcal{A} t}\|_E \cdot \| \Sigma_\text{s}\|_E \\
    & \leq  \chi e^{-\xi \tau} \|\Sigma_\text{s}\|_E.
\end{align}
The quantity $\|\Sigma_\text{s}\|_E$ can be evaluated by the following arguments: after changing to the dimensionless coordinates, the covariance matrix is
\begin{equation}
    \Sigma_\text{s} = 
    \begin{pmatrix}
        \gamma^2 M^{-1} A^{-1} & 0\\
        0 & \mathbb{I}
    \end{pmatrix}.
\end{equation}
By definition,
\begin{align}
    \|\Sigma_\text{s}\|_E^2 &= \max_{y,p} \frac{ M \gamma^{-2}y^\intercal \Sigma_{\text{s},\tilde{y}} A \Sigma_{\text{s},\tilde{y}} y  + p^\intercal p}{M \gamma^{-2} y^\intercal A y + \ p^\intercal p} \\
    & = 
    \max_{y,p} \frac{ y^\intercal  A^{-1}y  + p^\intercal p}{M^2 \gamma^{-4} y^\intercal A y + \ p^\intercal p} \\
    & = 
    \max\left\{1, M^{-2}\gamma^{4}\|A^{-1}\|_A^2\right\}.
\end{align}
We then see that
\begin{equation}
    \|\Sigma_\text{s}\|_E = \max\{1, \gamma^{2} M^{-1} \alpha_\text{min}^{-1}\}.
\end{equation}
By the assumption that the system is in the underdamped regime, $2M \gamma^{-2}\alpha_\text{min}^{-1} >1$, so we finally obtain
\begin{equation}
    \|\Sigma_\text{s}\|_E = \gamma^{2}M^{-1}\alpha_\text{min}^{-1} =\gamma^{2}M^{-1}\kappa \|A\|^{-1}.
\end{equation}
By the same manipulations as in the overdamped case, we arrive at
\begin{align}
\left\|\braket{\bar{y}\bar{y}^\intercal}\right\| &\leq 
\frac{2 M\sqrt{\kappa}}{\tau^2\gamma^2 \beta}\int_{t_0}^{t_0 + \tau}dt''\int_{t_0}^{t''} dt' \left\|G_{\text{s}}(t''-t') \right\|_E\\
&\leq 
\frac{2M\sqrt{\kappa}}{\tau^2\gamma^2 \beta}\chi  \left\|\Sigma_\text{s}\right\|_E\int_{t_0}^{t_0 + \tau}dt''\int_{t_0}^{t''} dt' e^{-\xi(t''-t')}\\
&=
\frac{2\sqrt{\kappa}}{\xi\tau^2\beta\|A\|} \chi  \int_{t_0}^{t_0 + \tau}dt'' 1-e^{-\xi(t'' - t_0)} \\
& \leq 
 \frac{2 \sqrt{\kappa}}{\xi\tau^2 \beta \|A\|} \chi  \int_{t_0}^{t_0 + \tau}dt''1 \\
 &= 
 \frac{2\sqrt{\kappa} \chi \tau_{\text{r(UD)}}}{ \beta \|A\|\tau} 
\end{align}
The exact same reasoning employed in the overdamped case now gives the following statement: In order to have
\begin{equation}
    \left \|A \bar{x} - b \right\| \leq \varepsilon_x \|b\| 
\end{equation}
with probability at least $P_\varepsilon$, it is sufficient to require that
\begin{align}
    \tau \geq  \frac{2\sqrt{\kappa} \chi \|A\| d}{ \beta \|b\|^2\varepsilon_x^2(1-P_\varepsilon)} \tau_{\text{r(UD)}}.
\end{align}
We note that by the same reasoning as was applied in the overdamped case, we may state the above as an energy-time tradeoff
\begin{align}
    \mathcal{E}\tau \geq  \frac{2\sqrt{\kappa} \chi  d}{  \varepsilon_x^2(1-P_\varepsilon)} \beta^{-1}\tau_{\text{r(UD)}}.
\end{align}

\subsection{Ergodicity of Covariance Matrix}
Repeating the steps from before, we evaluate 
\begin{align}
    \sum_{ij}\text{var}(\overline{y_i y_j})
    &\leq \frac{2d(d+1)}{\tau^2}\int_{t_0}^{t_0 + \tau} dt'' \int_{t_0}^{t_0 + \tau} dt'\|G_{\text{s},y}(t'' - t')\|^2\\
    & =  \frac{4d(d+1)M^2\kappa}{\gamma^4 \beta^2 \tau^2}\int_{t_0}^{t_0 + \tau} dt'' \int_{t_0}^{t''} dt'\|G_\text{s}(t'' - t')\|^2 \\ 
    & \leq \frac{4d(d+1)M^2\kappa}{\gamma^4 \beta^2 \tau^2} \int_{t_0}^{t_0 + \tau} dt'' \int_{t_0}^{t''} dt'e^{-2\xi(t''-t')} \left\| \Sigma_\text{s}\right\|_E^2 \\
    &=
\frac{4d(d+1)\kappa^3}{\beta^2\|A\|^2 \tau^2 } \int_{t_0}^{t_0 + \tau} dt'' \int_{t_0}^{t''} dt'e^{-2\xi(t''-t')}  \\
    &=
\frac{4d(d+1)\kappa^3}{\beta^2  \|A\|^2\tau^2 \xi} \int_{t_0}^{t_0 + \tau}dt'' 1 - e^{-2\xi(t''-t_0)}\\ 
& \leq \frac{4d(d+1)\kappa^3}{\beta^2  \|A\|^2\tau \xi} \\
& = \frac{4d(d+1)\kappa \|\Sigma_{\text{s},y}\|^2\tau_{\text{r(UD)}}}{\tau }
\end{align}

Then repeating the same steps as in the overdamped, we find that in order to ensure that $\|\overline{y y^\intercal} - \Sigma_\text{s}\|_F/\|\Sigma_\text{s}\|_F \leq \varepsilon_\Sigma$, it is sufficient to require that
\begin{equation}
    \tau \geq \frac{4 \kappa d(d+1)}{(1-P_\varepsilon)\varepsilon_\Sigma^2}\tau_\text{r(UD)}.
\end{equation}

\subsection{Asymptotic scaling with condition  number}
The results derived for an underdamped system are only valid insofar as the system is really in the underdamped regime, i.e. when $\xi < \omega_\text{min}$, where $\xi = \gamma/ 2M$. However, if the parameters of the system are fixed,  then as $\kappa$ is increased (and the smallest eigenvalue $\alpha_\text{min}$ decreases), the system is eventually not in the underdamped regime anymore, and would begin to behave as an overdamped system. In this sense, the above results for an underdamped system cannot be interpreted as precisely ``asymptotic" in $\kappa$, as they are invalidated when $\kappa$ tends to infinity. Using the definitions $\omega_j = \sqrt{\alpha_j/M}$, and $\kappa = \|A\|/\alpha_\text{min}$, we see that the underdamped treatment is valid when
\begin{equation}
    \frac{\gamma}{2M}< \sqrt{\frac{\|A\|}{\kappa M}},
\end{equation}
so
\begin{equation}
    M > \frac{\gamma^2 \kappa}{4\|A\|}
\end{equation}
Then, because $\tau_\text{r(UD)} = 2 M / \gamma$, we have
\begin{equation}
    \tau_\text{r(UD)}>   \frac{\gamma \kappa}{2\|A\|}.
\end{equation}
Therefore, as the averaging time $\tau_\text{r(UD)}$ is proportional to $\gamma \kappa$, we include a factor of $\kappa \gamma$ in the expressions for asymptotic time-complexity in the underdamped case.
\section{Determinant Estimation}
Suppose we are given a series of work measurements $W_1 \dots W_N$ resulting from changing a potential from $U_1$ to $U_2$. Importantly, the system must be allowed to come to equilibrium once per measurement (so $N$ times in total). It has been shown, via the Cramér-Rao bound \cite{cramer1999mathematical}, that if $\Delta \hat{F}$ is \emph{any} asymptotically unbiased estimator of the difference in equilibrium free energies between the two potentials, then its variance is bounded below \cite{shenfeld2009minimizing}
\begin{equation}
    \text{var}\left(\Delta \hat{F} \right)\geq \frac{\mathcal{L}^2}{N},
\end{equation}
where $\mathcal{L}$ is the Fisher-Rao length of the path in parameter space between the two equilibrium states. Explicitly, the Fisher-Rao length of a path through the space of probability distributions is
\begin{equation}
    \mathcal{L} = \int_0^\tau dt \sqrt{\int dx \frac{\dot{f}_t(x)^2}{f_t(x)}}.
\end{equation}
For Gaussian distributions sharing the same mean, an explicit formula has been found for the minimal Fisher-Rao length between two distributions. In particular, if one of the Gaussian distributions has covariance matrix $\beta^{-1}A^{-1}$ and the other has covariance matrix $\beta^{-1} \|A^{-1}\|\mathbb{I}$, and both share the same mean, then the minimal Fisher-Rao distance is given by \cite{nielsen2023simple}
\begin{equation}
  \mathcal{L}_{\text{min}}  = \sqrt{\frac{1}{2} \sum_{j=1} \ln( \lambda_j(A\|A\|^{-1}))^2},
\end{equation}
which is bounded by
\begin{align}
  \mathcal{L}_{\text{min}} &\leq  \sqrt{\frac{1}{2} d \left|\ln( \kappa^{-1})\right|^2} \\
  &\leq  \sqrt{\frac{1}{2} d \ln( \kappa)^2} \\
  &=\sqrt{\frac{d}{2}}\ln(\kappa).
\end{align}
Therefore the minimal variance of an unbiased estimator is
\begin{equation}
    \text{var}\left(\Delta \hat{F} \right)\geq \frac{d\,\ln(\kappa)^2}{2N \beta^2},
\end{equation}
Moreover, this lower bound can be attained in the limit of large $N$ by the Bennett Acceptance Ratio (BAR) estimator or its multistate generalization (MBAR). The free energy difference between equilibrium states of the device is
\begin{equation}
    \Delta F = F_2 - F_1 = \frac{1}{2}(b_1^\intercal A_1^{-1}b_1-b_2^\intercal A_2^{-1}b_2) + \frac{1}{2\beta} \ln \left( \frac{\left|A_2\right|}{\left|A_1\right|}\right).
\end{equation}
Apparently, if $b_1 = b_2 = 0$ and $A_1 = a_1 \mathbb{I} $ then
\begin{equation}
    \ln\left(\left|A_2\right|\right)= 2\beta \Delta F +  d\, \ln(a_1). 
\end{equation}
Therefore the variance of an unbiased estimator of the log determinant $\ln(\left|A_2\right|)$ is lower bounded as follows
\begin{equation}
    \text{var}(\widehat{\text{LD}}(A_2)) \geq \frac{2 d\,\ln(\kappa)^2}{N}
\end{equation}
where we write $\text{LD}$ for the log determinant and $\widehat{\text{LD}}$ for the estimator of the log determinant. To achieve absolute error of at most $\delta_\text{LD}$ with probability $P_\delta$, according to Chebyshev's inequality we need a number of samples given by
\begin{equation}
    \text{var}(\widehat{\text{LD}}(A_2)) \leq \delta_\text{LD}^2 (1-P_\delta),
\end{equation}
which can be guaranteed (assuming the estimator is optimal) by requiring
\begin{equation}
    N> \frac{2 d\, \ln(\kappa)^2}{\delta^2_\text{LD}(1-P_\delta)}.
\end{equation}
As the equilibration does not require a change of the mean of the distribution but just the covariance matrix, the equilibration time is, in the overdamped case
\begin{equation}
\frac{1}{2} \kappa \tau_\text{r}\ln\left(2\kappa\varepsilon_{\Sigma 0}^{-1}\right),
\end{equation}
for some sufficiently small $\varepsilon_{\Sigma 0}$. The total time of the optimal protocol would then be approximated by
\begin{equation}
\tau \approx \frac{ d\, \kappa \ln(\kappa)^2\ln(2 \kappa\varepsilon^{-1}_{\Sigma0})}{\delta^2_\text{LD}(1-P_\delta)} \tau_\text{r} = O(d \, \kappa \ln(\kappa)^3).
\end{equation}
In the underdamped case, the equilibration time would be
\begin{equation}
\frac{1}{2} \tau_\text{r(UD)}\ln\left(\chi^2 \kappa^{3/2}\varepsilon_{\Sigma0}^{-1}\left[\frac{1}{4\zeta_\text{max}^{2}}+1\right]\right),
\end{equation}
for a total time of
\begin{equation}
\tau \approx \frac{ d\,\ln(\kappa)^2}{\delta^2_\text{LD}(1-P_\delta)} \ln\left(\chi^2 \kappa^{3/2}\varepsilon_{\Sigma0}^{-1}\left[\frac{1}{4\zeta_\text{max}^{2}}+1\right]\right)
\tau_\text{r(UD)} = O(d \, \ln(\kappa)^3).
\end{equation}

\begin{figure}
    \centering
    \includegraphics[width=0.5\linewidth]{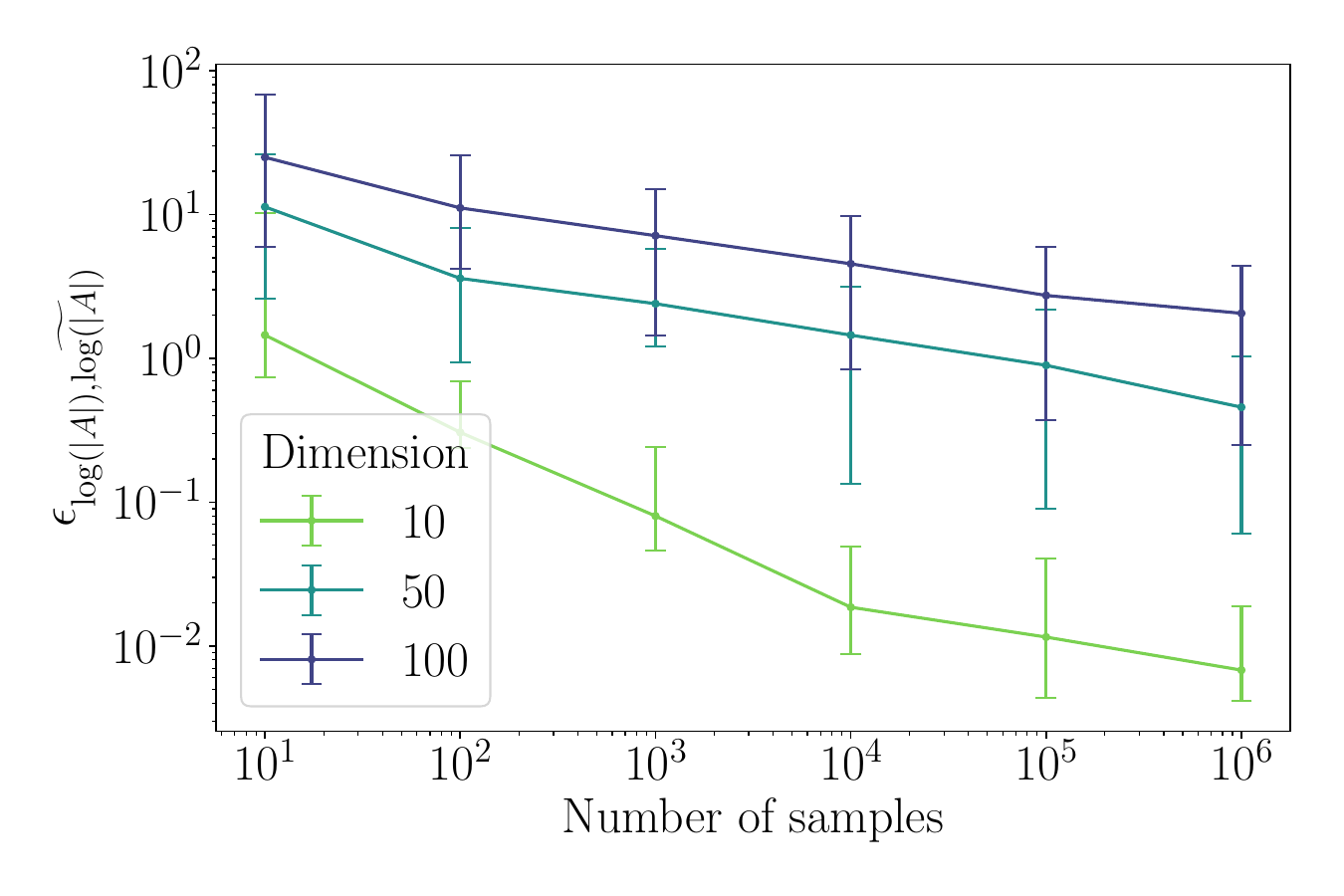}
    \caption{\textbf{Convergence of determinant estimation using the Jarzynski estimator.} This shows relative error of the determinant estimate with increasing numbers of samples used in the Jarzynski estimator. Solid lines show the median and the error bars show the interquartile range over 10 different $A$ matrix realizations.}
    \label{fig:det}
\end{figure}

\subsection{Numerical Results}\label{app:det_numerics}

To explore the performance of our thermodynamic algorithm for estimating the log determinant of a matrix we numerically explore using a Jarzynski estimator of the free energy difference in the simplest case. Here the potential is initialized using an identity matrix and this is instantaneously switched to the matrix for which we desire to evaluate the log determinant. We explore the performance of the estimated log determinant when using a Jarzynski estimator with different numbers of samples. Recall the free energy can be estimated using
\begin{equation}
    e^{-\beta \Delta F}\approx \overline{e^{-\beta W}}  \equiv \frac{1}{N}\sum_{j=1}^N e^{-\beta W_j}.
\end{equation}
 This in turn is then used to estimate $\ln\left( \left| A_2\right|\right)$ with the relation
\begin{equation}
\ln\left(\left| A_1\right|\right) \approx 2 \ln\left(\overline{e^{-\beta W}}\right)  + \ln\left( \left| A_2\right|\right).
\end{equation}
In Fig.~\ref{fig:det} we show the relative error $\epsilon_{\log(|A|), \widetilde{\log(|A|)}}$ of the log determinant obtained for several dimensions, estimated with increasing numbers of samples. This shows estimating the determinant in this fashion is a valid protocol.

\section{Thermodynamic Algorithm for Solving the Lyapunov Equation}

In this section we present our general algorithm for solving the Lyapunov Equation. Note that our matrix inversion algorithm from the main text is a special case of our general Lyapunov Equation algorithm. 

In what follows, we assume we have access to a device with a controllable noise source such that the covariance matrix of the noise term may be chosen to be an arbitrary symmetric positive definite matrix. We do not include the linear $b^\intercal x$ term in the potential, and therefore obtain the following overdamped Langevin equation
\begin{equation}
\label{eq:odl-lyapunov}
    dx = -\frac{1}{\gamma}A x dt + \mathcal{N}\left[0, \frac{2}{\gamma \beta } R\, dt\right],
\end{equation}
where $R$ is symmetric and positive definite.
In this case, the stationary distribution has mean zero and covariance matrix $\Sigma_\text{s}$, which is a solution to the Lyapunov equation
 \begin{equation}\label{eq:Lyapunov}
     A\Sigma_\text{s}+ \Sigma_\text{s} A^{\intercal} = 2\beta^{-1}R.
 \end{equation}
We propose the following protocol for solving the Lyapunov equation.
\medskip
\begin{tcolorbox}[title={Lyapunov Equation Protocol},breakable]
\begin{enumerate}
\item Given two symmetric positive definite matrices $A$ and $R$, set the potential of the device to
\begin{equation}
    U(x) = \frac{1}{2} x^\intercal A x,
\end{equation}
and the noise term in the overdamped Langevin equation to $\mathcal{N}\left[0,2\gamma^{-1}\beta^{-1} R dt\right]$ at time $t=0$. That is, the system evolves under the dynamics of Eq. \eqref{eq:odl-lyapunov}.

\item Choose equilibration tolerance parameter $\varepsilon_{\Sigma 0} \in \mathbb{R}^+$, and choose the equilibration time
\begin{equation}
\label{eq:UDL-eq-time}
t_0 \geq \widehat{t}_0,
\end{equation}
where $\widehat{t}_0$ is computed from the system's physical properties, Eq.~\eqref{eq:t0-tau-lyapunov-odl}. Allow the system to evolve under its dynamics until $t=t_0$, which ensures that $\left\|\Sigma - \beta^{-1} A^{-1} b\right\|/\|\beta^{-1}A^{-1}\| \leq \varepsilon_\Sigma$.
\item  Choose error tolerance parameter $\delta_\Sigma$ and success probability $P_\delta$, and choose the integration time
\begin{equation}
    \tau \geq \widehat{\tau},
\end{equation}
where $\widehat{\tau}$ is computed from the system's physical properties, Eq.~\eqref{eq:t0-tau-lyapunov-odl}. Use analog multipliers and integrators to measure the the time averages
\begin{equation}
    \overline{x_i x_j} = \frac{1}{\tau^2} \int_{t_0}^\tau dt\, x_i(t) x_j(t),
\end{equation}
which satisfies $\|\overline{x x^\intercal} - \Sigma_\text{s}\|_F \leq \delta_\Sigma$ with probability at least $P_\delta$.
\end{enumerate}
\end{tcolorbox}
\medskip

The timing parameters for the Lyapunov equation protocol are, for the overdamped case,
\begin{align}
\label{eq:t0-tau-lyapunov-odl}
\widehat{t}_0 =
\frac{1}{2}\kappa\tau_\text{r}\ln\left(\frac{\|\Sigma_0 - \Sigma_\text{s}\| + \kappa \beta^{-1} \|A\|^{-1}
}{\delta_{\Sigma 0}}\right),\quad
    \widehat{\tau} = \frac{4 \kappa d(d+1)}{(1-P_\varepsilon)\varepsilon_\Sigma^2}\tau_\text{r},
\end{align}
For the underdamped case, $\widehat{t_0}$ would be somewhat different, but $\widehat{\tau}$ would be the same, because the behavior of the equilibrium correlation function does not depend on the noise, so the same result derived for the matrix inverse protocol is applicable. Note that Eq. \eqref{eq:t0-tau-lyapunov-odl} only vaguely determines the equilibration time, as the target covariance matrix $\Sigma_\text{s}$ is not known beforehand. The corresponding equilibration time $\widehat{t_0}$ for an underdamped system could also be evaluated in principle; however, this would only result in a similarly vague expression, which is anyway not necessary to determine the asymptotic time-complexity scaling of the algorithm, so it is not pursued here. Moreover, the relative error cannot be bounded as straightforwardly as was done for the other protocols given that there is no explicit formula for the target covariance matrix. For this reason, we have used absolute error as the error tolerance in the above protocol.

\end{appendix}

\end{document}